\documentclass{aa}

\usepackage{graphicx}
\usepackage{txfonts}
\usepackage{subfigure}
\usepackage[nonamebreak]{natbib}
\bibpunct{(}{)}{;}{a}{}{,} 
\usepackage{longtable}

\newcommand{\ie}{\textit{i.e. }}

\newcommand{\eg}{\textit{e.g. }}

\newcommand{\ha}{\mbox{H$\alpha$}}
\newcommand{\hb}{\mbox{H$\beta$}}
\newcommand{\hd}{\mbox{H$\delta$}}
\newcommand{\oiii}{\mbox{[O{\sc iii}]}}
\newcommand{\oii}{\mbox{[O{\sc ii}]}}
\newcommand{\nii}{\mbox{[N{\sc ii}]}}

\newcommand{\Rv}{\ensuremath{R_{virial}}}
\newcommand{\Den}{\ensuremath{\Sigma_5}}
\newcommand{\avg}[1]{\ensuremath{\langle #1 \rangle}}
\newcommand{\ew}[1]{\ensuremath{W_0(#1)}}

\newcommand{\rdos}{\ensuremath{r_{200}}}

\begin{document}
  \title{The galaxy populations from the centers to the infall regions in $z\approx 0.25$ clusters\thanks{Table \ref{T:Objects}
is only available in electronic form at the CDS via anonymous ftp to cdsarc.u-strasbg.fr  (130.79.128.5) or via \newline http://cdsweb.u-strasbg.fr/cgi-bin/qcat?J/A+A/ }}

  \author{M. Verdugo\inst{1},  B. L. Ziegler\inst{1}\inst{,2}\inst{,3}\thanks{Visiting astronomer of the 
German-Spanish Astronomical Center, Calar Alto, operated by the Max-Planck-Institut f\"ur Astronomie,
Heidelberg, jointly with the Spanish National Commission for Astronomy.} \and B. Gerken\inst{4}$^{\star\star}$
  }
  
  \offprints{M. Verdugo,\\ mverdugo@astro.physik.uni-goettingen.de}
  
  \institute{Institut f\"ur Astrophysik G\"ottingen, 
    Georg-August Universit\"at G\"ottingen,
    Friedrich-Hund-Platz 1, 37077, G\"ottingen, Germany\\
    \email{mverdugo@astro.physik.uni-goettingen.de}
  \and 
European Southern Observatory, Karl-Schwarzschild-Strasse 2,
85748, Garching bei Muenchen, Germany
\and
 Argelander-Institut f\"ur Astronomie, Universit\"at Bonn
  \and
  Oxford Astrophysics, Department of Physics,  University of Oxford, Keble Road, Oxford, OX1 3RH, UK.
}
  \date{}

  \abstract 
{In the local universe, the relative fractions of galaxy types differs in galaxy clusters in comparison to the field. Observations at higher redshift provide evidence that cluster galaxies evolve with lookback time. This  could be due either to the late assembly of clusters, which is predicted by bottom-up scenarios of structure formation, or to cluster-specific interaction processes.}
{To disentangle various effects, we explore the evolutionary status of galaxies from 
the center of clusters out to their infall regions in  $z\approx0.25$ clusters.}
{We conducted a panoramic spectroscopic campaign with MOSCA at the Calar Alto observatory.
We acquired low-resolution  spectra of more than 500 objects. Approximately 150 
of these spectra were of galaxies that are members of six different clusters, which differ in intrinsic X-ray luminosity.
The wavelength range allows us to  quantify the star formation activity by using the \oii\ and the \ha\ emission lines. This activity 
is examined in terms of the large-scale environment expressed by the clustercentric distance of the galaxies 
as well as on local scales given by the spatial galaxy densities.}
{The general decline in star-formation activity  observed for galaxies inside nearby clusters is also seen 
at $z\approx0.25$. A global suppression of star-formation is detected in the 
outskirts of clusters, at about $3\Rv$, where the galaxy densities are low and the  intra-cluster  
medium is very shallow.  Galaxies with ongoing star-formation have similar activity, 
regardless of the environment. Therefore,  the decline of the star-formation activity inside the investigated clusters 
is driven mainly  by the significant change in the fraction of active versus passive populations. This suggests
 that the  suppression of the star-formation activity occurs on short timescales.   
We  detect a significant population of red star-forming galaxies whose colors are consistent with the red-sequence of 
 passive galaxies. They appear to  be in an intermediate  evolutionary stage between active and passive types.} 
{Since a suppression of  star-formation activity is measured at large clustercentric distances and low 
projected densities, purely cluster-specific phenomena cannot fully explain the observed trends. Therefore, as 
suggested by other studies, group preprocessing may play an important role in transforming galaxies
before they enter into the cluster environment. Since models predict that a significant fraction of galaxies observed in
 the outskirts may have already transversed through the cluster center and intracluster media, the effects of ram-pressure
stripping cannot, however, be neglected; this is, in addition, true because ram-pressure stripping may even be effective, 
under certain conditions, inside group environments.}

   \keywords{galaxies: clusters: general -- galaxies: evolution --  galaxies: stellar content -- galaxies: distances and redshifts}

   
   \authorrunning{Verdugo et al.}
    \titlerunning{Galaxy populations in the infalling regions of $z\approx 0.25$ clusters}
   
   \maketitle

   
\section{Introduction} 
\label{S:intro}
  
The study of the galaxy population inside clusters dates back to \citet{Hubble1936}, 
who noted that cluster of  galaxies  are  dominated  by elliptical and lenticular galaxies, 
and the surrounding field by spirals. Several modern studies have quantified this effect 
(\eg \citealt{Dressler1980,Goto2003}), which is now known as the morphology-density relation. 
It has been suggested that spiral galaxies are being transformed into S0s by cluster-specific
processes. Further evidence is provided by \citet{dressler97}, who noted that the fraction of S0 
galaxies decrease strongly at moderate redshifts  with spiral galaxies filling the gap.

Since galaxy types correlate strongly with spectral properties, similar behaviors  have been found. 
 for colour and spectroscopic data.   For example,  \citep{BO78} noted an increase in the blue-galaxy 
fraction  inside clusters at intermediate  redshifts. This result has been confirmed by many subsequent  studies 
(\eg \citealt{kodamabower01,Ellingson2001}).  Also colors are  a strong function of environment 
(\citealt{Balogh2004B}) and according to the SDSS-based studies of \citet{Hogg2003} and \citet{Blanton2005}  
broad-band colors correlate more strongly with environment than morphology,
breaking in part the degenerate effect of  different physical properties 
and indicating that the processes that change the stellar population properties are acting 
on different timescales than those that transform the galactic structure. 

However, colors can be unreliable indicators of current star formation. Galaxies may have already 
shut down their star-formation activity  and still show blue colors as evidence  of  previous activity 
(\eg \citealt{Kauffmann1996,Ellingson2001}). Although,   models 
predict that when a galaxy quenches its star-formation it  moves
onto the red-sequence quite rapidly ($\sim$400\,Myr, \citealt{Harker2006}). 
Evidence of  this is provided by the strong  bimodality  observed in galaxy colors 
(\eg \citealt{Balogh2004B}), which cannot be simply explained otherwise.

The question about the environmental dependence of  galaxy  physical properties  can be 
addressed  by studies that use   more reliable indicators, such as emission lines. 
Those studies find strong correlations between  star-formation  activity and galaxy environment 
(\eg  \citealt{balogh99,lewis02,Gomez2003,Pimbblet2006,Haines2007}). 
Furthermore, these relations do not appear to depend on the 
mass of the system in which  the galaxies are  embedded  (\citealt{popesso2007}).

Since  the hierarchical mass assembly with time is a natural prediction of  
$\Lambda$CDM cosmologies,  it is obvious to link the decline of the volume-averaged 
star-formation rate  (\citealt{Hopkins2004} and references therein) and the galaxy evolution in general 
to the growth of  structure. However, the  relative importance  of the different  processes that act, is not yet clear. 

Observations suggest that, at least, two  different phenomena are  required. 
One  process acts on the stellar populations to terminate the star-formation activity and another 
process changes the galaxy structure. Ram-pressure stripping (\eg \citealt{quilis2000}) is known to be 
very effective in removing the galaxy cold gas and thus quenching the star-formation activity, but only 
works under  special conditions  present in cluster cores where the intra-cluster 
gas density and the relative galaxy velocities are high. The softer variant of ram-pressure stripping, 
strangulation or starvation  (\eg \citealt{Bekki2002}),  removes  the thin gaseous halo present around 
galaxies, and the star-formation continues until the remaining disk gas is consumed.  

Other possible mechanisms are galaxy-galaxy merging and low-velocity galaxy interactions 
that trigger an episode of high star-formation,  which consumes  a high fraction of  gas in a 
short time and may stripp the remaining via gravitational shocks and feedback processes 
(\eg \citealt{Larson1978,Bekki2001b}). This may provide  explanation 
to modern observations where the  decrease of star formation activity has been detected already 
at very low galaxy densities (\citealt{lewis02,Gomez2003}).
 However, other mechanisms are necessary to explain the change in morphology. 
Mergers are  known  to be  efficient in changing late-type galaxies into  ellipticals 
(\citealt{Toomre1977,Hernquist1992}), but the relative velocities  must be low, which 
is not the case in clusters. But the galaxy structure can be changed 
on  longer timescales by  harassment \citep{moore98} due to high-velocity encounters  
between cluster galaxies (see also \citealt{Gnedin2003}).

Despite  the accumulation of observational evidence over the  years, the link 
between the growth of  structure with time and galaxy evolution remains elusive and 
the fundamental questions  remain unanswered. How rapidly and significantly is 
supressed the star-formation activity in infalling galaxies? What  exactly is the  environmental 
dependence of the star-formation activity? Is it suppressed  mainly due 
to local or global processes? What is the predominant mechanism?

Studying clusters at  higher redshift may provide new clues about the processes  involved, 
because the global star-formation activity was higher in the past and 
clusters show  at all redshift much lower  activity when compared with the surrounding field  
(\eg \citealt{balogh99}). Models also predict that in the past the galaxy-infalling rate must have been higher  (\eg \citealt{Bower1991}).  The processes at work must therefore have been increasingly more effective 
at increasingly higher redshift, and at higher redshift the probability of observing the processes in action, increases.

Several studies at higher redshift have  focused on the central parts of clusters 
 (\eg \citealt{balogh99,balogh02,Poggianti2006}), but, as studies at $z\approx0$
show, the relation between star-formation activity and density is already discernible
at  low galaxy densities, inside the infalling  regions where the galaxies, which are infalling from the field, 
may begin to experience the influence of  cluster, and  interactions become more frequent. 

Even in the distant universe, clusters of galaxies project a large solid angle, and wide-field 
observations are therefore required. The contamination due to foreground and 
background objects is larger, 

We report the results of a project   to study   galaxy evolution
from the  infalling regions to the cluster centers,  covering projected radial distances out to 
4  virial radii for six clusters at  $\avg{z}\approx 0.25$. First 
results for  two clusters were already published by  \citet{gerken04}. In Sect. \ref{S:obs}
we describe the observations as well as the method used to measure the important 
parameters of the galaxies.  In Sect. \ref{S:clusters} we describe cluster identification
and other general properties  including the environmental definition.  
In Sect. \ref{S:fields} we describe in detail each observed field. In Sect.  \ref{S:results} we show 
the main results,discussing their implications  in Sect. \ref{S:origin}. In 
Sect. \ref{S:redSF} we explore some properties of the star-forming population. 
Our summary and conclusions are provided in Sect. \ref{S:conclusions}.

Throughout this paper, we use a cosmology of $H_0=70$\,km~s$^{-1}$Mpc$^{-1}$, 
$\Omega_m=0.3$ and $\Omega_\Lambda=0.7$.

   \begin{table*}[t]
     \begin{minipage}[t]{\textwidth}
       \begin{center}
        \caption{Main parameters for the cluster sample. The cluster denominations  come from \citet{vik98} (VMF) 
and \citet{gilbank04} (XDCS).  Coordinates are  given with respect   to the X-ray centroid.  X-ray fluxes are taken from
 \citet{mullis03}.  \Rv\ is the virial radius and $\sigma$ the velocity dispersion.  $N$ is the number of  members
 identified  in each cluster.}
	 \begin{tabular}{l l l l l l l l l l l}
	 
	   & & & & & & & & & \\ \hline\hline 
	   Field  & Cluster & Alternative name & RA & DEC & $z$  & f$_X$ &  L$_{X,bol}$  & $\sigma$ & \Rv & $N$\\
	              &       &          &               &       &   &  [10$^{-14}$ ergs/s cm$^2$] & [10$^{43}$ ergs/s] & [km/s] & [Mpc]  &  \\
	   \hline
R220 & VMF194  & RX\,J1729.0+7440 & 17:29:02 & 74:40:46  & 0.210 & 17.3 & 5.01 & 282$\pm$52  & 0.742 & 8  \\
     & XDCS220 & cmJ172333+744410\footnote{\citet{gilbank04}} & 17:23:33 & 74:44:10 & 0.261 & 0.3 & 0.14 & 621$\pm$271\footnote{This  value is likely overestimated, since the cluster does not seem to be in dynamical equilibrium, see text for details}  & 1.535 & 15 \\
R265  & VMF131  &  RX\, J1309.9+3222 & 13:09:56 & 32:22:31 & 0.294 & 9.0  & 6.03 & 476$\pm$110 & 1.132 & 29 \\
      & VMF132  &  RX\, J1313.2+3229 & 13:11:13 & 32:28:58 & 0.247 & 46.7 & 24.5 & 774$\pm$150 & 1.945 & 18\\
R285  & VMF73   &  RX\, J0943.5+1640 & 09:43:32 & 16:40:02   & 0.254 & 23.1 & 12.3 & 661$\pm$65  & 1.647 & 44 \\
      & VMF74   &  RX\, J0943.7+1644 & 09:43:45 & 16:44:20  & 0.180 & 21.2 & 4.79 & 481$\pm$79  & 1.313 & 34 \\
	   \hline\\
	 \end{tabular}
	 \label{T:maindata}
       \end{center}
     \end{minipage}
   \end{table*}
   
   \section{The data}
   \label{S:obs}
   
   \subsection{Cluster selection}
   \label{SS:sel}
   
The sample was selected from the X-ray Dark Cluster Survey   (XDCS, \citealt{gilbank04}) 
whose aim  was to compare X-ray and optical identification  algorithms of clusters. 
For this purpose, deep, optical imaging of RIXOS fields (\citealt{Mason2000}) was acquired, 
which were imaged in the X-ray by the ROSAT Position Sensitive Proportional  Counter (PSPC). 
Some of the X-ray data were also analyzed  by  \citet{vik98},  and later by 
\citet{mullis03}, from which the  X-ray fluxes were taken. 
The XDCS provides us with $V$ and $I$-band photometry taken 
with the Wide Field Camera (WFC) at the Isaac Newton telescope (La Palma, Spain). 
This camera has a field of view (FOV) of  34$\times$34\,arcmin.

We selected for follow-up spectroscopy three fields containing, in projection, two clusters each, thus 
increasing the probability of targeting a cluster member. The clusters have a wide range of
 X-ray luminosities  and probably different evolutionary states.  They are at similar redshifts, 
making them good candidates to probe evolution uniquely due to environmental effects at a 
cosmological epoch with look-back times of $\sim$3.0\,Gyr.

A summary of the cluster properties can be found in Table \ref{T:maindata}. Details  of how 
the different quantities were  calculated are described in the forthcoming  sections.

   \subsection{Observations}
   \label{SS:obs}

The spectroscopy was performed with the multi-object spectrograph  
MOSCA mounted at the 3.5\,meters telescope at Calar Alto 
Observatory\footnote{CAHA, Centro Astronomico Hispano Alem\'an.} (Spain).
These observations were carried out in two runs, from  
10 to 15 February and 20 to 24 March, 2002. Each field  was 
observed using  7--8 slit masks, each covering $\sim$11$\times$11\,arcmin FOV, 
therefore the original  WFC fields are adequately covered by the spectroscopic observations. 
 Each MOS mask contains 20 to 30 slits of $\sim$25\,arcsec of length   
to subtract the sky accurately. 

We used the low-resolution grism {\sc green\_\small{500}}, which encompasses a wide  
wavelength range, from 4300 \AA\ to 8200 \AA, allowing us to study both the \oii$\lambda$3727 
and the  \ha\ emission lines, at the targeted redshifts, which are  critical to study star-formation 
activity  in galaxies. The grism provides a spectral resolution of  $R\sim10-15$\,\AA, which corresponds
in the rest-frame to $8-12$\,\AA, for our slit width of 1\,arcsec.

The exposure times ranged between one  and three hours depending on the 
apparent magnitudes of the objects selected. In a few cases these   times were increased 
to account for  variations in the weather.  The magnitude distribution of   the final sample
in Fig. \ref{F:selfunc}.  The selection of  objects for  spectroscopy was based only on 
their $I$-band magnitudes to avoid any  color bias.  Additional restrictions were imposed by 
 masks geometry.

In  total, 537 spectra were acquired. For our analysis, we included in addition 21 spectra from our 
previous  projects ``Low X-ray luminosity clusters''   \citep{balogh02} in the R265 field and 
``X-dark cluster survey'' \citep{gilbank04}  in the R220 field. This was possible because 
those spectra were observed using a similar  instrumental setup. However, we reexamined  
all spectra  to be able to apply the same criteria for the whole sample. Finally, we found that  
297 spectra were suitable for analysis (see below for the precise criteria used).


\subsection{Data reduction}
   \label{SS:datared}

Our data reduction procedures were described in \citet{gerken04} and can be summarized 
by the following steps: Bias subtraction, extraction of  individual  slits, correction of 
the distortion induced by   the focal reducer  in MOSCA, flat-fielding, sky subtraction,  
extraction of the one-dimensional spectra, wavelength calibration, and combination 
of the individual exposures.


All of these tasks were performed within the {\sc midas}\footnote{{\sc midas}, the 
Munich Image Data Analysis  System is developed and maintained by the  European Southern 
Observatory (ESO)} environment, interactively, using custom-made routines. Each spectrum 
was visually-inspected  to detect  pecularities that may affect the measurements.


  \subsection{Individual redshift determination}
   \label{SS:redshift}

Individual galaxy redshifts were determined by  fitting a Gaussian profile to a  set of prominent emission 
and absorption lines (\oii$\lambda$3727, CaK, G-band, \hb, \oiii$\lambda$5007, Mgb, Fe5335, NaD and \ha).  
For each galaxy, we defined the galaxy redshift to be the mean value of the individual line redshifts;  we note 
that not all lines were always visible in each galaxy spectrum.  
The redshift  error was the standard deviation of  the redshift determined from  at least 
four clearly-identifiable lines. 

We  assigned to each spectrum a number representative of its quality,  based on how clearly 
the lines could be seen  in comparison to the continuum noise, how many lines were visible or 
whether the lines were contaminated by artifacts.  Spectra designated  with a quality 0 (zero) 
were of the highest quality and those with 7 (seven) the poorest. Spectra of a  quality  value of  above 
3 (three) were, in general, considered not trustworthy and were excluded from our final analysis.

Redshifts and other parameters for individual  galaxies are provided in 
Table \ref{T:Objects}, which is only available \emph{online}.

   \subsection{Quality  and completeness}
   \label{SS:QC}

We recognize that to assign a quality number based on  eye perception may be highly subjective. 
The main risk is an  over-representation of star-forming galaxies, since emission lines are    
easily visible and identifiable. Those galaxies have a greater chance to be included in the final sample, 
although they can be systematically fainter  than passive galaxies. To test for a presence a a bias, 
an accurate estimation of the  continuum noise is required.

Each spectrum  was normalized  by a polynomial fit to the continuum, in the range of interest, 
from \oii\ to \ha. In the normalized spectra, the standard deviation was calculated, using a  
3-$\sigma$ clipping  algorithm over five iterations, as an estimation of the continuum noise.

\begin{figure}[t]
     \centering
     \includegraphics[bb=25 160 350 400,width=0.48\textwidth,clip]{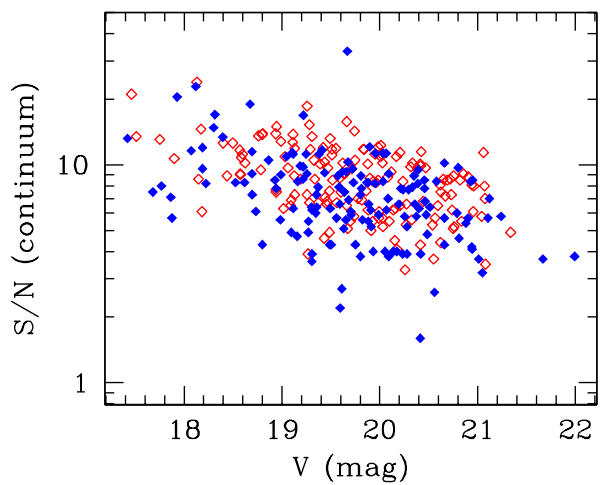}
     \caption{$V$-band apparent magnitude versus continuum signal-to-noise ratio   
as measured in Sect. \ref{SS:QC}. Open red diamonds  are galaxies without emission lines, 
whereas filled blue diamonds are galaxies with at least one emission line.}
     \label{F:signal2noise}
     \end{figure}

The algorithm used to fit the polynomial ignored emission lines and other  small-scale prominent 
features, such as sky-line residuals and telluric lines.  We present in Fig.  \ref{F:signal2noise} 
our measurements of continuum noise as a function of V-band apparent magnitude, which is a
 measure of the total flux. Although the selection of  the objects for observing was  made using 
$I$-band   magnitudes,  the $V$-band  magnitudes provide a more accurate measure of the galaxy  
continuum in the spectroscopic wavelength range of interest  and provides a good estimation of the total flux.  
No significant difference  in the distribution of star-forming versus passive galaxies was observed, with the exception
of two faint  star-forming galaxies.

  \begin{figure}[t]
     \centering
     \includegraphics[bb=25 170 390 420,width=0.5\textwidth,clip]{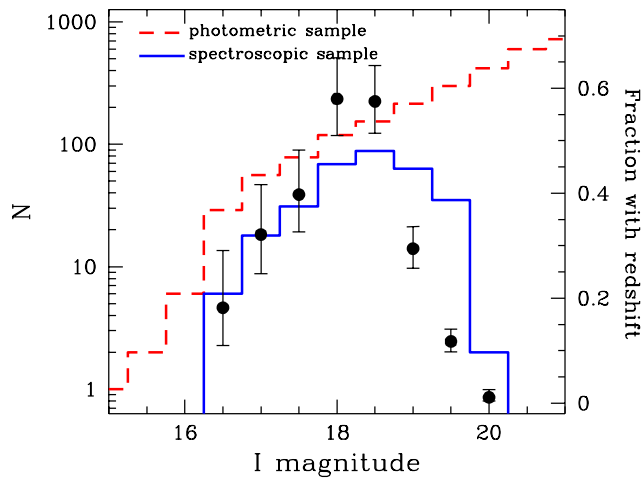}
     \caption{The combined selection function for the whole sample. The histograms    
    show the $I$-band magnitude distribution   for the photometric (dashed red line)   
    and spectroscopic (blue solid line).  The points show the fraction of galaxies for which 
we derived redshifts.  The error  bars are Poisson distributed errors (\citealt{gehrels86}).}
     \label{F:selfunc}
   \end{figure}

\subsection{Selection function}
\label{SS:selfunc}

In all fields, only a fraction of the galaxies below our spectroscopic limit  ($I\approx19.5$\,mag) was observed. 
Therefore, selection effects may be present and  need to be corrected. This is achieved by constructing 
a selection function. However, as part of the fields  were covered by a different number of slit masks 
(some had only one, others two), and the galaxy distribution is not uniform across the field, we developed two 
selection function for each field  taken into consideration these effects. 

The individual selection functions were calculated by counting the number of objects with successful spectroscopy 
(\ie  reliable redshifts)  versus  the number of photometrically detected objects up to the spectroscopic limit  
($I\approx19.5$) inside  the areas covered  by the corresponding spectroscopic masks, in different 
magnitude bins.  No background correction was applied, because  we only needed to know the relative 
number of photometrically and spectroscopically  observed galaxies to evaluate the success of our 
spectroscopy (see also Sect. \ref{SS:redshift}). The resulting  functions were applied to the cluster 
galaxies in the form of weights to the statistical properties of the cluster galaxies. 
 The combined selection function is shown in  Fig. \ref{F:selfunc}. However, some tests have 
shown us that the results depend little on the weighting   applied and are robust against 
other considerations.

   \subsection{Equivalent widths and star-forming galaxies}
   \label{SS:EW}
 
We use equivalent widths  (hereafter EWs) as a measure of the line strengths of the absorption and emission lines. 
We measured EWs automatically using  a custom-made routine,  which automatically corrects for
the effects of cosmic expansion.  In the case of \oii\ and \ha, which  are used as tracers of 
ongoing star formation, we adopted the definition given   by \citet{balogh99}. 
We adopt the  convention that typical  emission lines are shown 
with positive values when detected, but also, that typical absorption lines (\eg \hd) are 
positive in absorption.  

The \ha\ definition used, effectively isolates the targeted line from the  
adjacent \nii\ (which was also measured).  Each spectrum  was visually inspected  to find out 
whether any lines  fell into the prominent telluric  bands (A \& B), were affected by sky-subtraction  residuals 
or by artifacts  in the spectra. In some cases,  the lines were flagged and  not  used in subsequent analyses.

Usually the minimum EW that could be reliably measured was 5\,\AA\  
(see \citealt{balogh02} for a demonstration based on similar data),   therefore galaxies with equivalent widths
$W_0>5$\,\AA, either in \oii\   or \ha\ (or both), are considered star forming galaxies. 
We will show later in this paper (in Sect. \ref{S:redSF} and in the Appendix \ref{S:SFR}) that this classification is robust 
and  physically meaningful. 

\subsection{Absolute magnitudes}
\label{SS:absmags}

We use the  software {\sc kcorrect} \citep{blanton2007} to calculate  k-corrections and thus 
absolute magnitudes for galaxies in our spectroscopic sample. This code is based
on the latest stellar population models of  \citet{Bruzual2003} and photoionization
models of \citet{Kewley2001}. As a byproduct of the k-correction, the code also derives stellar masses, 
which will be used in Sect. \ref{S:redSF}.

The fields R265 and R285 were also imaged by SDSS\footnote{Sloan Digital Sky Survey, http://www.sdss.org} 
(\citealt{York2000}), therefore, we can use the advantage of multicolor photometry. 
Unfortunately, the remaining field (R220) was not observed by the SDSS  
and we have to use the available $V$ and $I$-band magnitudes  provided by \citet{gilbank04} and therefore,
larger uncertainties are expected in the calculations. However, we can test the accuracy of the magnitudes
by comparing the results obtained using the two-band photometry and the multi-band photometry in the other two fields.
For our analysis, we obtained  $B$, $V$ and $R$ rest-frame  absolute magnitudes (in the Vega system 
using Johnson-filter definitions). 

We found  scatters of $\sim$0.2\,mag and offsets of $\sim$0.15\,mag between the
magnitudes obtained in either way. The offsets depend on redshift and can be corrected  using  
a linear fitting. The scatter is in agrees  with  values found  by \citet{Blanton2005} for  
transformations between different filter systems. These differences are small and hardly change 
the conclusions in this study.

We selected  the original absolute magnitudes calculated  using the SDSS photometry for  the R265  and R285 fields 
and  applied the redshift correction for the galaxies in  R220 to only the magnitudes derived using the 
$V$ and $I$-band photometry. All apparent magnitudes were  corrected for Galactic extinction  
using the maps of \citet{schlegel98}. No correction for internal absorption   was attempted, since we do 
not have information, in many cases, about galaxy inclination, and the Balmer decrement cannot 
be used in all cases because \hb\ is rarely detected for  emission lines galaxies, and  uncertainties for passive galaxies
will remain. No important differences were found between the absolute magnitude 
distributions for the field and  cluster sample. 

The stellar masses were tested against the formulae of \citet{Bell2005} using our restframe $B$ and $V$-band
magnitudes. We found  deviations only at the high mass end. Since the {\sc kcorrect} code 
is reliable in predicting magnitudes between the SDSS and our system, we preferred to use 
its data outputs.

\section{The clusters}
\label{S:clusters}
  
 \begin{figure}[t]
     \centering
   \includegraphics[width=0.5\textwidth]{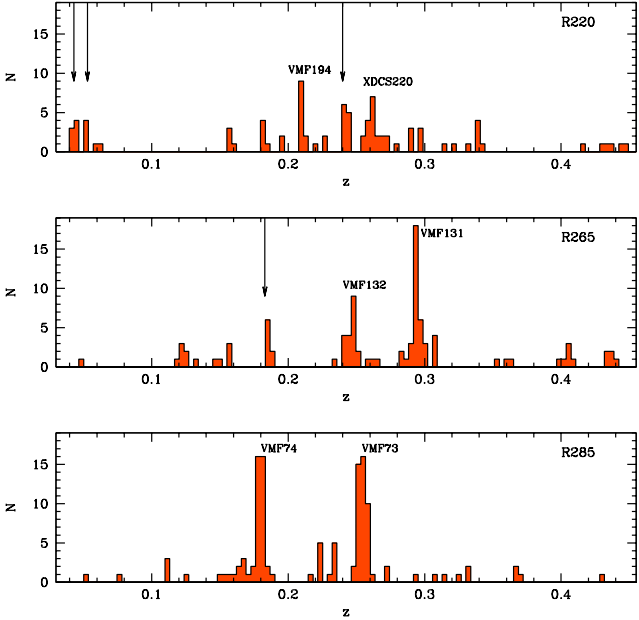}
   \caption{Redshift distribution of the targets in the three fields, with the cluster names marked.
   The small arrows mark the position of the group candidates   (see Sect. \ref{S:fields}).}
   \label{F:zfield}
   \end{figure}

\subsection{Cluster membership}
\label{SS:membership}

In each field, the redshift distribution was analyzed to detect  prominent structures. 
The clusters studied had already known redshifts, with  the exception of those in the 
R220 field whose redshifts were unclear  (see  Sect. \ref{S:fields}  for details), but were confirmed. 
The mean  cluster redshift ($z$) and velocity dispersion ($\sigma$) were calculated  
using the  bi-weight estimators of  \citet{beers1990} and iteratively 
excluding galaxies beyond 3-$\sigma$  of the mean redshift until the solution converged. 
We applied a bootstrapping technique to check the stability of the results and  
calculate the errors in the velocity dispersion. The results can be found in Table \ref{T:maindata}
and the redshift distribution in Fig. \ref{F:zfield}.

\subsection{Galaxy colors}
   \label{SS:galcolor}

 \begin{figure}[t]
     \centering
     \includegraphics[width=0.5\textwidth]{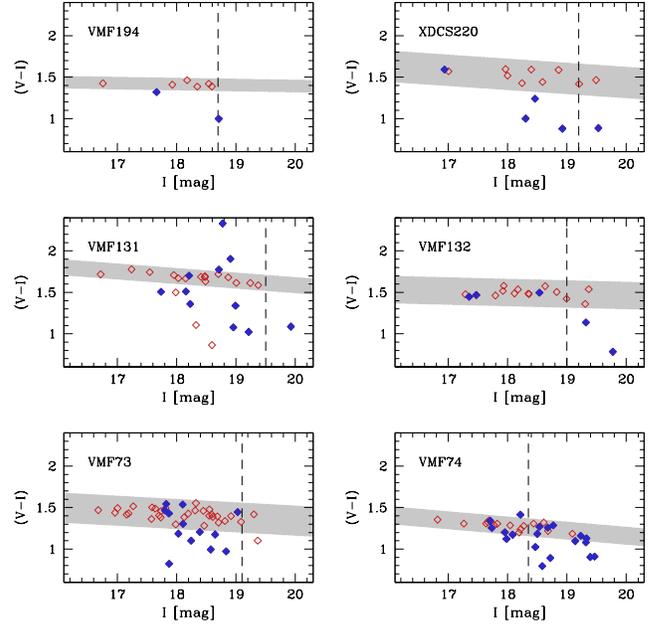}
     \caption{Color-magnitude diagrams of the members of the   six observed clusters. 
Filled blue diamonds are star-forming galaxies, whereas open red diamonds are 
passive galaxies.  The shaded areas are defined by the 3-$\sigma$ deviation
of the least squares fits to the passive galaxies. The vertical dashed line mark 
$M_I\approx -21.4$ used in the density calculation (see Sect. \ref{SS:den}). We note the red star-forming  
galaxies belonging  to the red-sequence and even redder in some  of the clusters.}
     \label{F:cmd}
   \end{figure}

We use the spectroscopic information to separate the galaxy population. Galaxies 
with emission lines are considered star-forming and those without  emission, passive  
(see Sect. \ref{SS:EW}). Plotting the $V-I$ color versus $I$-band magnitude (Fig. \ref{F:cmd}) for cluster
galaxies shows that all clusters have clear red-sequences. Only few cluster galaxies 
have blue colors but no emission lines.

The distribution of the passive galaxies, the red-sequence,  is well described by simple least-squares
 fits. The weighted mean dispersion  of the red-sequences   is $\sigma \approx 0.05$ mag, 
which is  the typical error in the photometry.  All galaxies  redder than the lower 
3-$\sigma$ limit are considered red galaxies, and blue otherwise.  Given this criterion,  
we note the existence of a population of red  star-forming galaxies belonging to the 
red-sequence and even redder. More striking is the high
number of those galaxies belonging to the cluster VMF74. Some of the
characteristics of this sub-population will be described in Sect. \ref{S:redSF}.

\subsection{X-ray luminosities}
\label{SS:Ocluster}


The X-ray luminosities of the intracluster medium and cluster velocity dispersions 
are indicators of cluster masses.  The  correlation between these two parameters
has been extensively studied (\eg \citealt{markevitch98,david93,xuewu2000}) and
 is interpreted as a sign of dynamical equilibrium,  even though the large scatter in the local
relation indicates deviation from this equilibrium. Nevertheless, later studies
have found that cluster masses derived from using independent  methods, including gravitational weak-lensing, 
correlate with relative small scatter (\eg \citealt{Hicks2006}), solving a long-standing
controversy. 

\begin{figure}[t]
     \centering
     \includegraphics[width=0.5\textwidth,bb=20 160 360 400,clip]{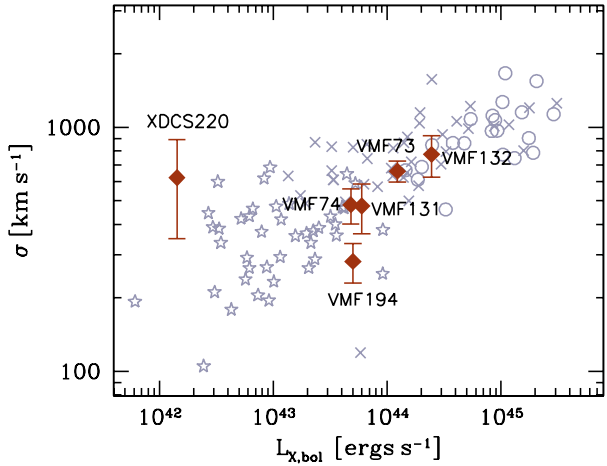}
     \caption{Bolometric X-ray luminosity plotted against   velocity dispersion. Open circles
     \citep{markevitch98},  crosses \citep{david93} and stars \citep{xuewu2000} represent    
       the $L_X$-$\sigma$ relation for local clusters. The six clusters    studied here are plotted as diamonds 
      with error bars in the   velocity dispersion.}
     \label{F:lxsig}
   \end{figure}

In Fig. \ref{F:lxsig}, we plot the bolometric X-ray luminosities against the derived velocity dispersions. 
The clusters follow  the local  $L_X-\sigma$ relation, with the notable exception of
XDCS220, which is underluminous for its velocity dispersion. This cluster displays a tail in the
redshift  space,  which complicates  the calculation  of the velocity dispersion  and implies, therefore,
that it is  likely overestimated. 

In addition, VMF194 is  peculiar, because it has a velocity dispersion that is too low for its X-ray luminosity. This effect
may come from two different sources. First, $\sigma$ may be underestimated due to selection effects 
given the  low number of members identified. Second, we detect a  background group at 
$z\approx0.24$   of a relative large  velocity dispersion (see Sect. \ref{SS:r220}), 
which may have contaminated the X-ray measurements.  Nevertheless, this cluster
does not appear to be so extremely offset from the $L_X-\sigma$ relation as XDCS220. 

With the exception of XDCS220, the cluster X-ray luminosities and velocity dispersions 
are similar to those of Virgo, A496 and Coma clusters  (\eg \citealt{david93,Rines2003}) 
and thus are  expected to be clusters that are as equally massive. 

\subsection{Virial radius}
\label{SS:Rvir}

From the results shown in the previous section, it is possible to assume that the clusters sampled in this study
are in general in dynamical equilibrium\footnote{It is probably not true for XDCS220, however for 
the sake of comparison it will be assumed that it is. VMF194 is also peculiar, but the differences may 
arise  from another sources.}  and therefore, the virial theorem is applicable. The radius within which 
the virial mass is estimated to be contained is called the virial radius. According to the 
observationally-calibrated derivations  of \citet{Carlberg1997c}, \Rv\  is defined  as the distance where the mean 
inner  cluster density is 200 times  the critical density $\overline\rho(z)$ of the Universe   
and it is  also called \rdos. Its relation to the   velocity dispersion $\sigma$ is given by

\begin{equation}
  \Rv=\rdos= \frac{\sqrt{3}}{10} \frac{\sigma}{H(z)}
  \label{E:Rv}
\end{equation}

\noindent where  $H(z)=H_0\sqrt{\Omega_m(1+z)^3+\Omega_\lambda}$ for  a Hubble
constant of $H_0$=70\,km\,s$^{-1}$\,Mpc$^{-1}$, $\Omega_\lambda=0.7$ and  $\Omega_m=0.3$.

Since \rdos\ (\Rv) characterizes the size of clusters following the assumption of a universal mass profile, it
is useful as an environmental  indicator of mass density, given the clustercentric distances of cluster galaxies.
Therefore the distances of galaxies to the center of the cluster are normalized by the
respective cluster virial radius, allowing the entire sample to be combined
into a single cluster, increasing the statistical significance of our analysis and reducing the effects of
cluster-to-cluster variations. 

 
\subsection{Projected density}
   \label{SS:den}

\begin{figure}[t]
     \centering
     \includegraphics[bb=20 155 600 430,width=0.5\textwidth,clip]{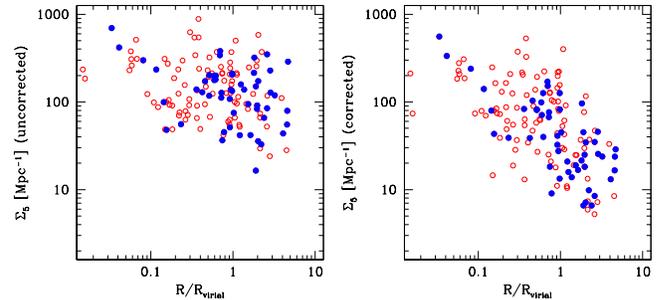}
     \caption{Relation between virial radius and projected  density before  and after correction 
for field contamination.  The open  red circles are  passive galaxies, whereas filled blue circles  
are star-forming galaxies.}
     \label{F:Rvdens}
   \end{figure}

\begin{figure*}[t]
     \centering
\includegraphics[width=0.49\textwidth]{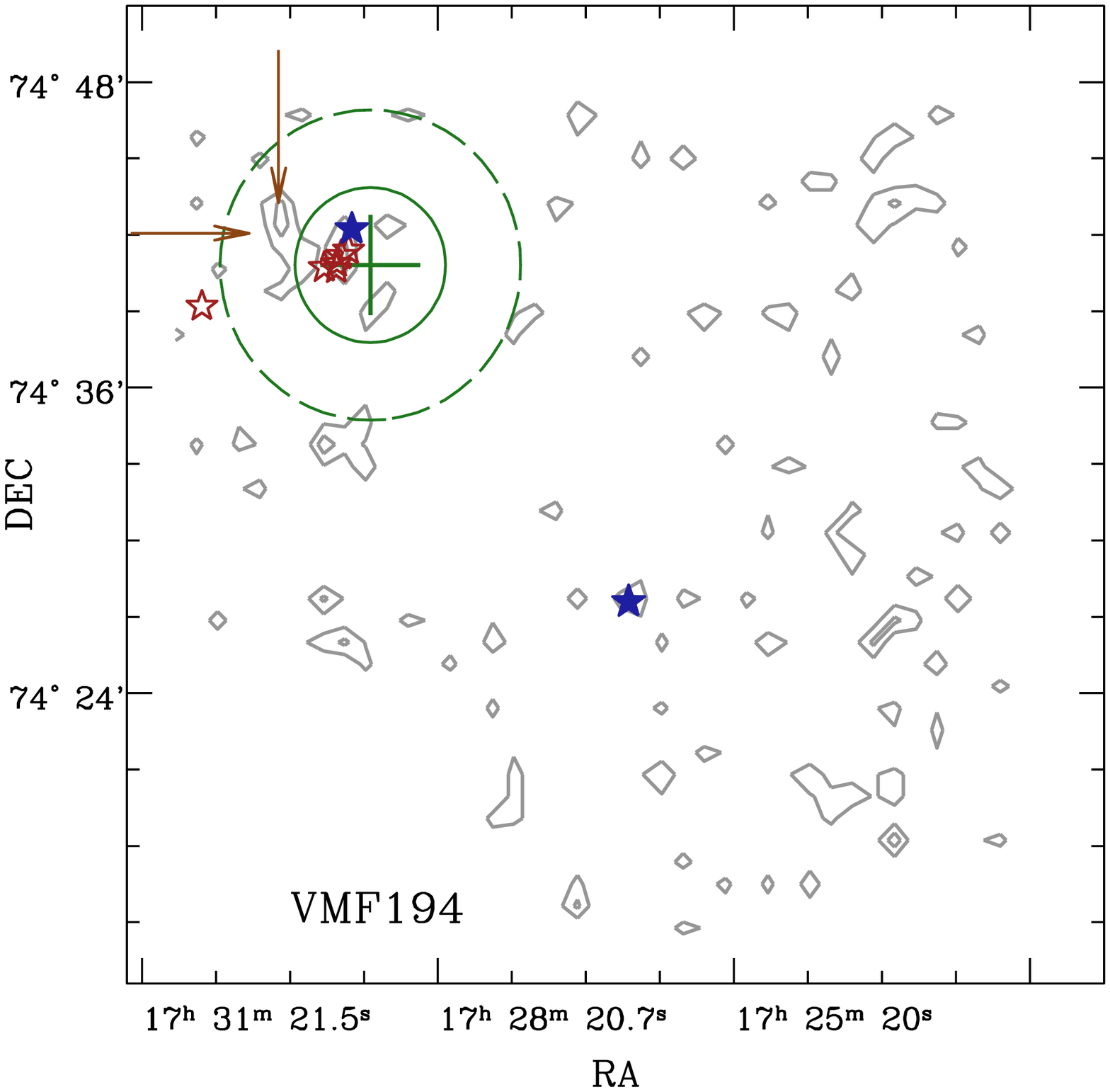}
\includegraphics[width=0.49\textwidth]{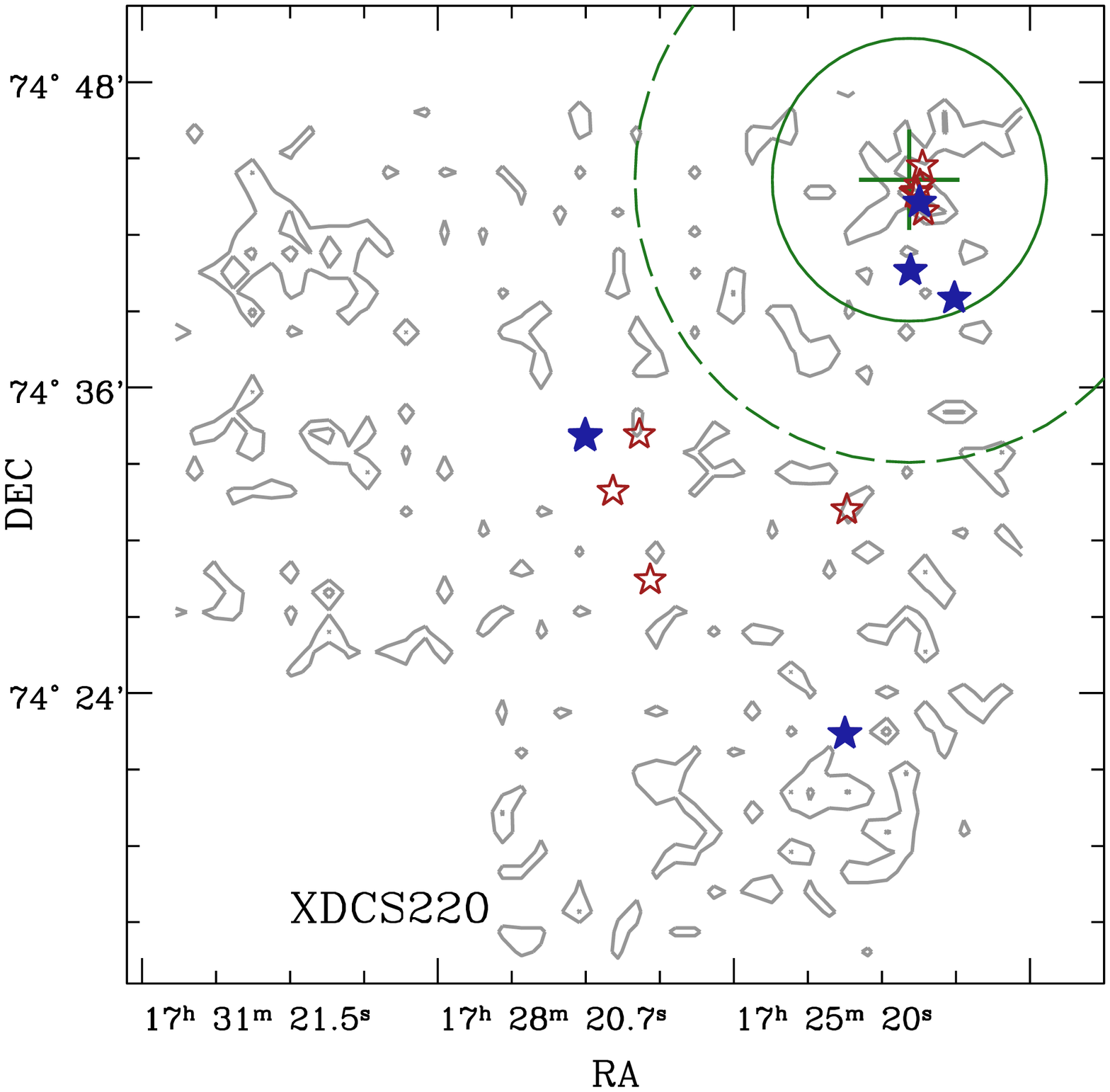}
\includegraphics[width=0.49\textwidth]{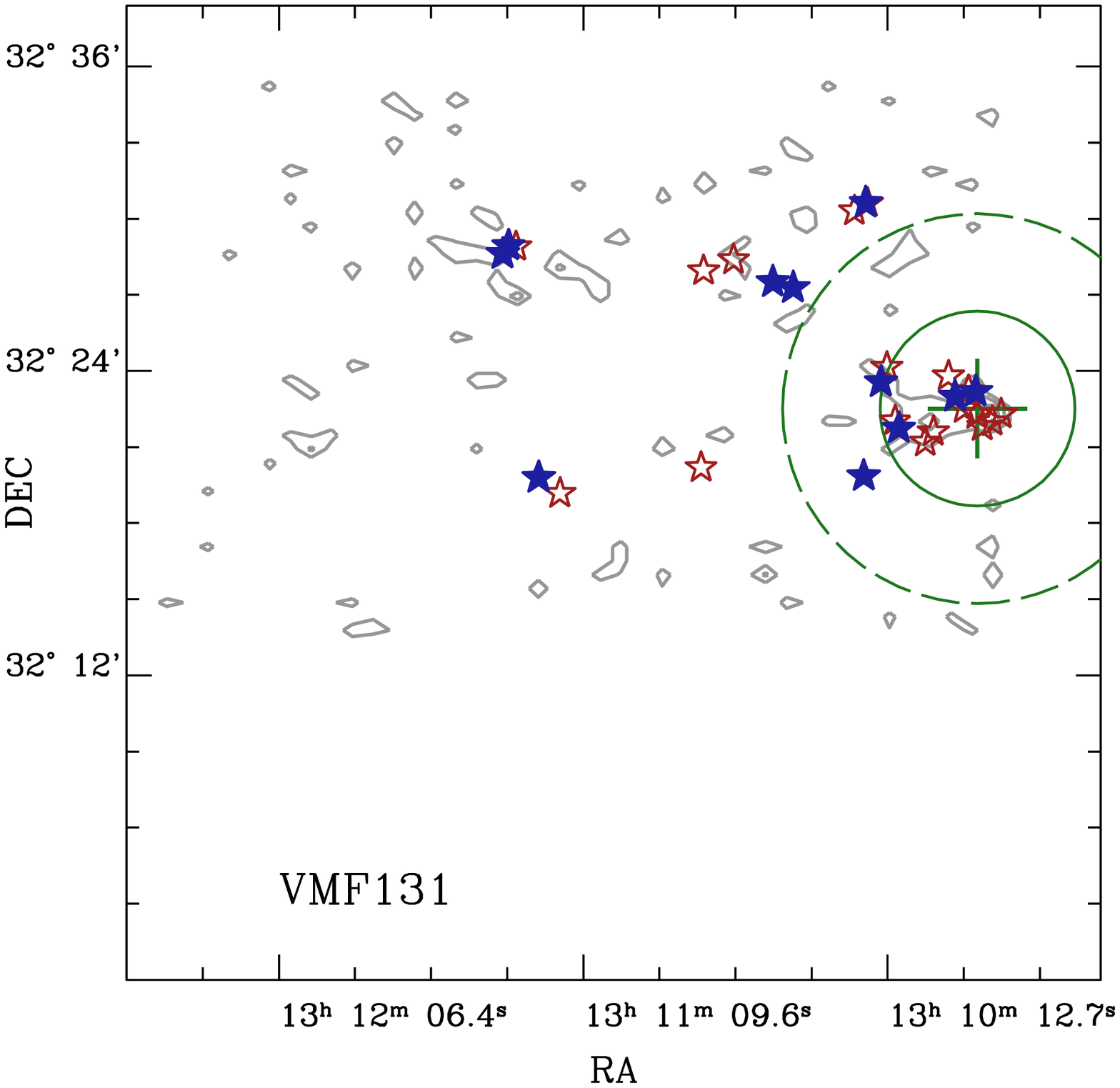}
\includegraphics[width=0.49\textwidth]{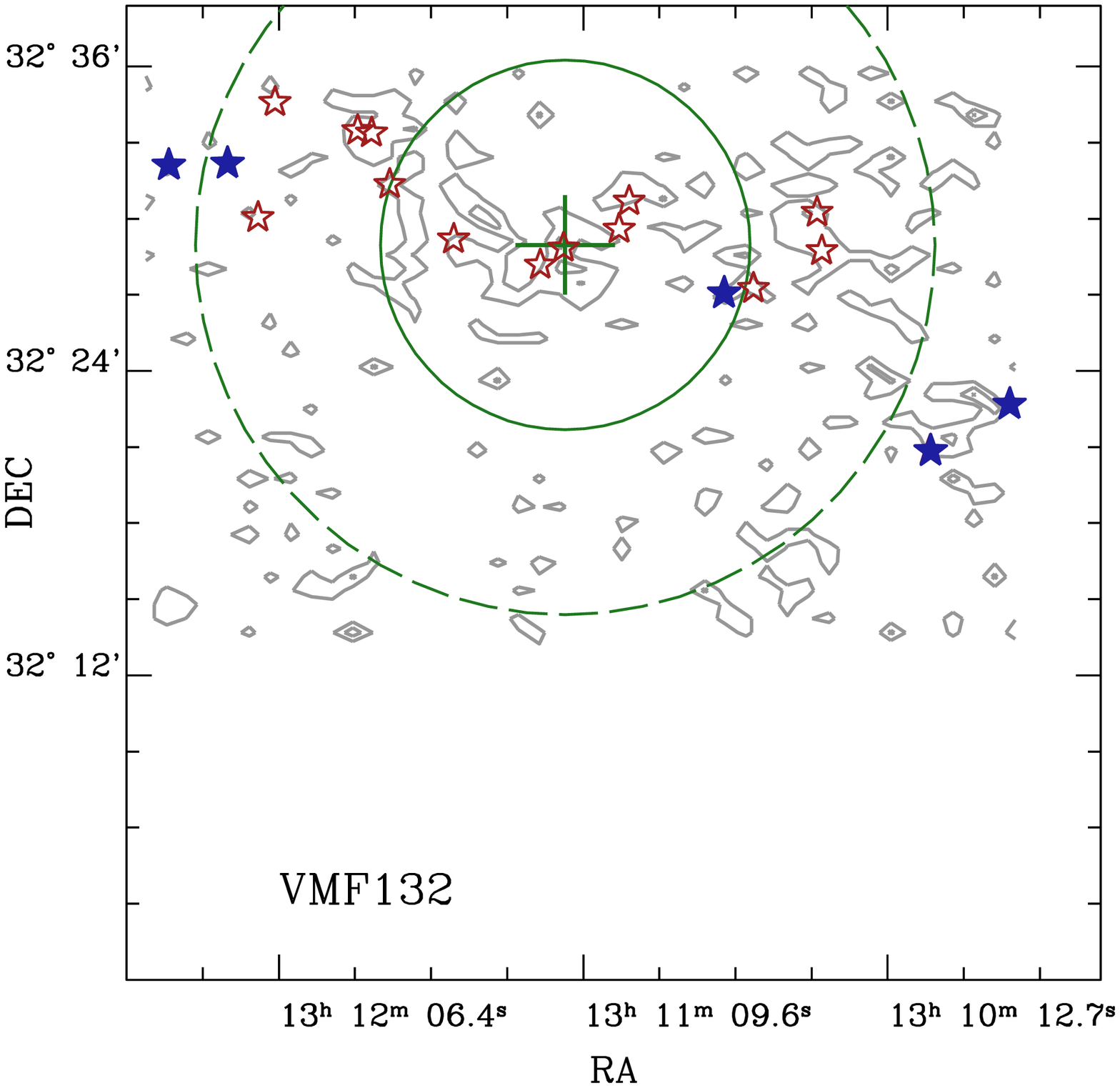}
\caption{Representation of the fields of the observed clusters as 
indicated by the names on the individual figures. Only cluster members are shown. 
Blue filled and open red symbols represent star-forming  and passive galaxies  respectively. 
The cluster centers are marked with  large vertical crosses and the large concentric
circles represent one and two virial radii respectively. The contours are the density maps of all galaxies
with colors compatible with the red-sequence of the respective cluster (see Sect. \ref{S:fields}). The arrows in the VMF194
plot indicate the position of a rich background group (see Sect. \ref{SS:r220}).}
\label{F:clusters1}
 \end{figure*} 

\addtocounter{figure}{-1}


\begin{figure*}[t]
     \centering
\includegraphics[width=0.49\textwidth]{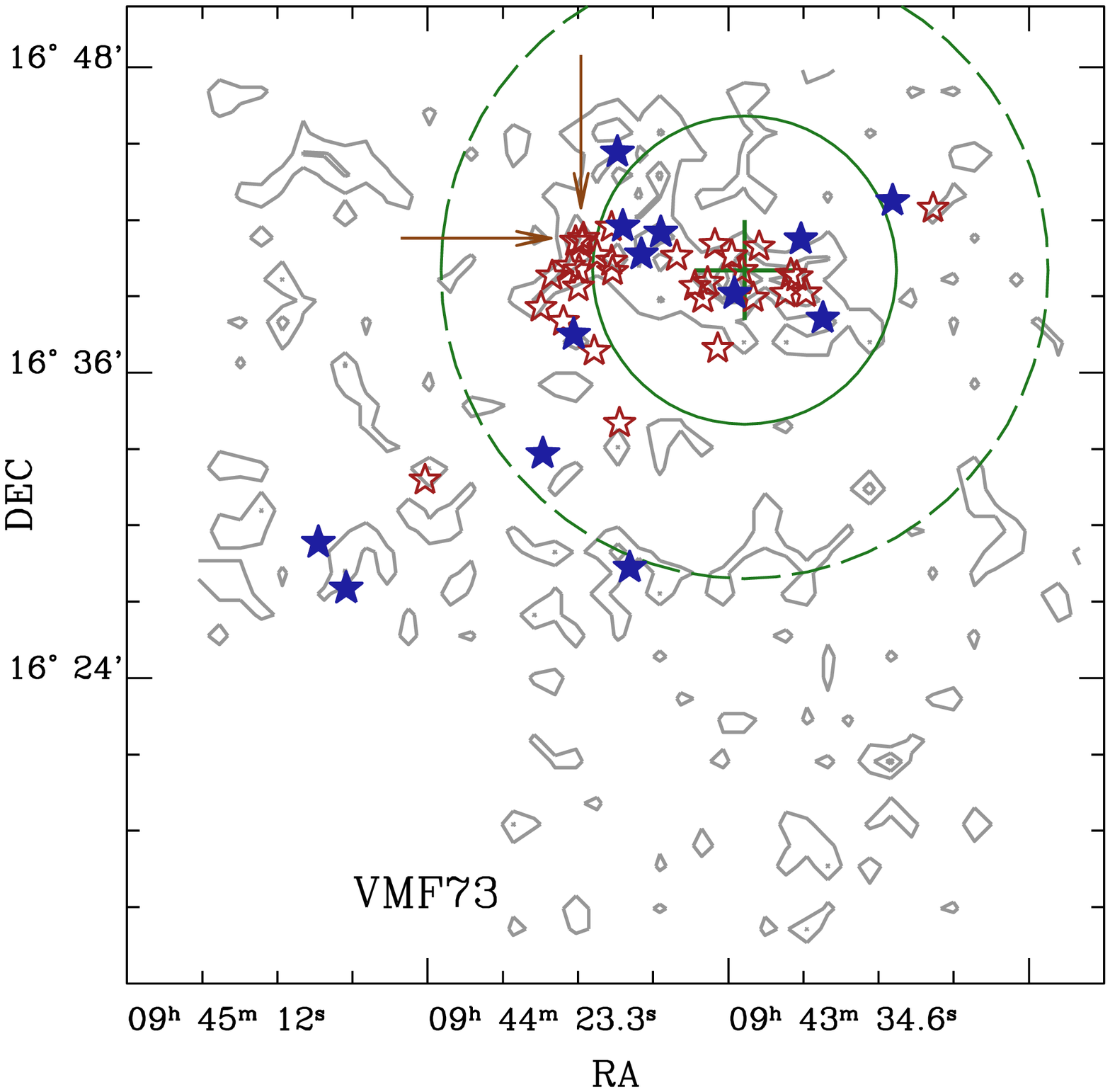}
\includegraphics[width=0.49\textwidth]{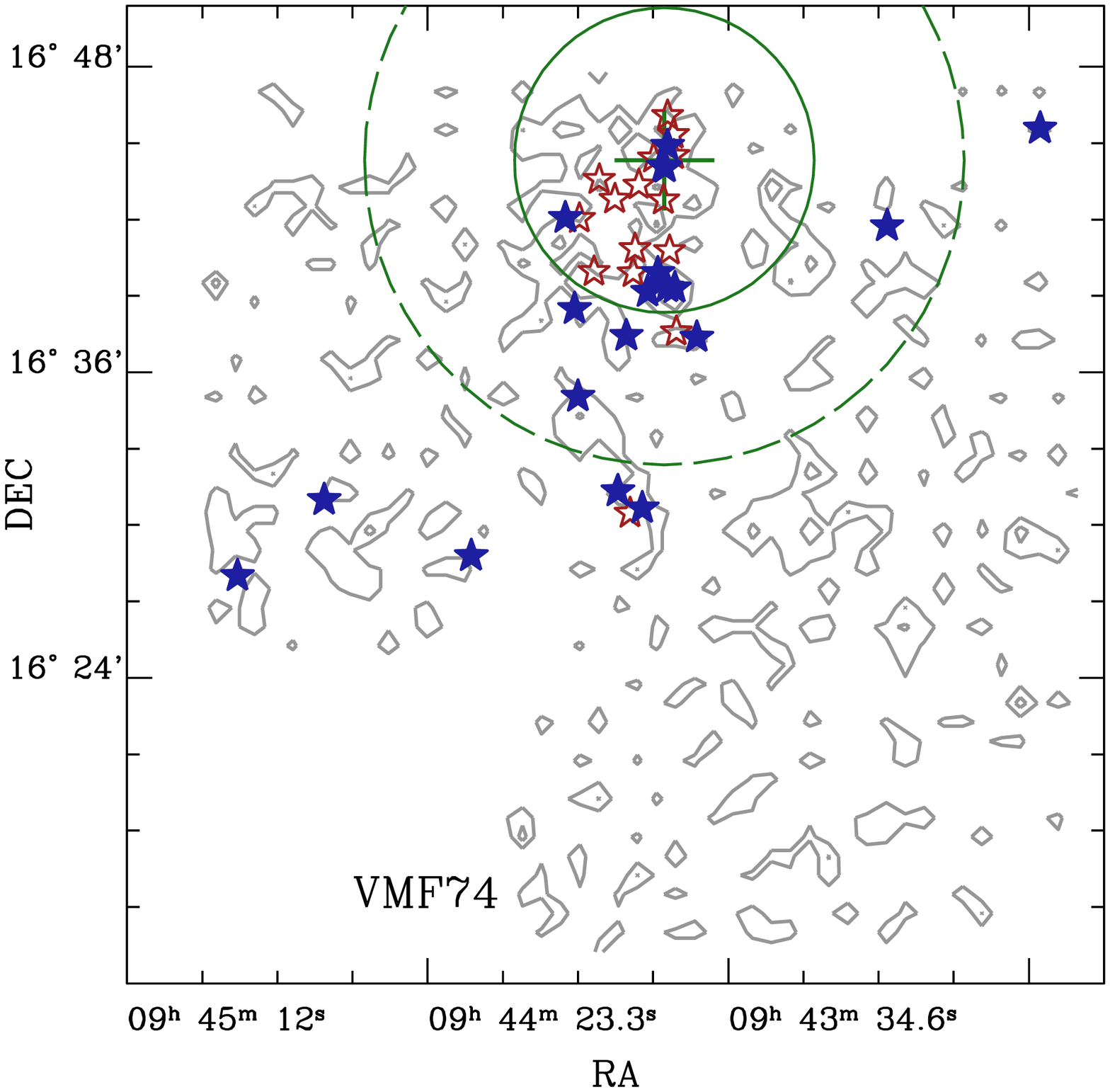}
\caption{continued. Representation of the observed clusters as indicated by the names on the individual figures.
Only cluster members are shown. Blue filled symbols are star-forming galaxies and open red are 
passive ones.  The arrows in VMF73 show the position of the X-ray structure detected by \citet{Rasmussen2004} 
(see Sect. \ref{SS:r285}).}
\label{F:clusters2}
   \end{figure*}

Another common  indicator of environment is the local number projected (2-D) density of galaxies.
Its calculation does not make any assumption about the physical properties of the clusters, but other precautions
 must be taken. First, the galaxy number density is a function of luminosity.  The spectroscopic limit of 
$I\approx19.5$\,mag corresponds to $M_I\approx -21.4$ for the furthermost 
cluster ($z\approx0.3$) and $M_I\approx -20.2$ for the closest one ($z\approx0.2$),
taking in consideration the typical k-corrections (see \citealt{Fukugita95}). 

For each cluster, the photometric catalog was divided using an apparent magnitude that 
corresponds on average, to the luminosity limit of  the most distant cluster,  
which translates into an apparent magnitude cut of  $I\approx18.3$ at $z=0.18$ (see Fig. \ref{F:cmd}).

The projected density is defined by the area that encircles the fifth nearest  neighbor to this galaxy, which is
referred as  \Den. However, significant foreground and background contamination is expected and must be corrected 
 before completing  any statistical analyses. In the literature, several methods of different complexity 
are  described to deal  with this problem.  Most  of them subtract a value (local or global) from the calculated density, 
making different assumptions.  However, those methods often  yield unphysical values  (\ie negative numbers) 
for the density estimates. Our case is   even more complicated, because we do not only have 
field contamination, but also contamination from the other projected cluster. Therefore, 
we  chose another approach using in combination   the photometric and spectroscopic data-set.

If the \emph{true} number density of galaxies in a certain region  of the cluster is $N$ 
(unknown) and the observed is $M$  (determined from the photometric catalog and including 
the contamination),  one has a relative fraction of $f=N/M$. From the spectroscopic data set,
we know that there are $n$ galaxies belonging to the cluster and $m$ is the  number of total observed 
galaxies in the same area with secure redshifts. Since  the   selection  was performed randomly 
(based only on $I$-band magnitudes), we can assume that we have  the same fraction  expressed 
now by  $f=n/m$, thus we can correct the  observed value $M$, multiplying  it by $n/m$, obtaining $N$.

The areas used to make these corrections are larger than the areas considered by the individual 
density calculations.  They encircle always 10 galaxies with secure redshifts, and we count  
the number of cluster members versus the non-cluster galaxies. Having a high-filling factor 
helps to the statistical reliability of  this simple method, because the areas sampled 
will have smaller physical sizes and thus  smaller deviations from the  local density. The results of the 
correction can be seen in Fig. \ref{F:Rvdens}. After this process, a correlation between  
virial radius and projected density becomes evident. 

We would like to emphasize that the densities calculated here  are not  
directly comparable to those calculated elsewhere,  because the  magnitude cuts and  approaches to subtract 
the background  vary between different authors. 

Finally, galaxies fainter than  the individual cluster magnitude cut were not included in the composite
cluster; this reduced the final sample size to $\sim$120 galaxies. We note that, many  of the galaxies excluded 
are member of  the VMF74 cluster.


\section{Description of the fields}
\label{S:fields}

We describe each field, providing detail in particular of the general cluster properties, candidate groups, and cluster substructure. Each cluster is represented separately in Fig. \ref{F:clusters1},
with different symbols for star-forming and passive galaxies.  The large concentric circles represent one 
and two virial radii respectively calculated according to Eq. \ref{E:Rv}. 

The contours show the distribution of all galaxies down to $I=23$\,mag with  colors 
similar to the respective red-sequences (see Fig. \ref{F:cmd}). They provide some information 
about the spatial distribution  of galaxies without spectroscopy. Since 
the CMR for ellipticals has little scatter, the structures 
are probably at \emph{similar} redshifts. This technique has been  successfully used by other studies
to detect substructures around clusters (\eg \citealt{Kodama2001,Tanaka2005}). In this case, however, 
it is not possible to firmly state the significance of those structures because only
the $V-I$ color, provided by \cite{gilbank04}, is used and the red-sequences of each projected cluster 
have similar colors  (see Fig. \ref{F:cmd}). The use  of the SDSS multi-color photometry 
does not help because their uncertainties are larger at faint luminosities and the red-sequences become 
completely blended. Therefore, the contours plotted in each figure must be taken only as informative. 
Nonetheless,  many of the spectroscopically  identified members are actually
associated with structures that show up using this simple color cut.


   \subsection{R220}
   \label{SS:r220}

The R220 field is a very complex field. There is, first, a  larger number of objects than in the other fields.
This is maybe due to its lower galactic latitude. Our photometric catalog was cleaned
of star-like objects, but , the separation is not perfect and many of our slits unintentionally contained  stars,
losing the advantage of having an extra mask for this field (8 instead of 7). 
The redshift distribution  also looks  more  complex (see Fig.  \ref{F:zfield}), 
with a number of  associations besides the two clusters. 

The cluster VMF194 was found to be difficult to confirm  optically by  \citet{vik98}
and collaborators. According to \citet{gilbank04}, the proposed cluster corresponds to  
``a very extended  X-ray emission and the galaxy over-density is similarly  extended''. 
Here, VMF194 at $\avg{z}=0.210$ (see Table \ref{T:maindata}) 
was  unequivocally detected, but the data  obtained showed that the cluster has a surprisingly low velocity
dispersion  for its X-ray luminosity (see  Fig. \ref{F:lxsig}).  Three additional galaxies  have  
redshifts that imply cluster membership, according to the previously-measured 3-sigma limits;
 these galaxies are located, however, at large clustercentric radii
 ($>7 \Rv$).  When they are included,  the velocity dispersion  does not change  substantially, and 
thus they were excluded as  members, but not included in the field sample.
   
At an angular distance of $\sim$4.4\,arcmin of VMF194 (\ie almost overlapping positions), we detect a clump of galaxies at  
redshift $\avg{z}=0.243$. This clump also shows up in the spatial  distribution: 8 out of the 11 galaxies 
are clustered in an area  smaller than $\sim 0.3\times 0.7$\,Mpc .  The velocity dispersion of this group is 
$\sigma=401 \pm 74$\,km/s, indicating that it may be quite massive.  No red-sequence  is detected  and 4 out of the 8  galaxies,  
show star-forming activity. This group may have been  the cause of confusion in all previous studies in this field. 
In fact, the concentration of galaxies is more prominent for this group than for VMF194 when selected
by a  color cut (see Fig. \ref{F:clusters1}).
 
We confirm the presence of the  cluster at $\avg{z}=0.261$ detected by  \citet{gilbank04}, 
which is  there referred  as  XDCS\,cmJ172333+744410 (it is called here XDCS220 for short).
We confirm the  redshift calculated there.  This cluster has a very low X-ray luminosity
and passed  undetected in the X-ray analysis of  \citet{vik98} and \citet{mullis03}. 
It  displays  a large  velocity dispersion  (see Table \ref{T:maindata}), 
which  is probably overestimated  because of the existence  of a tail in redshift space. 
Excluding  members that are located at  large clustercentric distances  does not change the biweight estimate of the velocity 
dispersion. We conclude that  it is a real feature of the cluster, which is probably  in the process  
of relaxing  or has an extended structure along the line of sight.  This cluster shows a  clear red-sequence  
and 5 out  14 galaxies show ongoing star-forming activity. 

Two other group candidates were found (see Table \ref{T:groups}), one at  $\avg{z}=0.04293$ 
($\pm$390\,km/s) with 6 members in 1\,Mpc$^2$   (or 5 in $0.3\times 0.7$\,Mpc), all being star-forming galaxies, and 
the other is at $\avg{z}=0.05274$ ($\pm$126\,km/s), with four members in   $0.3\times 0.4$\,Mpc. 

   \subsection{R265}
   \label{SS:r265}

The central parts of the cluster VMF131 were  previously observed  by  \citet{balogh02} as part 
of their low luminosity X-ray cluster  project, where it was known  as CL1309+32, using the same 
instrument and setup;   we have therefore added their data into our study. Since, it is the most distant
 cluster studied, we were able to detect members up to clustercentric distances of $R>4\Rv$.
The color contours shows little substructure around the cluster
but the central overdensity is clearly visible in Fig. \ref{F:clusters1}.

The cluster VMF132 is the richest cluster in our sample and has the  largest velocity 
dispersion and thus the largest  virial radius,  occupying  a large proportion of the field. 
In spite of this, the galaxy concentration is  clearly irregular when color cuts are applied   and 
only sparse structures are detected.  

An extended group was also detected at $\avg{z}=0.186 \pm 0.001185$   (349\,km/s) 
with 8 members in an area of $1\times 2$\,Mpc, or  $0.7\times 1.5$\,Mpc if one excludes one galaxy.

\begin{table}[t]
 \centering
  \caption{Main parameters for the groups candidates for our fields. Their identification codes show
the average positions of the members. Mean redshifts ($z$) and average deviations  
are shown as velocities ($\sigma$) . The biweight estimators
were used only in groups with at least 8 members. The group number identify  member galaxies in the 
online table.}
 \begin{tabular}{ccccl}
 \hline\hline
 Group &   ID              &  $\avg{z}$  & $\sigma$            &  $N$    \\
               &                                   &                       &  [km s$^{-1}$] &              \\ \hline
1 & r220\_1J 172604+742830    &  0.053          &  126                     &  4          \\
2 & r220\_2J 172518+742844    &  0.043          &  390                     &  6          \\
3 & r220\_3J 172958+744204    &  0.243          &  401                     &  8 (11)  \\
4 & r265\_1J 131030+322840    &  0.186         &  349                     &  7           \\ \hline
 \end{tabular}

 \label{T:groups}
 \end{table}

\subsection{R285}
\label{SS:r285}

The two clusters present in this field almost overlap in their positions on  the sky 
(angular separation $\sim$5\,arcmin, see Fig. \ref{F:clusters2}). In addition,  we placed more
 masks in the central  parts of the clusters,  which led to a higher success rate compared with the 
other fields.   The  cluster VMF73  at $z=0.254$ has the largest number of members  identified ($N=44$).
Most of the identified members of this cluster are located inside 1 \Rv, in an elongated structure  
running  approximately in the East-West  direction. In fact, when galaxies are selected by the colors  
of the red-sequence   this structure is clearly visible. Unfortunately, the foreground cluster (VMF74) 
has a CMR with very similar colors (see Fig. \ref{F:cmd}) and it is not possible to separate clearly both clusters using 
this technique.

The cluster center is approximately at the middle  of this structure, but in one extreme, 
at a distance $\sim 1\Rv$,  a compact group  (100$\times$100\,kpc$^2$) of bright,  passive galaxies is found. 
Their positions coincide with the extended X-ray source XMMJ0943.9+1641 detected by \citet{Rasmussen2004}.  
The X-ray flux of this  structure is $f_{X,1-2\mbox{\scriptsize\,keV}}=3\times10^{-14}$ erg cm$^{-2}$ s$^{-1}$ 
(Rasmussen, private communication), which yields an X-ray luminosity 
$L_{X,bol}=1.38\times10^{43}$ erg s$^{-1}$,  assuming that the  X-ray structure is  
associated with the VMF73 cluster.  This structure may be the center  of a large, newly 
infalling group of galaxies, although no peculiarities  were detected in the redshift distribution. 

The cluster VMF74 has a surprisingly large number of star-forming members: 19 out of 34, and many of them 
have colors similar to the red sequence (see Fig. \ref{F:cmd}). It  is also the closest of the clusters studied with a 
mean redshift of $z=0.18$.  The spectroscopically-identified members are also distributed in a 
elongated structure in an almost North-South direction,  although less clear than in VMF73. It also shows  
up using color cuts. The cluster center lies at the northern extreme of this structure. 

According to the $XMM$--Newton  X-ray analysis of   \citet{Rasmussen2004}, both VMF clusters do not exhibit 
peculiarities and are fairly typical for their masses.

   \subsection{Field sample}
   \label{SS:field}

The field sample consist of all galaxies between $0.15<z<0.35$, with at   least 6-$\sigma$ 
of distance in the redshift space from the clusters.  We included the galaxies belonging to the 
suspected groups.  Since the sample is built using the same observations any
comparison is straightforward. The same redshift-dependent magnitude cuts have 
been applied, which yields 97 galaxies used for direct comparison. 
Throughout this paper, many quantities will be compared with those of this subset.


 
\section{Analysis of the composite cluster}
\label{S:results}

 \subsection{Galaxy colors and environment}
   \label{SS:galcolors}

   \begin{figure}
     \centering
     \includegraphics[bb=30 165 325 420,width=0.49\columnwidth,clip]{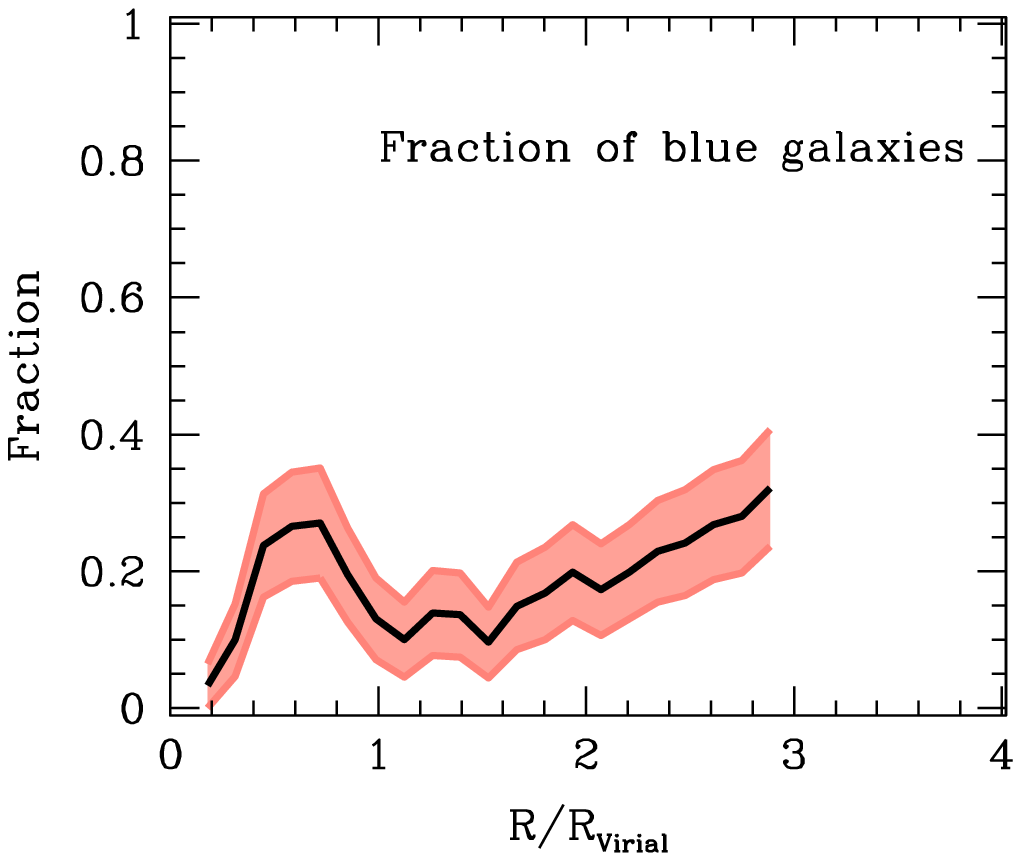}
     \includegraphics[bb=30 165 325 420,width=0.49\columnwidth,clip]{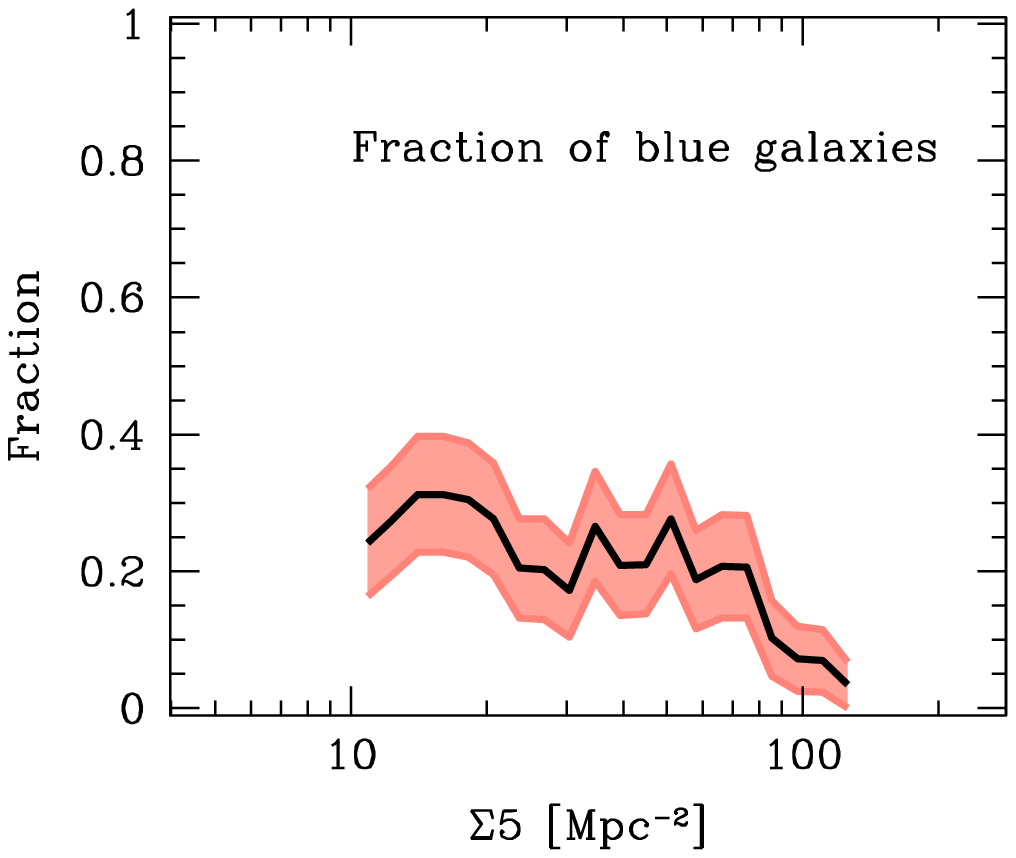}
     \caption{Fraction of blue cluster galaxies (as defined in  Sect. \ref{SS:galcolors}) 
against normalized   virial radius and projected density.}
     \label{F:bluefrac}
   \end{figure}

In  Fig.  \ref{F:bluefrac}, we plot the fraction of  blue galaxies  (as defined in Sect. \ref{SS:galcolor})  
against our environment indicators.  We observe an increase in the  fraction in both cases towards 
large radius and low density regions, however  a notable  peak inside 
$R<1\Rv$  is observed.  In the high-density regions, the fraction remains low and is statistically 
similar for clusters at those redshifts  (\eg \citealt{Ellingson2001}). 

\begin{figure*}[t]
     \centering
      \includegraphics[bb=20 155 325 420,width=0.3\textwidth,clip]{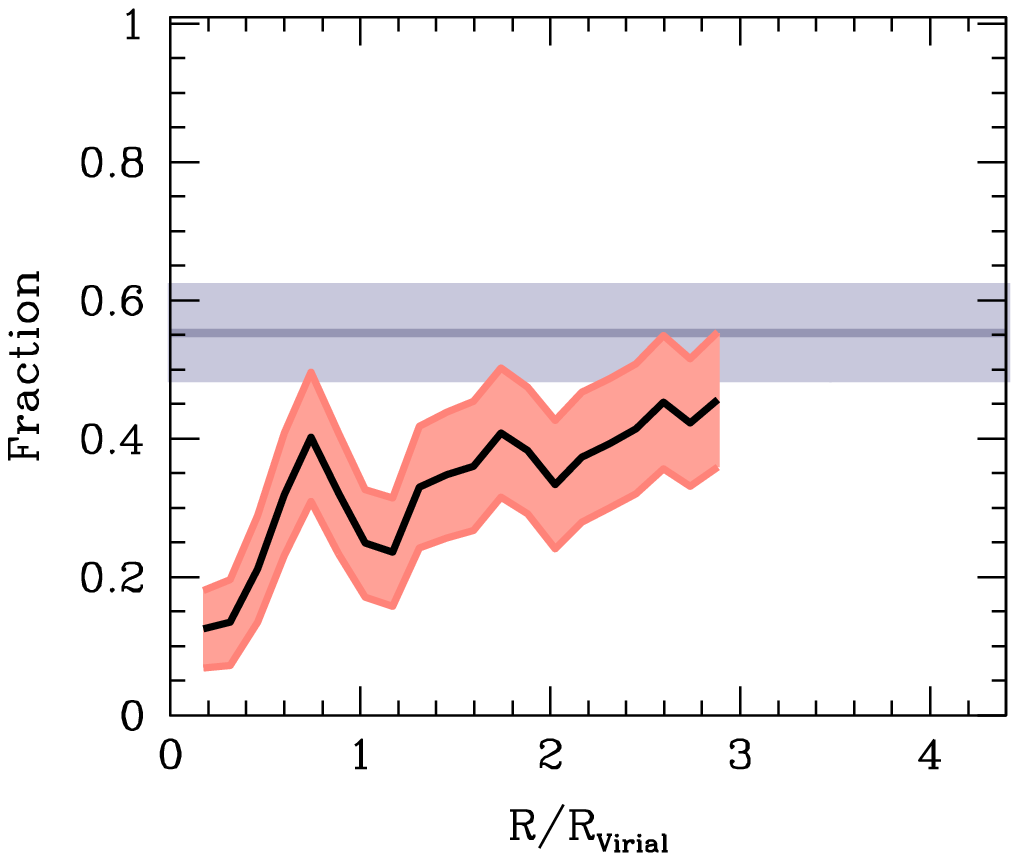}
      \includegraphics[bb=20 155 325 420,width=0.3\textwidth,clip]{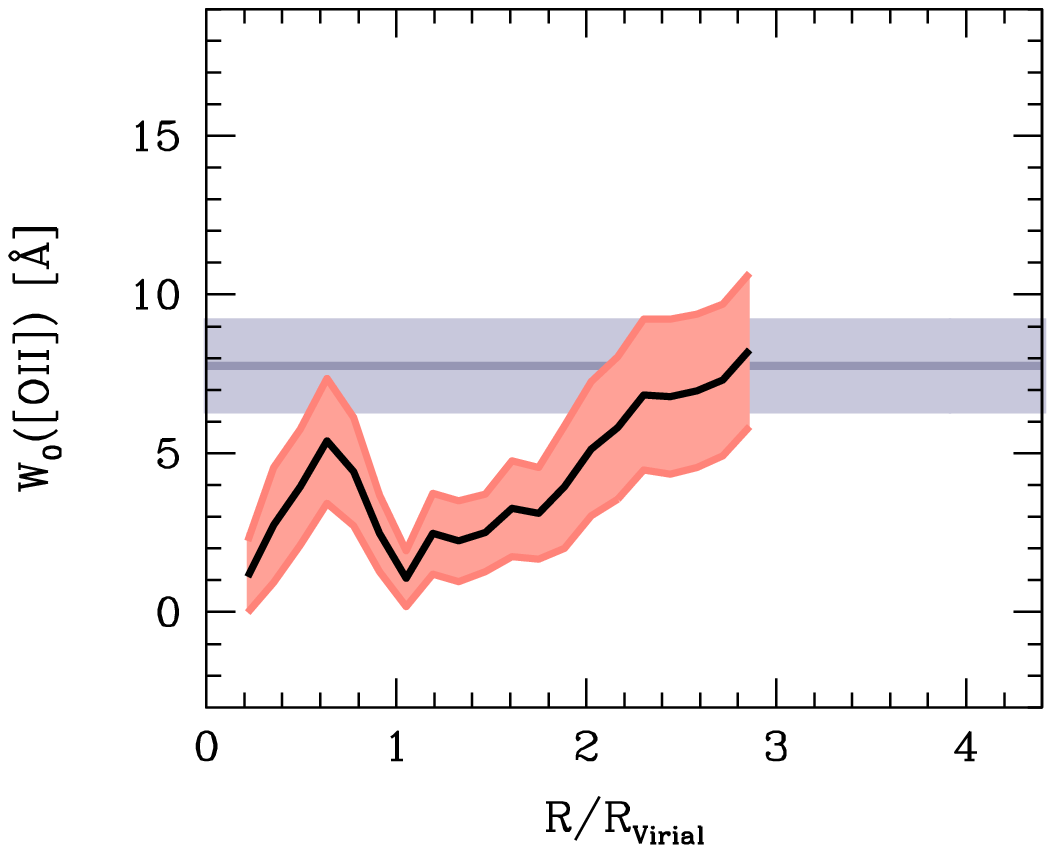}
      \includegraphics[bb=20 155 325 420,width=0.3\textwidth,clip]{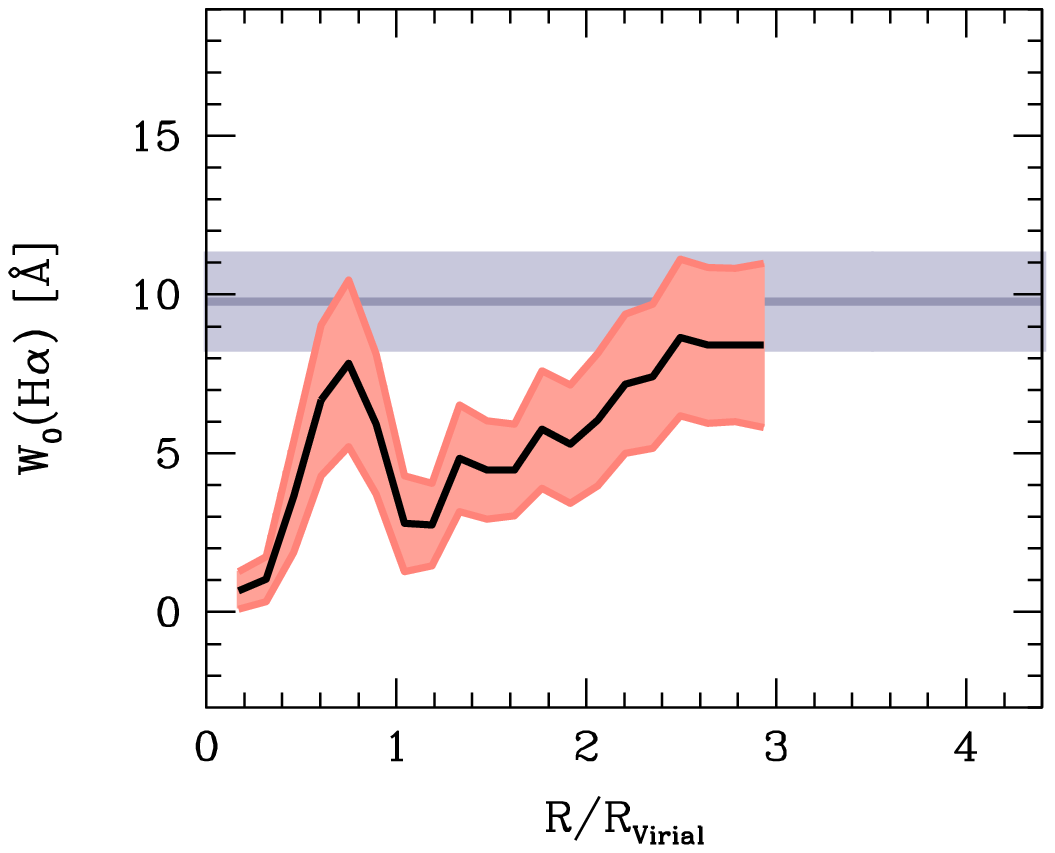}
      \includegraphics[bb=20 155 325 420,width=0.3\textwidth,clip]{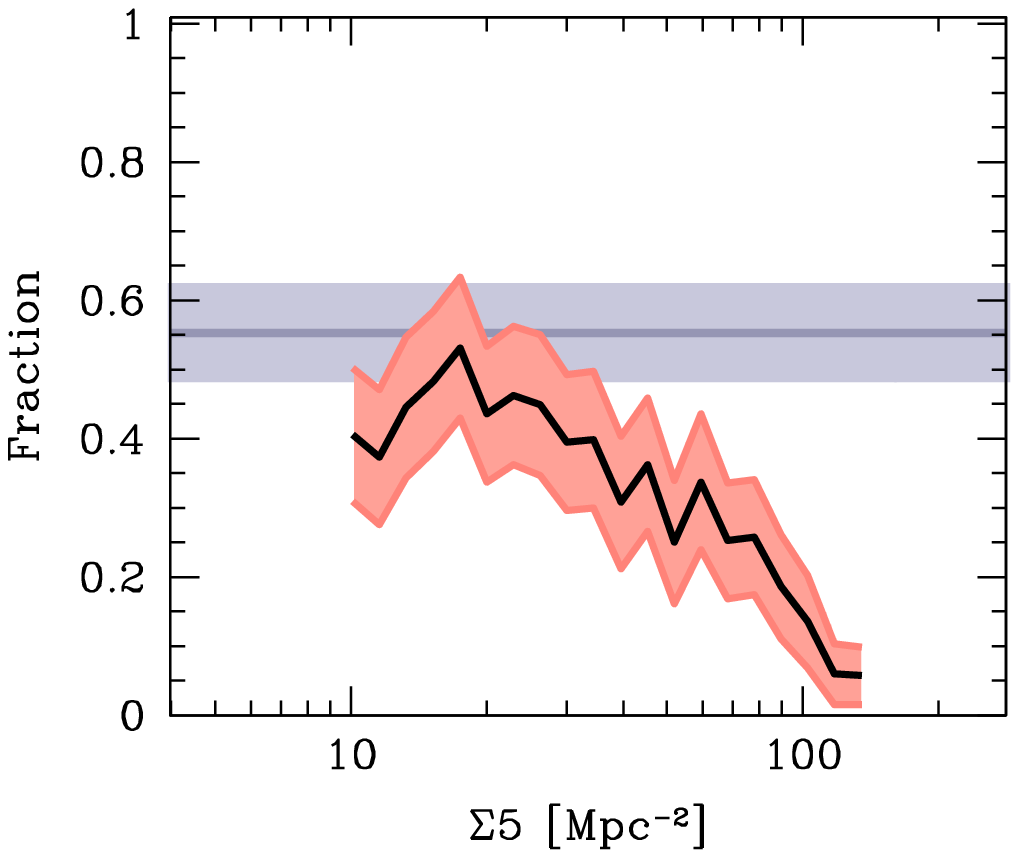}
      \includegraphics[bb=20 155 325 420,width=0.3\textwidth,clip]{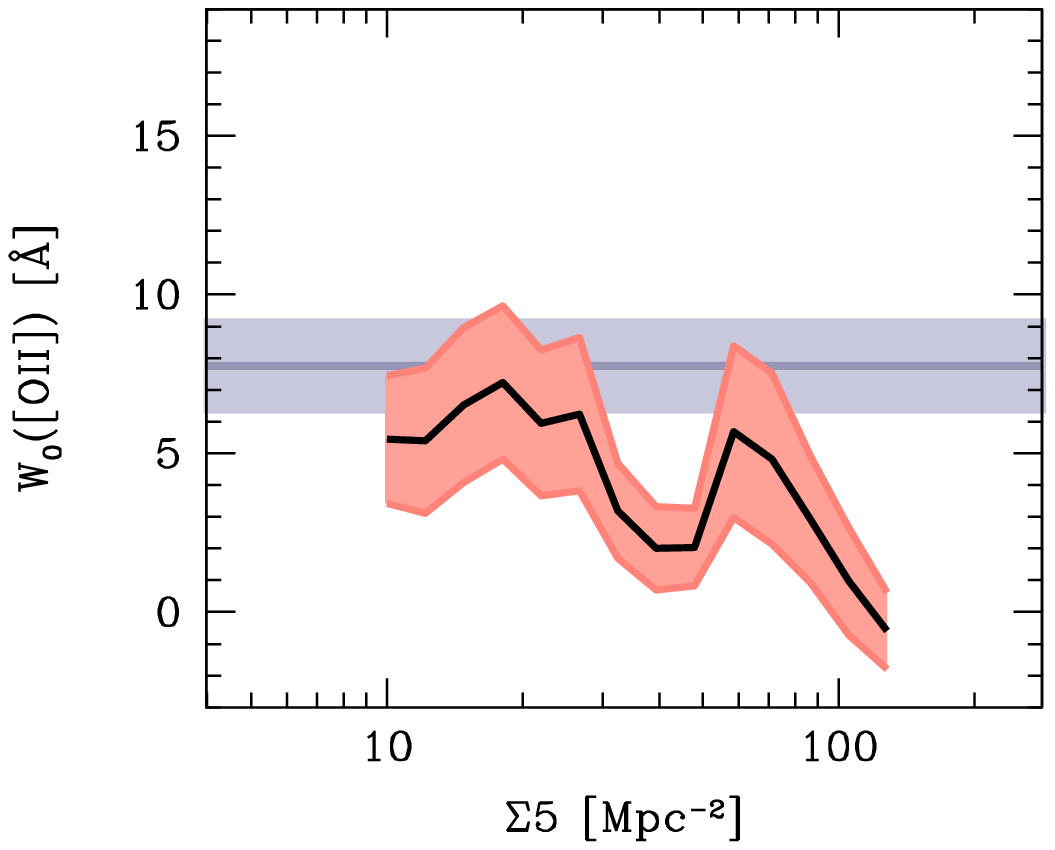}
      \includegraphics[bb=20 155 325 420,width=0.3\textwidth,clip]{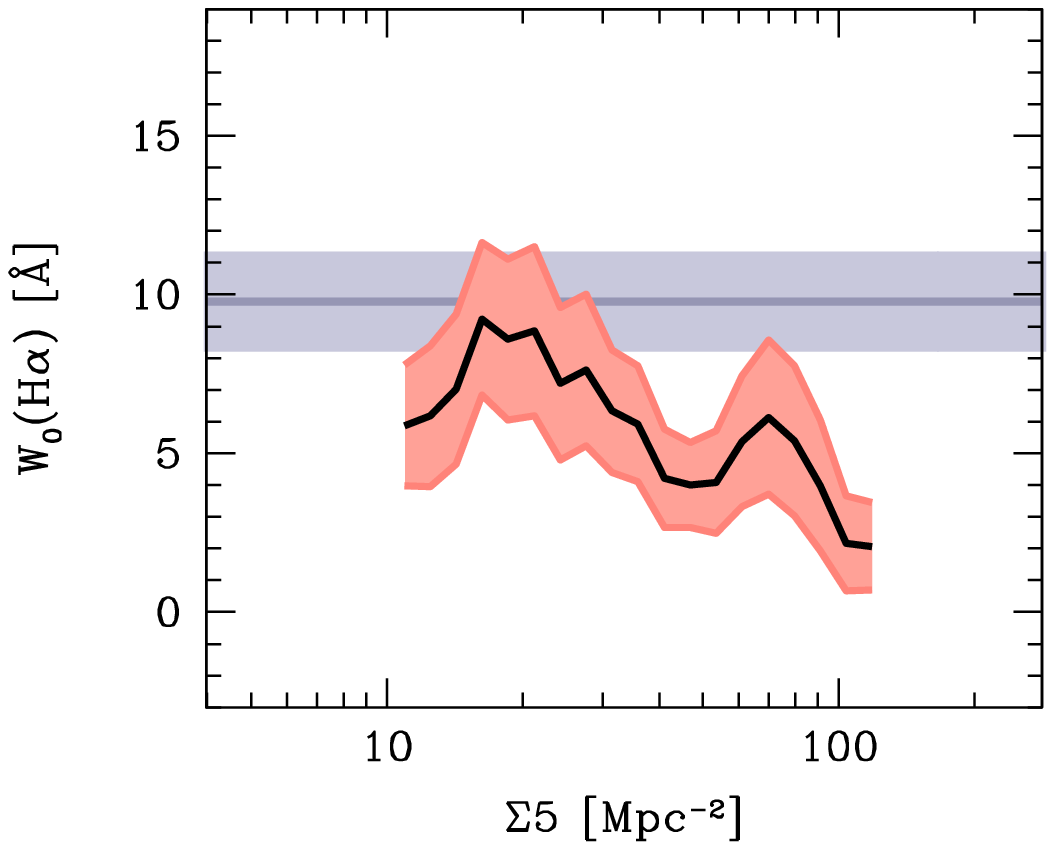}
     \caption{Fraction of star-forming galaxies (left panels) and  mean EWs of \oii\ (middle panels) 
	and \ha\     (right panels)  against normalized clustercentric distances  (top panels) and 
       projected densities to the 5$^{th}$ neighbor (\Den, bottom panels), plotted as the thick, 
       solid, black lines. The shaded areas around the curve in light blue are  the standard deviations of the
bootstrapped values.  The  horizontal areas show the  field values for galaxies between $0.15<z<0.35$.}
     \label{F:SFstats}
   \end{figure*}

The  shapes of those trends are  similar to the fraction calculated using emission lines 
as   indicators of  star-formation activity (Fig. \ref{F:SFstats}), which should not  be surprising since bluer 
colors often reflect the presence of young stellar populations.  However,  there is  an important 
fraction of star-forming galaxies with  red colors, and in principle it may break down the previous 
relation. They only appear to affect the fraction value, \ie the blue fraction is lower than the star-forming fraction at 
 fixed clustercentric distances and densities, but not  the shape of the trends. We  investigate  this further 
in  forthcoming sections.  

The fraction of blue galaxies was calculated  over the  nearest $N$ galaxies to each 
point in the plane, \ie inside  a moving box  containing a fixed number of objects, centered 
 on each galaxy.   Making the number $N$ too small increases the noise; making it
too large shortens the dynamical range covered, because this method truncates the extremities
of the lists. It was found that using the nearest 15--25 points is a good compromise between
spatial coverage and stability. 

To check the statistical significance, a bootstrap technique with 2000 iterations was 
applied to each value,  taking the  mean and the standard deviation of the bootstrapped values 
(checking previously if the distributions are compatible with Gaussian) as the final values and their errors, 
respectively.

Noise can increase or decrease as one includes  more or fewer points in
the calculations, but the overall shapes of the curves do not change, as for 
the case of choosing arbitrary bins.  This is particularly important in small samples and 
eventually under the effects of substructure.  The bootstrapping method helps to characterize the confidence region. 
This procedure is applied in all similar statistical analyses   in this work
As a final visual procedure, the lines were smoothed with simple spline fits; however, this procedure, however, 
only erases  local scale variations. 

   \subsection{Star formation activity and environment}
   \label{SS:SFenv}

We investigate further the dependence of the star-formation activity
on environment based on emission lines, which are sensitive to the ionizing radiation
coming from the newly-formed hot stars.  We plot the weighted fraction of star-forming galaxies 
(as defined in Sect.  \ref{SS:EW})   and the mean of \oii\   and \ha\ EWs in Fig. 
\ref{F:SFstats},  against normalized clustercentric distance and projected density, respectively. 
The field value is shown as a horizontal area in the plots. 

We observe that the star-formation activity is strongly suppressed in the 
cluster cores with less than 20\% of the galaxies forming stars. This fraction increases
steadily up to $\sim$50\% at $R\approx3\Rv$, but it does not reach clearly the field value of  $\sim$56\%.
of star-forming galaxies. This field  fraction is typical for those redshifts 
(see \citealt{hammer1997,balogh99,Nakata2005}). Although each of these authors used different cuts 
to define the star-forming population, the derived values agree within the statistical uncertainties.

However, the increase in the star-formation activity with radius is irregular. In a similar way to
the fraction of blue galaxies, we observe a peak at $R\sim 0.6 \Rv$ in both, star-forming fraction
and mean EWs. Only outside of 1\Rv, those indicators start to increase again.  
The explanation for this peak is discussed in  Sect. \ref{SS:bycluster}.

The mean  fraction of star-forming galaxies increases linearly towards low-density regions and 
reaches the field value only within the uncertainties.  The mean EWs of \oii\ and \ha\ 
follow  similar trends, but they also display  a peak at $\Den\sim60$ galaxies\,Mpc$^{-2}$.
The  mean EWs of those lines display similar values, which are slightly lower for \oii, even though
that in the local universe the typical relation is $\ew{\oii}\approx0.4\ew{\ha}$ (\citealt{Kennicutt92}).

The previous  trends indicate that the quenching of the star-forming activity starts 
at slightly larger clustercentric distances and lower  projected density that those sampled 
here. 

Several studies in the local universe  have found  that the star-formation activity reaches the field value 
 approximately  at clustercentric distances $\sim$2\Rv\  and projected densities around $\sim$1 galaxy Mpc$^{-2}$.
(\eg \citealt{lewis02,Gomez2003,Rines2005}). Those results  are compatible with the results found here, although those 
low densities are not reached in this study, but the clustercentric distances are,  
and we still observe slight star-formation depletion at distances $R>2\Rv$. 
As the field star-forming fraction in the local universe is much lower ($\sim$35\%, see \eg \citealt{Rines2005}), 
the radial trend found in this study is, therefore, steeper, indicating that the suppression 
of the star-formation activity in clusters  at $z\sim0.25$ was more effective, because the star-forming fraction in the 
internal regions of clusters is similar at all redshifts (\eg \citealt{balogh99,Nakata2005}).

\citet{Pimbblet2006} studied a sample of 11 clusters between 0.07$<$z$<$0.16, 
with  quite good coverage outside of  1\Rv. Alhough, the results point towards similar conclusions as the studies at $z\sim0$, 
the break in the star-formation activity appear to be  shifted  slightly towards
higher densities, an effect that we cannot confirm nor exclude, although in our case the fraction approaches
to the field value at $\Den\approx$10 galaxies Mpc$^{-1}$, but the differences on the density calculation hamper
direct comparisons. 

At higher redshifts, most studies have been focused on the central regions of clusters 
(\eg \citealt{balogh99,balogh02,Balogh2002}). Our results complement those studies, 
sampling  similar clusters at larger clustercentric distances, with focus on the cluster-field 
interface. It  also bridges the studies  being performed by deep surveys 
 which have  focused mainly on  low-density regions (\eg \citealt{Elbaz2007,Cooper2007,Franzetti2007}).

   \section{Origin of the trends}
   \label{S:origin}

To explore the origin of  trends described in the previous section, we split the 
sample into different subsamples according to various criteria. 
 
   \subsection{The star-forming population}
   \label{SS:onlySF}

We analyze first the properties of the star-forming population only, defined
to be  the galaxies with equivalent widths $W_0>5$\,\AA. The dynamical range of radius and 
galaxy densities is smaller because the subsample is smaller than the original sample. 

\begin{figure}[t]
     \centering
     \includegraphics[bb=25 155 325 420,width=0.49\columnwidth,clip]{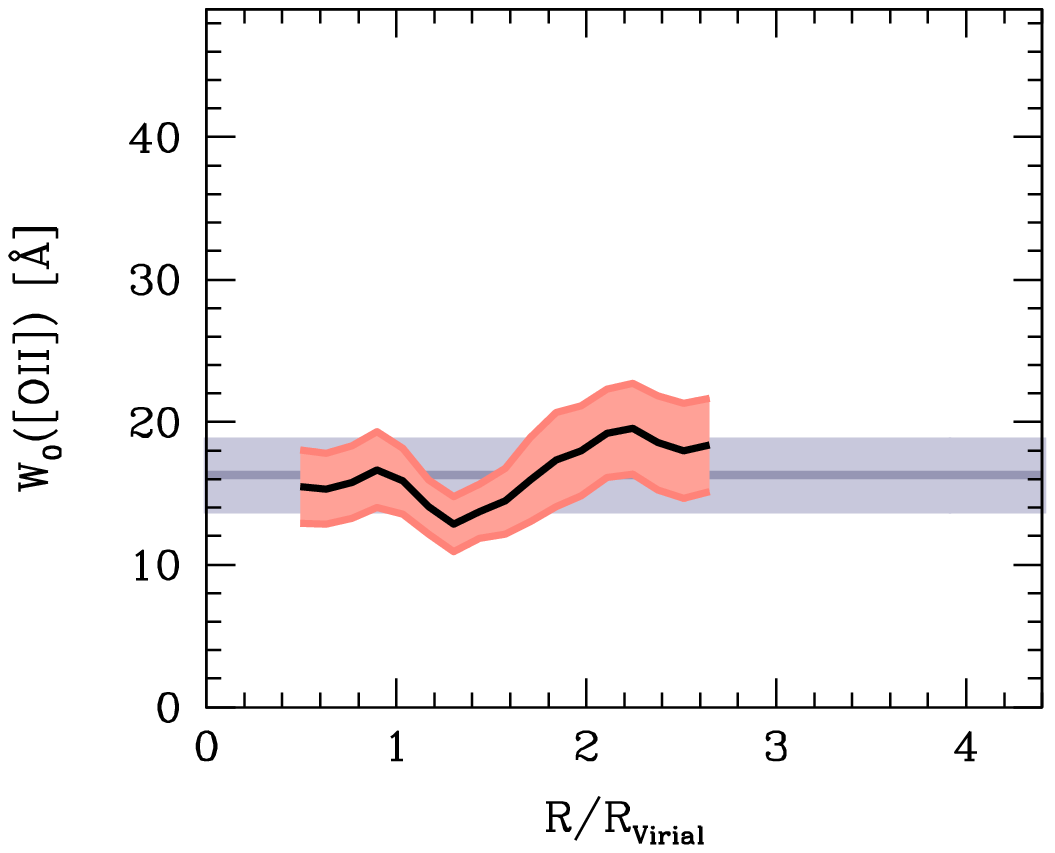}
      \includegraphics[bb=25 155 325 420,width=0.49\columnwidth,clip]{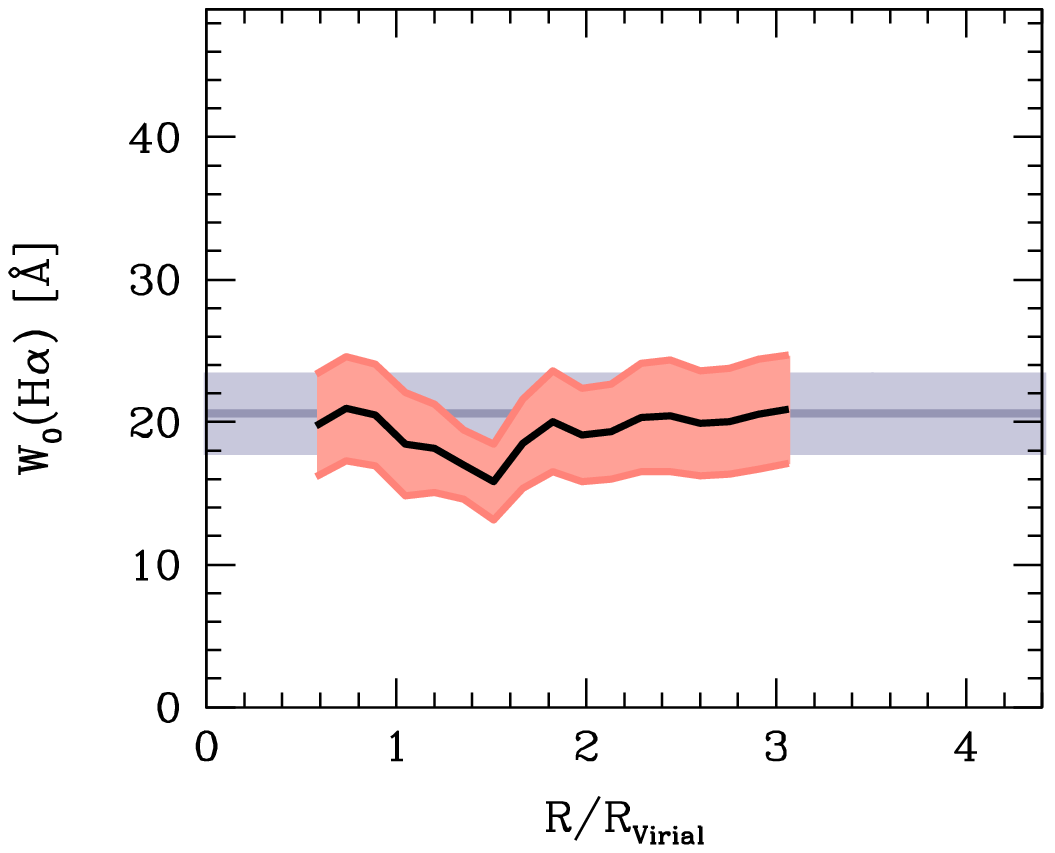}
     \includegraphics[bb=25 155 325 420,width=0.49\columnwidth,clip]{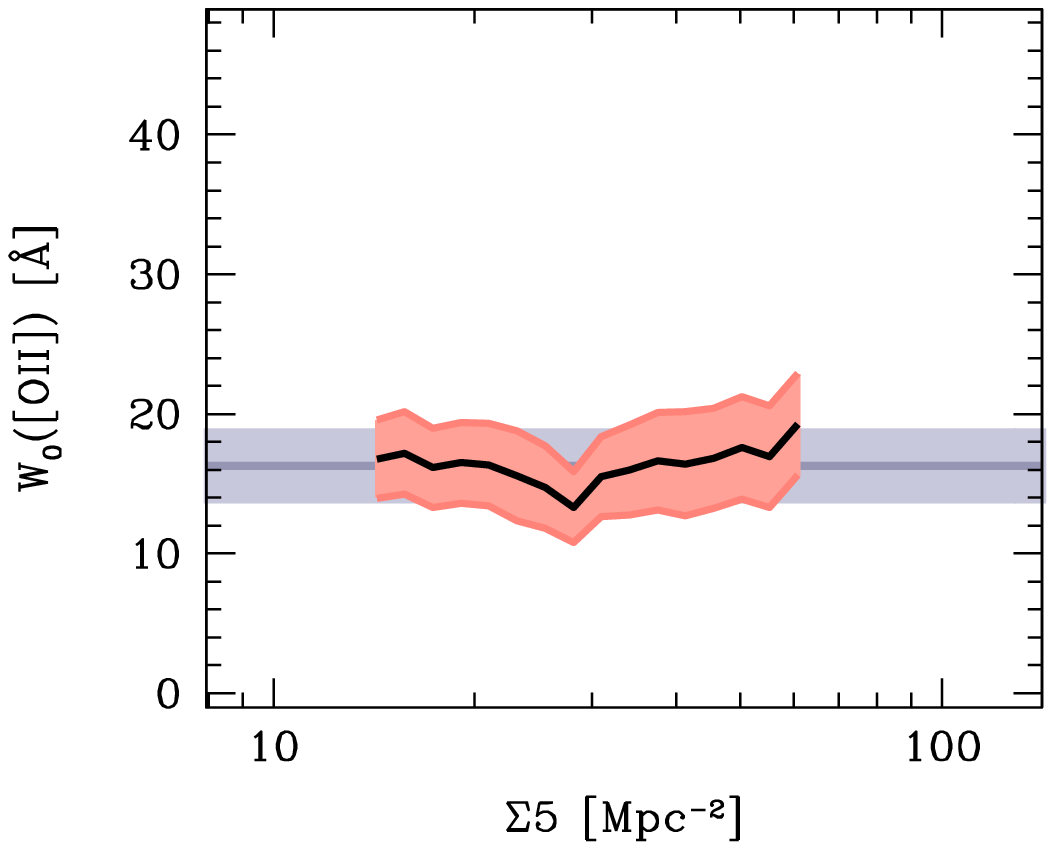}
        \includegraphics[bb=25 155 325 420,width=0.49\columnwidth,clip]{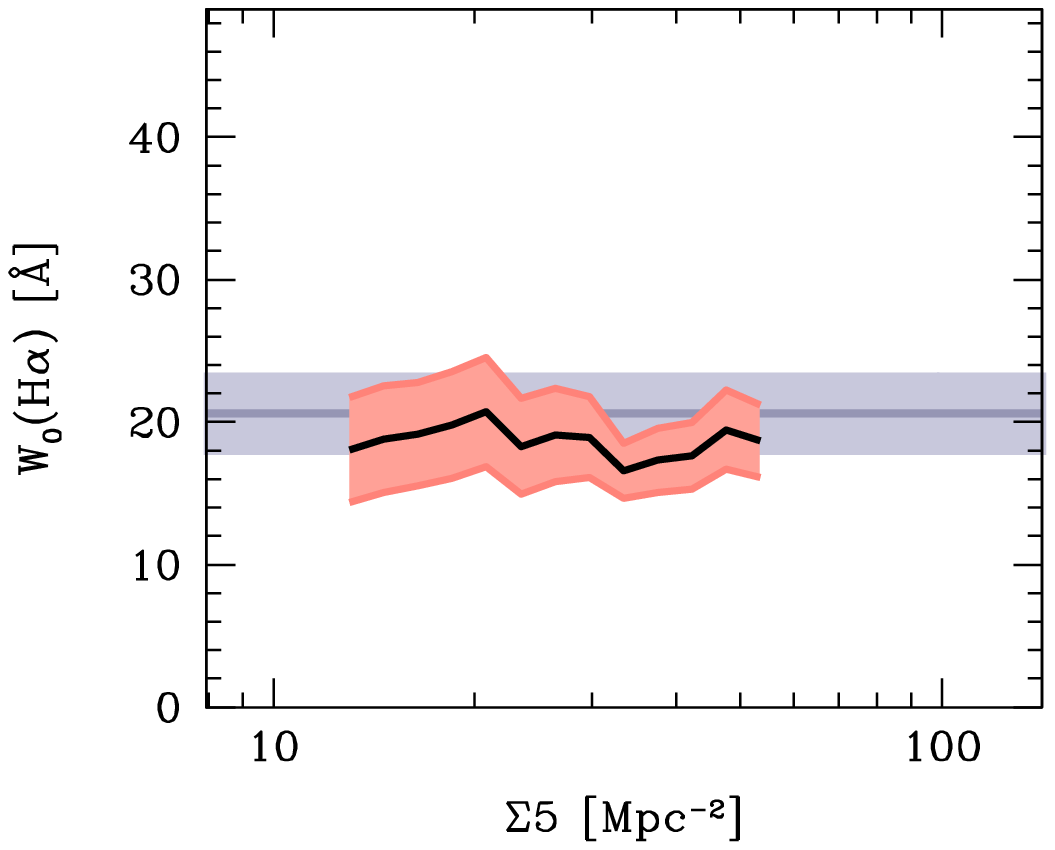}
     \caption{Similar to figure \ref{F:SFstats} but now analyzing the distribution
       of the star-forming population only (\ie $\ew{\oii,\ha}>5$\AA).}
     \label{F:onlySF}
   \end{figure}

The mean EWs (Fig.  \ref{F:onlySF}) remain stable over 
a wide range of clustercentric distances and density values and  are statistically similar to 
those found for field star-forming galaxies, which imply that the  
populations do not differ substantially. This leads to the conclusion that the trends seen in Fig. 
\ref{F:SFstats} are driven only by the change in the relative numbers of star-forming and passive
galaxies in different environments.

This result is similar to the findings of \citet{Balogh2004} and \citet{Rines2005} at $z\approx0$ 
who found that  the mean \ha\ EWs display a similar distribution for star-forming
galaxies located in ``high'' and ``low'' density environments. 

This behavior of the active galaxy population, together  with the strong bimodality, observed in 
colors (\citealt{Balogh2004B}) and EWs (\citealt{Haines2007}), detected
in large local surveys  favors mechanisms that trigger a rapid evolution  between galaxy subtypes.

 \subsection{Subsamples according to membership}
   \label{SS:bycluster}
 
\begin{figure*}[t] 
  \centering
  \includegraphics[width=0.4\textwidth]{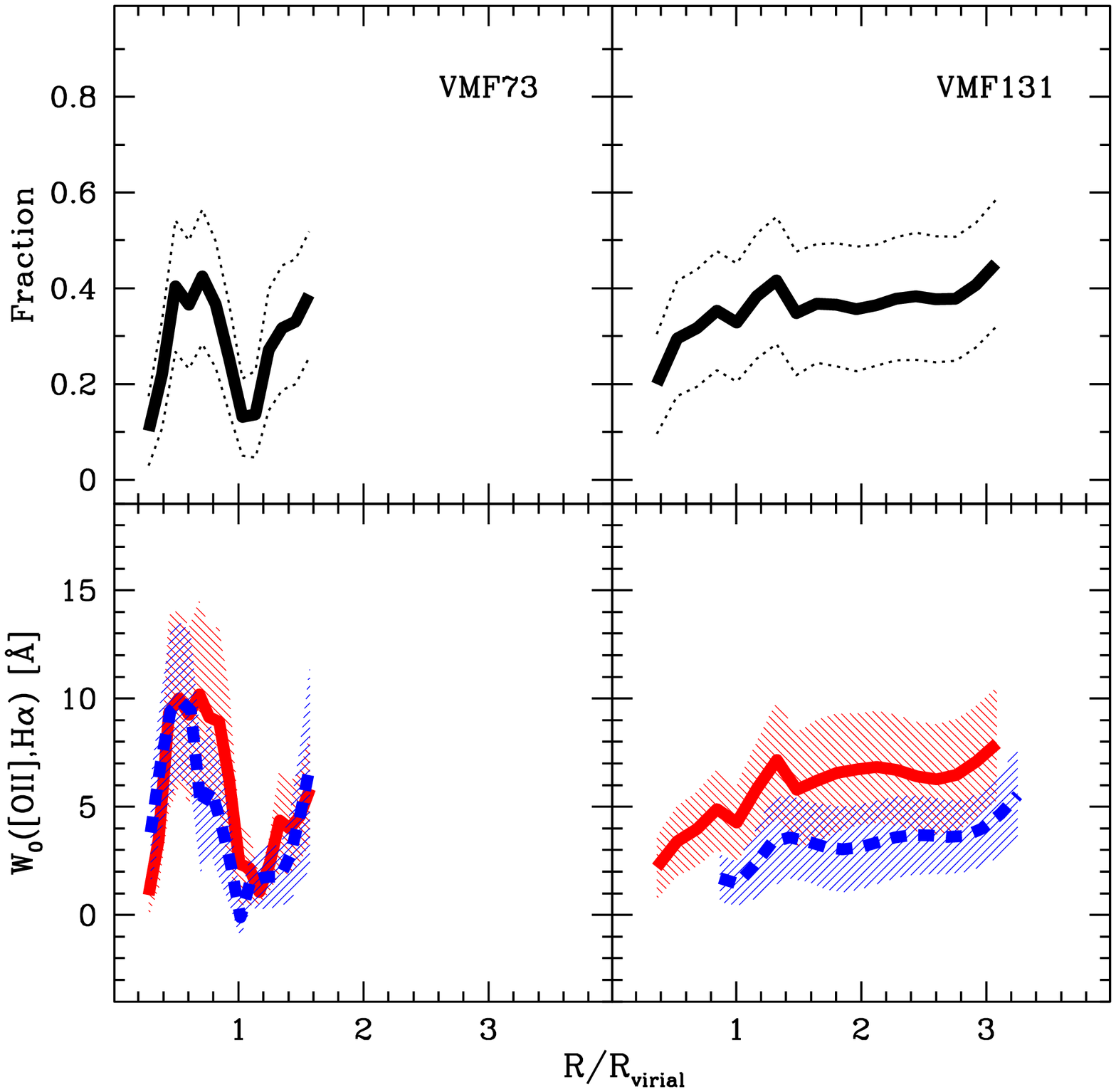}
 \includegraphics[width=0.4\textwidth]{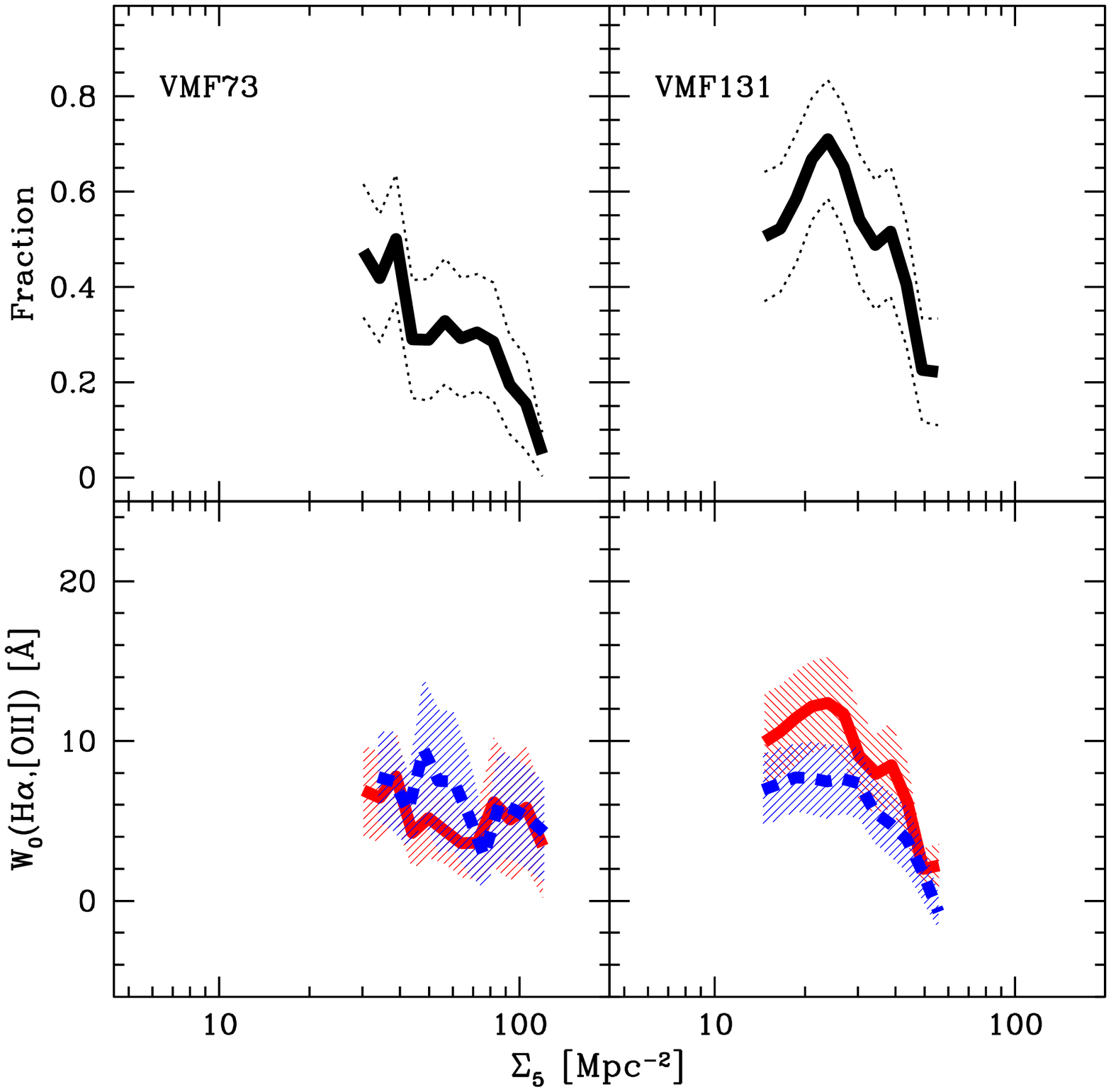}
  \caption{Fraction of star-forming galaxies and mean equivalents widths against
normalized cluster distance and projected density for the clusters VMF73 and VMF131
as depicted in the respective panels. In the bottom panels, 
dashed blue lines are for \avg{\ew{\oii}}  and solid red lines for \avg{\ew{\ha}} 
   bottom panels the  dashed  blue line represent the mean \oii\ EWs 
and the solid red line  the \ha\ ones. The respective 1-$\sigma$ 
are marked as the hashed areas in the bottom panels and thin dotted lines 
in top panels. }
  \label{F:Rbycluster}
\end{figure*}

Given  the relative small sample and some unusual features in the composite cluster, 
we  investigate  the influence of individual clusters on the final measurements for the composite cluster. Since 
two clusters, VMF73 and VMF131, account for an important fraction of the data used in the composite cluster,
we   investigate them individually.  Here, given the smaller number of galaxies, we 
are forced to use fewer data points in our statistical analyses, which increases the level of  noise. 

The results can be seen in Fig. \ref{F:Rbycluster}.  We note striking differences 
between the clusters, especially in the radial distribution. The trends for the cluster VMF73  show 
  peaks inside  1\Rv. Therefore, we conclude that   the peaks detected in the   global trends are exclusively 
due to this cluster. The  existence of this peak,   or rather the depletion at $\sim 1\Rv$ 
is likely an effect of a secondary  structure in this cluster (see Sect. \ref{SS:r285}), because 
the radial gradient is the combination  of both substructures. This can be taken as 
additional evidence that the X-ray structure detected by \citet{Rasmussen2004} 
actually belongs to the cluster. It  may  form  part of an infalling group and clearly has a noticeable 
effect on the galaxy population  of this cluster. Additional effects may arise from the geometrical configuration 
of the  cluster at $R<1\Rv$, given its elongated galaxy concentration.  Those features passed unnoticed 
in the previous analysis of \citet{gerken04} as the fixed bins used there effectively erased the detail.

VMF131 shows, on the other hand, a modest but steady increase in its star-forming
activity towards larger clustercentric distances.   This cluster is quite well studied at large radii. 
Thus, the general trends of the composite cluster at these distances are  very dependent on it.

Since density probes environment independently of the cluster geometry, cluster substructure does not
affect, in principle, the correlations. Nevertheless, we observe that the trends for these 
two clusters are quite different. VMF73 shows a  sharp increase in the fraction 
of star-forming galaxies towards lower projected densities 
but a modest increase in their overall activity, as measured
by their EWs.   VMF131 displays an increase in 
its fraction of star-forming members and  the average star-formation activity is similarly increased.

The scatter of the galaxy population inside  clusters has been already noted, it does not however depend strongly on  their X-ray 
luminosity nor velocity dispersion according to \citet{popesso2007}, although \citet{Poggianti2006}
find both a weak correlation of galaxy properties with cluster mass and evolution of the correlation  with redshift. 
This scatter may be related to more subtle properties, such as cluster substructure, mass-assembly history 
and intra-cluster gas distribution, as  well as  the properties of the large-scale structure surrounding
the clusters.

 \section{The  case of the red star-forming galaxies}
   \label{S:redSF}

We already noted in Sect. \ref{SS:galcolor} the existence of a sub-population
of cluster galaxies with emission lines but  red colors. Twenty-five  out  of
56 star-forming galaxies belong to this population. 
Their average EWs  are   $\avg{\ew{\oii}}=14.8\pm 2.48$ \AA\ and 
 $\avg{\ew{\ha}}=19.9\pm 4.90$ \AA, respectively,  similar (within 1-$\sigma$ significance levels) to  the 
mean  star-forming population (see Fig. \ref{F:onlySF}). They do not seem to populate any special 
environment in the cluster, being more or less evenly distributed in radius and density, which explains
the similarity between the blue and the star-forming fraction (Fig. \ref{F:bluefrac} and \ref{F:SFstats}).
They also span the  full range of luminosities covered by this study. 

Galaxies with a red SED and star-formation activity have been routinely reported
at intermediate redshifts,  either in the field (\eg \citealt{hammer1997})  or in clusters 
(\eg  \citealt{Demarco2005}). In the case of the local Universe, a recent  paper by 
\citet{popesso2007} reports that red star-forming  galaxies constitute on average 
 25\% of the entire cluster population.  They suggest that those objects are in the process of 
evolution from late to early types.   \cite{Wolf2005}  identified hundreds in the field of  the supercluster
 A901/902 ($z\approx0.17$) based on the information content in the medium-band photometry 
of the COMBO-17 survey. They interprete  the color of those galaxies as a product of the combination 
of old stellar  populations and dust extinction.  Similarly,  \citet{Tanaka2007}  presented  indication
of  red galaxies  with younger stellar populations in  groups around a $z=0.55$ cluster. 
 They argued  that those red galaxies have truncated their star formation activity recently, on a short timescale, 
but that they host a large fraction of old stars in a addition to a  reasonable amount of dust. 

On the other hand, \citet{martini2002},  based on ROSAT X-ray data, reported an unexpectedly high 
fraction of AGNs in elliptical  galaxies in a massive  $z=0.15$ cluster,  which did not show optical 
signatures.  Although their sample is small, the fraction of obscured AGNs is similar to the 
fraction of blue galaxies identified in that cluster.  Furthermore, \citet{Yan2006} found that more 
than the  half of red galaxies in the SDSS-DR4 show emission lines, most of them 
consistent with being  low ionization nuclear emission-line regions (LINERs) . However, the 
LINERs  may not be  due only  to AGNs, for example  \citet{Sarzi2006} report \emph{extended} 
LINER-like emission in several early-type galaxies in their spatially-resolved spectroscopy. 
Therefore the question  is not clearly settled.

To decide whether those galaxies are AGNs or not, and to what degree  
our star-forming galaxies may be contaminated by nuclear activity,  we performed 
some tests based on the emission lines. We note that we may be unable to detect obscured AGNs.
 No galaxy shows signs of broadening typical of Seyferts 1, but  Seyferts 2 and 
LINERs may still be  present.  We calculate the ratios between emission lines 
(\oii, \hb, \oiii$\lambda$5007, \ha\ and \nii), where is possible since all lines are
 rarely  present altogether.  We conduct separate tests to check all possibilities. 

\begin{figure}[t]
     \centering
     \includegraphics[bb=20 150 470 470,width=0.5\textwidth,clip]{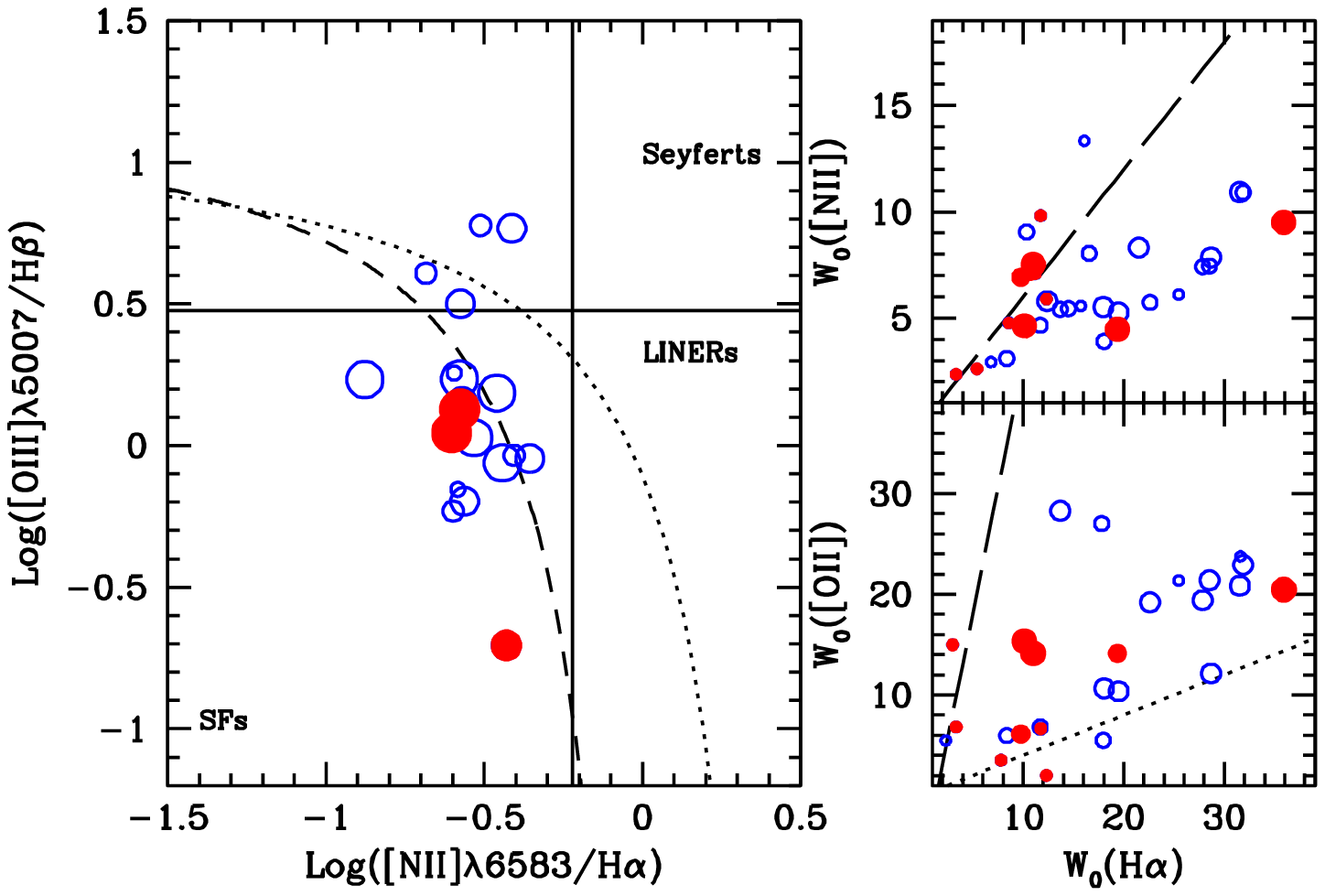}
     \caption{Line ratio diagnostic diagrams to identify AGNs. The left panel is the BPT  plane
showing the relation between four  emission lines. The dashed  and  dotted curves separate 
AGNs from star-forming galaxies  (see text). The vertical and horizontal  lines are 
the approximate separation between types. The lower right panel is the \oii--\ha\ diagram aimed 
to identify LINERs. The dotted line is the local Kennicutt's relation for star-forming galaxies, 
whereas the dashed line is the test to identify LINERs. The upper right panel is the relation 
between \nii\ and \ha\ EWs. Blue open circles are ``normal'' star-forming galaxies and 
red filled ones are the red star-forming galaxies. The size of the symbols is related to the 
confidence with which each index can be measured, the larger the better. }
     \label{F:AGN}
   \end{figure}

The first classical test  put the galaxies into the BPT plane  (\ie log(\oiii/\hb) vs log(\nii/\ha), 
\citealt{Baldwin1981}). Each par of lines are close enough to use the EWs instead 
of  the fluxes. We plot in Fig. \ref{F:AGN} all galaxies for which those indexes can be measured. 
The lines are the empirical separation between star-forming galaxies and AGNs of \citet{Kauffmann2003}:

\begin{equation}
\log\left(\frac{\oiii}{\hb}\right)=\frac{0.61}{\log\left(\frac{\nii}{\ha}\right)-0.05}+1.3
\end{equation}
\noindent and the theoretical predictions  of \citet{Kewley2001}
\begin{equation}
\log\left(\frac{\oiii}{\hb}\right)=\frac{0.61}{\log\left(\frac{\nii}{\ha}\right)-0.47}+1.19
\end{equation}

The separation between galaxy types is made using  $\oiii/\hb>3$ and  $\nii/\ha>0.6$, with the latter also used
independently  for all galaxies where these two lines are present, which occurred more often than in the 
case of the four lines test. 

The latest test was proposed by \citet{Yan2006}. It uses only the ratio between \oii\ and \ha\ 
EWs and  was aimed  mainly to detect LINERs.

 \begin{equation}
     W_0(\oii)>5 \cdot W_0(\ha)-7
     \label{E:AGN-O2}
\end{equation}

In total, 10 galaxies show some signs of  AGN activity with 6 being galaxies  classified as  ``red star-forming''.
Note that,  the emission-line data for all AGN candidates are positioned  close to the boundaries of the respective tests,  indicated
in Fig. \ref{F:AGN}, which means  that  their nuclear activity is rather low or composite.  The exclusion of these 
AGNs candidates does not affect the results shown in Fig. \ref{F:SFstats} and \ref{F:onlySF}, which is an expected
result because AGN frequency is  not correlated with environment (\citealt{Miller2003}).

As  noted before, in   Fig.  \ref{F:SFstats} and \ref{F:onlySF},  the mean EWs of   the \oii\ and the \ha\  lines  display similar values, even though  in  local samples the relation between the 
EWs of these emission lines follows the Kennicutt's law ($\ew{\oii}\approx 0.4\ew{\ha}$, \citealt{Kennicutt92}). 
This can be more clearly seen in the lower right corner of Fig. \ref{F:AGN}, where the  Kennicutt's relation is 
plotted. No clear explanation  has been found for this deviation, but  \citet{hammer1997}  reported 
the same effect in the Canada-France  Redshift Survey  galaxies at similar redshifts. 
They presented various hypotheses that  may apply to our work, such as, lower extinction, 
lower metallicities and contamination by AGNs. However,  we exclude the possibility of  here
a strong AGN contamination and most  \emph{normal} star-forming galaxies 
also present these ``unusual'' values.  The deviation is therefore probably caused by the lower 
metallicities  present in distant galaxies (see \citealt{Kobulnicky2004}),  
because the \oii-\ha\ ratio depends strongly on this parameter (\citealt{Jansen2001}).

 We can estimate the contribution of dust extinction using the \citet{Tully1985} extinction laws 
for disk galaxies. At $z\approx0.25$, the $V$ 
and $I$ filters correspond approximately to restframe $B$ and $R$-bands. At a given
inclination angle, the extinction in the $R$-band is  $\sim0.56E(B)$:  only disk galaxies with inclinations 
larger than 60$^\circ$ will therefore have a correction factor $E(B-R)>0.2$\,mag  (see Table 1 in  \citealt{Bohm2004}),
 a value sufficiently large to move their data-points away from  red-sequence.

Our \emph{ground-based} INT images do not allow us to securely classify the morphological properties
of our galaxies, since the  typical seeing of $\sim$1\,arcsec ($\sim$4\,kpc at $z=0.25$) represents approximately one 
scale-length for spiral galaxies  (\eg \citealt{Bamford2007}). Basic properties can however be obtained, as  
galaxies in our sample typically have an apparent size of 5--10\,arcsec. After examination, we find that out 
of the 25 ``red  star-forming'' galaxies,  11 are  clearly spirals, 11 appear bulge-dominated, two are irregular and  one 
galaxy, which is also an AGN candidate, shows signs of interaction. 
Out of  11 spirals, 8 are probably edge-on galaxies and the remaining 
three, face-on. 

Dust extinction can, therefore, explain the colors of only a fraction of the red star-forming objects, 
because dust properties at  $z\sim0.25$  do not differ much from those of the local universe, and highly-tilted galaxies 
can be easily distinguished.

\begin{figure}[t]
     \centering
     \includegraphics[width=0.5\textwidth]{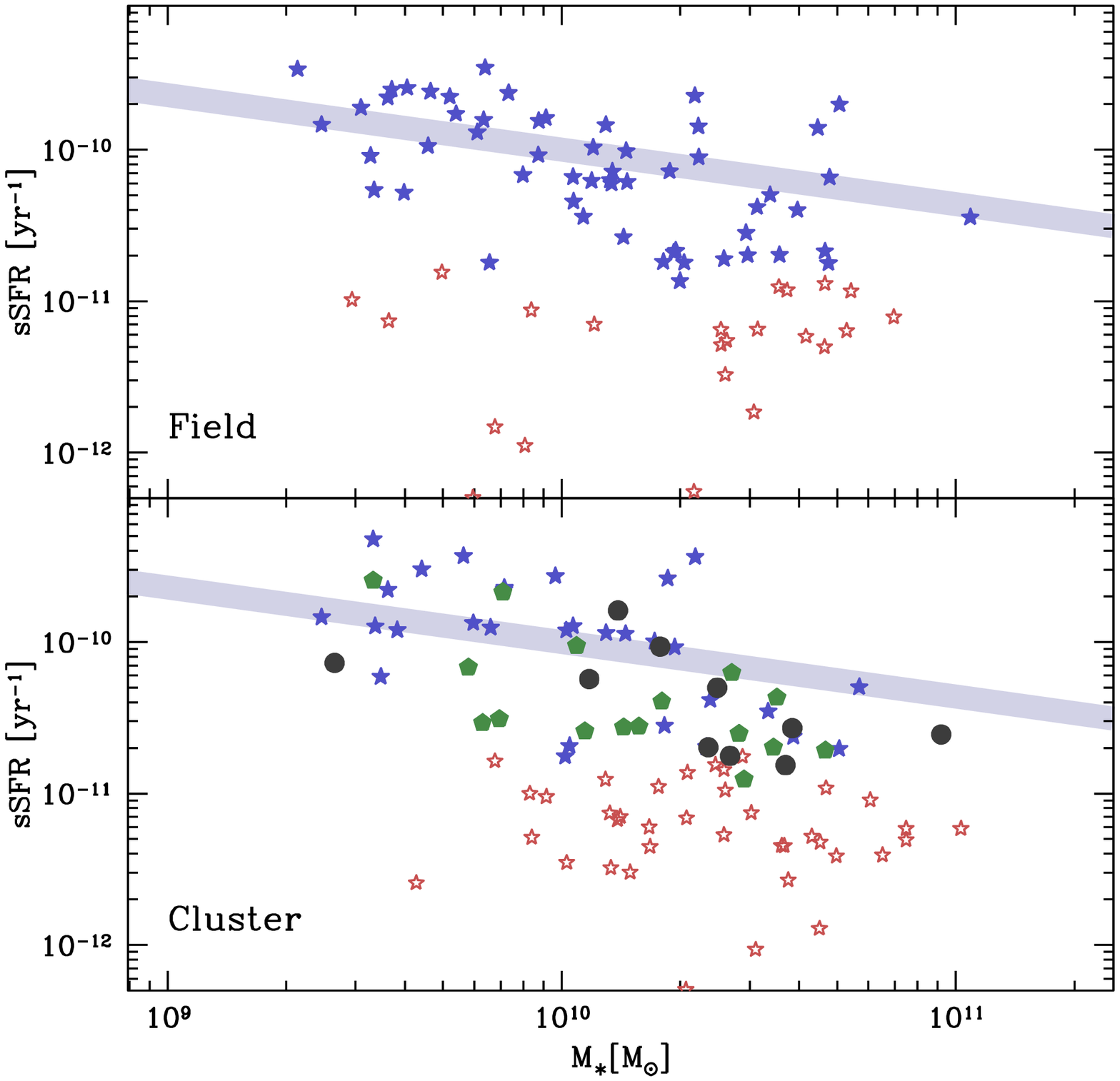}
     \caption{Specific star formation rates for field (top) and cluster (bottom) galaxies, 
versus stellar mass. Normal star-forming galaxies are plotted with blue filled stars and the red star-forming 
galaxies with green pentagons. The red open stars are passive galaxies with $0<\ew{\oii,\ha}<5$ 
and are shown for comparison. The black circles are the AGN candidates. The thick line is the local relation
from  \citealt{Salim2007}.}
     \label{F:SFRcolor}
   \end{figure}

The red star-forming galaxies appear to be  transition objects populating the ``green valley" 
(\citealt{Salim2007}) in Fig.  \ref{F:SFRcolor}, where we plot the specific star-formation (sSFR)
activity\footnote{The derivation of the sSFRs  is described in Appendix \ref{S:SFR}.}
 versus stellar mass. Most of the red-star forming galaxies (as well as some AGNs) 
are located in a ``transition region'' between  normal star-forming galaxies and passive ones\footnote{Here 
galaxies with  $5>\ew{\oii,\ha}>0$ are also plotted. Galaxies with negative EWs were not 
used  since the  calculations yield unphysical (\ie negative values) results.}. The mean sSFR for normal
star-forming galaxies is $\sim(1.08\pm0.65)\times 10^{-10}$\,yr$^{-1}$, whereas the red star-forming
galaxies have on average $sSFR\approx(2.4\pm0.6)\times 10^{-11}$\,yr$^{-1}$, about an order of magnitude
lower. The average upper limit for passive galaxies is  $sSFR\approx(4.8\pm3.3)\times 10^{-12}$\,yr$^{-1}$
because we do not include galaxies with unphysical star-formation rates.

We note that those galaxies may  also be present in the field, although we cannot clearly 
identify them,  given the uncertainties in the k-corrections  of $\sim$0.2\,mag, which are larger 
than typical red-sequence scatter. However, it can be seen that most of the normal field and cluster star-forming galaxies 
 are located in the upper part of 
this diagram, around the relation for local galaxies found in the UV-selected sample of \citet{Salim2007}.
We note that field and cluster star-forming galaxies are located in similar regions of this diagram, an additional 
indication that both populations are composed similar classes of objects. However, the red star-forming galaxies 
appear to be clearly offset from this relation. 

It is  interesting to note that the definition of a star-forming galaxy set at $\ew{\oii,\ha}\geq5$\,\AA\  not only 
has a observational sense but also a physical meaning and corresponds to a
 $sSFR\approx2\times10^{-11}$\,yr$^{-1}$, a rate sufficiently low to consider a galaxy as passive. 

It is part of the standard picture of galaxy evolution that objects in the blue cloud (here the 
star-forming sequence) slowly grow in stellar mass via gas  accretion over cosmic times. Mergers 
and other strong interactions can trigger star-bursts, displacing upwards the galaxies in the diagram 
(Fig. \ref{F:SFRcolor})  adding  a large amount of stellar mass in a short period of time (so moving rightwards).
 On the other hand, gas exhaustion or gas  removal by means of interactions or feedback processes can 
 lead  to a quenching of  star-formation that moves the galaxy downwards, towards the red sequence, 
where it can experience small episodes of star-formation,
accrete more gas and  move again into the blue cloud, or stay permanently there if the environment is hostile (as in the
galaxy clusters cores). 

In this picture, the red star-forming galaxies are located in an intermediate stage between 
the two main subtypes, with  lower but still appreciable amounts of star-formation.  The variation in the 
abundance of this population with cosmic time may provide  additional insights into the nature of  the stellar mass buildup,
although, a more careful treatment of AGN activity, dust extinction and stellar population is required to fully explain 
their nature. 


\section{Summary and conclusions}
\label{S:conclusions}

We have obtained MOSCA spectroscopy for 149 member galaxies in 6 clusters at $\avg{z}\sim0.25$,
out to large clustercentric distances. This sample is compared directly with 97 galaxies in the field.
The spectroscopic dataset is complemented with $V$ and $I$-band photometry in the three fields and
multiband photometry  from the SDSS in two of them.  The main findings can be summarized in the following.

\begin{enumerate}

\item The suppression of the star-formation activity can be detected at
large clustercentric distances ($R > 1 \Rv$) and low densities $\Den<10$ Mpc$^{-1}$,
in an environment where the cluster is supposed to have little influence. 
This result  agrees with similar results at 
redshift $z \approx 0$ based on the 2dF Galaxy Redshift Survey 
  \citep{lewis02} or SDSS \citep{Gomez2003}, where a critical value of
  density  was found, below which  the  environment  appear to begin to play a critical role. 
Although our density estimates are not directly comparable to these low-redshift studies, it is possible that
we did not reach this low threshold of $\Den\sim 1 \mbox{Mpc}^{-2}$ reported by those studies.
Our investigation  reached  star-formation activities  close to  those  found in the field, 
probing the transition between field and cluster  environment in the distant Universe. 
The decrease of the star-formation activity is smooth with increasing density, but a more complex
behavior was found when the radial dependence is studied, as it is strongly affected by substructure.

\item The trends in the star-formation activity measured by the mean \oii\ and \ha\ EWs are due  mainly  
 to a strong  decrease in the relative number of  star-forming galaxies towards higher densities and   
smaller clustercentric distances, rather than  a slow decline in the star-formation rates of galaxies.  
This finding favors violent   suppression of the star-formation activity.
 
\item Despite the importance of the overall trends, important differences are found
between the studied clusters.  The two most well-studied clusters were analyzed separately from all other clusters..
It was found that the shape of the star-formation gradients were quite different from each other..
This difference was more accentuated in the radial trends since the effects of 
substructure  could not be discerned in the assumed radial density profile. 

In the literature, many studies  have focused either  solely on one
usually well-sampled cluster (\eg \citealt{Kodama2001,Demarco2005,Sato2006}), 
or on a family of clusters (selected by X-ray luminosity,  redshift range, etc),  generally far less well-sampled,
which typically combine all data to create  a ``composite-cluster'' (\eg \citealt{balogh99,balogh02,Pimbblet2006}) 
in a similar way to our analysis here. However, our study indicates that many of 
the overall trends  may not be universal, but may be strongly related to the particularities of
the system that the individual galaxies belong to.  Therefore, the effects of the substructure should
not be neglected when analyzing the universality of star-formation-environment relation,
because each particular system may have different properties  (see also \citealt{Rines2005} for
a similar result at $z=0$).

\item   The clusters  show variations  not  only due to the substructure, but also
in their galaxy populations. For example, we  detected an important sub-population of
red star-forming galaxies in some  clusters, which have similar colors or are redder than the 
red-sequence.  The characteristics of this  population, as  measured by their environmental distribution, 
do not differ  much from the remainder of the emission-line  population. A fraction of them could be  
AGNs, but the AGN contamination is not larger than in the rest of the star-forming population. 
Nonetheless, all AGN candidates show relatively low activity. 

Dust may play  a role because some  galaxies  are clearly edge-on spirals. This effect may be
o present in other galaxies. It is, however, intriguing that some otherwise blue active star-forming 
galaxies  have the precise amount of dust to make them fall onto the narrow red sequence. 

These two effects together, however, are only able to explain  a fraction of this population. 

These galaxies are located in a transition zone, between normal star-forming galaxies 
and passive ones,  where galaxies appear to form stars at a relatively lower rate. They may be in the process of shutting 
down  their activity and/or they can contain  a relatively significant old stellar population  combined with 
a moderate amount of dust.  If these galaxies are truly transition objects, their abundance may provide 
important clues about the mass-assembly history as galaxies  grow in mass via accretion and merging 
and shut down their star-formation  over cosmic time  (\eg \citealt{Bell2005}).

\end{enumerate}

Our results favor mechanisms of strong star-formation suppression. Among the
preferred processes is ram-pressure stripping  (and other strong galaxy interactions 
within the intracluster medium).  This process can quench the star-formation on timescale  as short as 1\,Gyr,
which is the dynamical timescale of a cluster passage. Ram-pressure is very  effective in the central regions
of the clusters (\eg \citealt{Kapferer2007}).  We detect however star-formation depletion at clustercentric distances
as far as $\sim3\Rv$.  It is possible  that  many galaxies in the outskirts have already passed
through the denser intracluster media. In fact, models by \citet{Gill2005} predict that as many as  half
of the galaxies between 1--2.5\Rv\ may be ``bouncing" after a first passage (the ``backsplash'' scenario) 
and thus have experienced strong interactions in the inner cluster core for  a sufficient time to explain their passive nature. 
Therefore, ram-pressure stripping cannot  disregarded  as a  important mechanism, particularly because, since direct evidence 
of this process at work has been reported by some authors (\eg  \citealt{Boselli2006,Cortese2007}).

Other processes may be still acting, because  quenching  of star-formation is observed at distances larger than those predicted 
by the \citet{Gill2005} simulations. Also, their proposed ``backsplash'' population would only account for a fraction 
of the galaxy population in the cluster outskirts.  
Any other process that quenches that star-formation more gradually
(\eg starvation, harassment, etc) would  have been detected via enhancement or depletion
 in the star-forming population with environment, which is not the case. One possibility is
that other processes affect the star-formation on galaxies before they begin  to fall into the clusters, 
in groups and filaments embedded in the large-scale structure. In those
environments, several processes are thought to be effective in changing the galaxy stellar
populations.  Ram-pressure stripping may still be  effective in systems of lower masses
under certain certain conditions (\eg \citealt{Fujita2004,Hester2006}) and thus may contribute. 
Merger and tidal interactions in those environments 
are also likely and they can trigger  starbursts that consume  gas rapidly and strip
the remaining gas via  feedback mechanisms(\eg \citealt{Bekki2001b,Fujita2004}). 
This scenario is  compatible with the recent findings of  \citet{Tanaka2007} and  \citet{Haines2007}. 

It is important to note that every cluster is a particular entity of its own and it is likely that different 
processes are important. They can depend on  the cluster history and configuration,
as well as on the characteristics of the surrounding environment. These effects may influence the 
galaxy population that inhabit the clusters as shown recently by  \citet{Moran2007a}. 
This view is  supported here by the different star-formation gradients detected, due mainly to
substructure and the abundances in the galaxy population, with some clusters harboring 
an important fraction of red-star forming galaxies, which may be important in the general scheme
of galaxy evolution.

\begin{acknowledgements}

We thank to an anonymous referee for insightful suggestions that helped to improve this paper.
We would like to thank the Calar Alto local staff for efficient support during 
the observations and D. Gilbank for providing  INT imaging and photometry.
We thank K. J\"ager and A. Fritz for software and observational support.
We thank M. Balogh, C. Da Rocha and J. Rasmussen for helpful discussions.
This work has been financially supported by the Volkswagen Foundation (I/76 520)
and travel grants to Calar-Alto by DFG (ZI663/5-1).

\end{acknowledgements}


\bibliographystyle{aa} 
\bibliography{/users2/mverdugo/THESIS/paper/mverdugo} 


\begin{appendix} 

\section{Star formation rates}
\label{S:SFR}

All indicators of star-formation rates (SFRs) have their own bias   and systematics due to the
different  processes traced for each of them (for a recent review see \citealt{Moustakas2006}). 
In the optical, at least two effects are very important: extinction and metallicity. 
On the other hand, optical SFRs  calculation requires accurate flux calibration, 
that we lack. However we can still estimate SFRs using the EWs and the  
absolute magnitudes (calculated in Sect. \ref{SS:absmags}) as a proxy of the continuum flux.
Fortunately, our spectra cover both \oii\ and \ha\ lines, so both results can be compared.

For \oii\ derived SFRs, we can take the calibrated relation of \citet{Kennicutt92} 

\begin{equation}
SFR(\oii)= 3.4\times10^{-12} \left(\frac{L_B}{L_{B,\odot}}\right)\ew{\oii} E(\ha)\mbox{ [M$_\odot$ yr$^{-1}$] }
\label{E:SFR_O2}
\end{equation}

\noindent where $E(\ha)$ is the extinction at \ha, which according to the same paper is approximately 
1\,mag, and ($L_B/L_{B,\odot})=10^{0.4(M_B-M_{B,\odot})}$, with $M_{B,\odot}$= 5.48\,mag. 

For \ha, we also take the relation given by  \citet{Kennicutt92}:

\begin{equation}
SFR(\ha)=7.9*10^{-42}L(\ha)E(\ha)
\label{E:SFR_Ha1}
\end{equation}

\noindent with $E(\ha)=1$ mag, as indicated above. However, we do not have \ha\
fluxes, but we can estimate it from our $R$-band absolute magnitudes and
\ha\ EWs, since

\begin{equation}
W_0(\ha) \approx \frac{L(\ha)}{L_C}
\label{E:ew-Lha}
\end{equation}

\noindent where $L_C$ is the continuum luminosity in erg s$^{-1}\mbox{\AA}^{-1}$  
(see \citealt{lewis02}) and $L_C\approx L_R$.  For a $L^\star$ galaxy, 
$L_C =1.1 \times 10^{40}$ ergs s$^{-1}$, as determined by \citet{Blanton2001}, 
with $M_R^\star=-21.8$  mag, leaving:

\begin{equation}
L(\ha)=1.1\times 10^{40} \ew{\ha}10^{-0.4(M_R -M_R^\star)} \mbox{ [ergs s$^{-1}$]}.
\label{E:LHa}
\end{equation}

\noindent Therefore, finally we have

\begin{equation}
SFR(\ha)=0.079 \ew{\ha}10^{-0.4(M_R +21.8)} \mbox{ [ergs s$^{-1}$]}.
\label{E:SFR_Ha2}
\end{equation}

We obtained SFRs for all galaxies in which  either of these two lines is
measurable.  Both ways are likely
to have systematics and uncertainties, \oii\ because it is a calibrated relation
and doubts persist about its universality (\eg \citealt{hammer1997}). 
Also it is  strongly affected by dust and metallicity. In the case of \ha, the assumptions here made, 
introduce uncertainties about the accuracy of the flux. Therefore, we take the average of the SFR 
obtained from  \oii\ and \ha\ and when only one line  is present we take this value. We did
 obtain SFRs for galaxies considered passive, however those values probably
 have larger uncertainties, so their SFRs can be considered as an upper limit . We  
always make distinction of both populations based in the EW distinction (see Sect. \ref{SS:EW}). 
We did not attempt to obtain SFRs for galaxies with negative equivalent
 widths because they yield to  unphysical values, difficult to interprete if included. 

 \begin{figure}[t]
     \centering
     \includegraphics[bb=20 150 580 480,width=\columnwidth,clip]{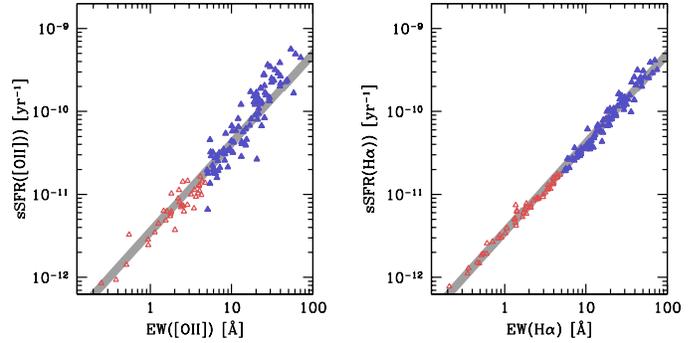}
     \caption{Specific star formation rates  based on \oii\ and \ha\  respectively   calculated using the relations
 indicated in the text. Blue filled triangles are galaxies with $\ew{\oii,\ha}>5$\,\AA, classified here as star forming.
The thick line is the fit to the \ha\ relation: $\log(sSFR)\approx1.07\log(\ew{\ha}-11.4$, which is plotted in the \oii\
panel for comparison. Note that the sSFR obtained with the \oii\ line may be slightly overestimated.}
     \label{F:SFR}
   \end{figure}
 
Using the stellar masses obtained with {\sc kcorrect} (see Sect. \ref{SS:absmags}) we obtained
the specific star formation rates (sSFR).  It is remarkable the strong correlation with  little
scatter between EWs and sSFR, obtained in either way (\ie \oii\ and \ha)
and the little scatter (albeit larger for \oii), as well as the similar values displayed
using both methods (see Fig. \ref{F:SFR}), despite the rough estimation made here. 
Also, it is  important that both indicators yield similar values as the \ha\ line becomes
inaccessible at larger redshifts and only \oii\ can be used. 
 
\end{appendix}

\Online

\onllongtab{3}{
\begin{scriptsize}
\begin{longtable}{lllllllrrrrrrr}

 \caption{Data for individual objects. We only present galaxies for which we obtained \emph{secured} redshifts 
(see section \ref{SS:redshift}).  Note that we include \emph{all} objects for which we were able to obtain the 
following parameters. Many objects  were excluded in the analysis in order to obtain a homogeneous sample  
(see sections \ref{SS:den} \& \ref{SS:field}).  
The columns are the following. Column (1): Object ID. Column (2): cluster, group or field membership. 
Columns (3) and (4): J2000 sky coordinates.  Column (6): redshift. Column (7): The $I$-band  magnitude. 
Column (8): The $V-I$ color. Columns (8) and (9): The absolute magnitudes in the restframe $B$ and $R$-bands.
Column (10): The logarithm of the stellar mass. Columns (11) and (12): The \oii\ EWs and  errors.
 Columns (13) and (14): The \ha\ EWs and errors. }
 \label{T:Objects}
\\ \hline  \hline
 (1)   & (2)                  &  (3)  & (4)      & (5)   &  (6)   & (7)      & (8)               & (9)       & (10)                  & (11)    &  (12)   & (13)    & (14) \\
 ID   & membership  &  RA  & DEC   & z   & I           & V-I       &   $M_B$  & $M_R$ & $\log(M_\ast)$ & \oii    & E(\oii) & \ha     & E(\ha)   \\
         &                        &         &             &     &  [mag]  & [mag] &     [mag]   &  [mag]  &   [$M_\odot$]  & [\AA] &  [\AA]  & [\AA]   & [\AA]  \\
\hline
\endfirsthead
\caption{Continued.} \\
 \hline \hline\\ 
(1)   & (2)                  &  (3)  & (4)      & (5)   &  (6)   & (7)      & (8)               & (9)       & (10)                  & (11)    &  (12)   & (13)    & (14) \\
 ID   & membership  &  RA  & DEC   & z   & I           & V-I       &   $M_B$  & $M_R$ & $\log(M_\ast)$ & \oii    & E(\oii) & \ha     & E(\ha)   \\
         &                        &         &             &     &  [mag]  & [mag] &     [mag]   &  [mag]  &   [$M_\odot$]  & [\AA] &  [\AA]  & [\AA]   & [\AA]  \\
\hline
\endhead
\\ \hline
\endfoot 
\\  \hline
\endlastfoot

r2211\_07  & vmf194  & 17:29:15.26  & 74:41:23.8 & 0.20998 & 18.59 & 1.388    & -19.63  & -20.94 & 10.02 &    9.66 & 0.75 &    0.75 & 0.23    \\
r2211\_08  & vmf194  & 17:29:19.77  & 74:41:11.2 & 0.21167 & 16.75 & 1.426    & -21.42  & -22.79 & 10.78 &    8.95 & 0.95 &    1.10 & 0.23    \\
r2211\_09  & vmf194  & 17:29:22.67  & 74:40:39.6 & 0.21187 & 18.17 & 1.462    & -19.96  & -21.37 & 10.22 &    0.43 & 0.48 &   -1.71 & 0.19    \\
r2211\_10  & vmf194  & 17:29:30.06  & 74:40:43.0 & 0.21107 & 18.54 & 1.422    & -19.64  & -21.01 & 10.06 &    0.18 & 0.55 &    1.13 & 0.24    \\
r2212\_06  & vmf194  & 17:29:13.37  & 74:42:13.8 & 0.20875 & 18.71 & 0.998    & -19.96  & -20.84 &  9.85 &  -32.93 & 0.63 &  -41.10 & 0.85    \\
r2212\_08  & vmf194  & 17:29:22.55  & 74:40:52.3 & 0.21080 & 17.92 & 1.409    & -20.27  & -21.62 & 10.30 &    3.93 & 0.60 &    1.12 & 0.28    \\
r2212\_23  & vmf194  & 17:30:44.85  & 74:39:11.8 & 0.20984 & 18.34 & 1.387    & -19.87  & -21.18 & 10.12 &    8.32 & 2.22 &   -0.87 & 0.36    \\
r2221\_14  & vmf194  & 17:26:24.27  & 74:27:35.2 & 0.20877 & 17.66 & 1.321    & -20.62  & -21.85 & 10.37 &   -0.02 & 0.69 &  -10.33 & 0.30    \\
r2221\_03  & xdcs220 & 17:26:50.67  & 74:34:04.6 & 0.26591 & 18.30 & 1.001    & -20.92  & -21.86 & 10.26 &  -39.80 & 0.67 &  -52.51 & 0.87    \\
r2221\_03b & xdcs220 & 17:26:50.87  & 74:34:08.8 & 0.26652 & 19.52 & 0.880    & -19.83  & -20.66 &  9.75 &  -61.16 & 1.72 &  -45.71 & 1.66    \\
r2221\_12  & xdcs220 & 17:26:11.11  & 74:28:26.6 & 0.26227 & 18.24 & 1.429    & -20.52  & -21.89 & 10.41 &   -4.34 & 0.85 &    1.46 & 0.61    \\
r2222\_03  & xdcs220 & 17:26:17.74  & 74:34:09.3 & 0.25712 & 18.85 & 1.587    & -19.68  & -21.19 & 10.17 &    0.21 & 0.39 &   -0.91 & 0.17    \\
r2222\_07  & xdcs220 & 17:26:33.92  & 74:31:55.8 & 0.26087 & 19.20 & 1.419    & -19.56  & -20.90 & 10.02 &    3.07 & 0.41 &    0.76 & 0.24    \\
r2231\_05  & xdcs220 & 17:24:11.04  & 74:31:12.1 & 0.26144 & 18.58 & 1.441    & -20.16  & -21.52 & 10.27 &    4.80 & 0.35 &    3.06 & 0.23    \\
r2241\_05  & xdcs220 & 17:23:24.91  & 74:44:42.8 & 0.26168 & 18.39 & 1.592    & -20.20  & -21.71 & 10.38 &   -0.02 & 1.23 &    3.42 & 0.52    \\
r2241\_07  & xdcs220 & 17:23:28.45  & 74:43:41.7 & 0.25977 & 17.00 & 1.572    & -21.59  & -23.09 & 10.92 &    5.64 & 0.45 &    1.65 & 0.17    \\
r2241\_09  & xdcs220 & 17:23:26.66  & 74:43:16.6 & 0.25991 & 16.93 & 1.592    & -21.64  & -23.16 & 10.96 &  -14.98 & 0.51 &   -3.00 & 0.20    \\
r2241\_10  & xdcs220 & 17:23:24.29  & 74:42:56.2 & 0.25953 & 17.99 & 1.518    & -20.64  & -22.10 & 10.52 &    8.83 & 0.64 &    2.30 & 0.20    \\
r2241\_15  & xdcs220 & 17:23:32.26  & 74:40:35.7 & 0.25489 & 18.46 & 1.241    & -20.43  & -21.61 & 10.26 &   -1.46 & 0.43 &  -12.41 & 0.24    \\
r2241\_18  & xdcs220 & 17:23:05.48  & 74:39:30.5 & 0.25451 & 18.38 & 2.567    & -19.55  & -21.63 & 10.43 &  -20.46 & 0.29 &  -35.91 & 0.28    \\
r2242\_06  & xdcs220 & 17:23:26.46  & 74:43:57.3 & 0.26169 & 19.49 & 1.465    & -19.23  & -20.63 &  9.91 &   -3.18 & 1.04 &    7.17 & 0.62    \\
r2251\_04  & xdcs220 & 17:24:12.22  & 74:22:23.8 & 0.26246 & 18.92 & 0.879    & -20.37  & -21.24 &  9.98 &  -37.27 & 1.09 &  -47.44 & 0.77    \\
xdc29\_04  & xdcs220 & 17:23:29.37  & 74:43:38.7 & 0.26160 & 17.96 & 0.870    & -20.63  & -22.15 & 10.55 &    0.72 & 0.58 &   -1.36 & 0.31    \\
r2621\_14  & vmf131  & 13:11:22.18  & 32:28:53.8 & 0.29902 & 19.24 & 1.613    & -19.75  & -21.26 & 10.19 &    0.98 & 0.30 &    0.35 & 0.20    \\
r2621\_15  & vmf131  & 13:11:23.60  & 32:28:56.3 & 0.29347 & 18.95 & 1.079    & -20.44  & -21.49 & 10.16 &  -23.51 & 0.17 &  -21.69 & 0.23    \\
r2621\_16  & vmf131  & 13:11:24.66  & 32:28:36.9 & 0.30015 & 18.99 & 1.340    & -20.24  & -21.50 & 10.23 &  -20.84 & 0.24 &  -31.54 & 0.33    \\
r2631\_20  & vmf131  & 13:10:41.55  & 32:28:23.1 & 0.29691 & 17.24 & 1.777    & -21.58  & -23.23 & 11.01 &    2.72 & 0.48 &   -1.85 & 0.34    \\
r2632\_02  & vmf131  & 13:10:16.57  & 32:30:36.6 & 0.29501 & 18.70 & 1.719    & -20.14  & -21.75 & 10.41 &    2.09 & 0.23 &   -1.66 & 0.16    \\
r2632\_02b & vmf131  & 13:10:16.86  & 32:30:38.0 & 0.29341 & 19.92 & 1.087    & -19.46  & -20.52 &  9.77 &  -22.93 & 0.46 &  -31.87 & 0.78    \\
r2632\_03  & vmf131  & 13:10:18.96  & 32:30:18.7 & 0.29380 & 18.47 & 1.693    & -20.38  & -21.97 & 10.49 &   -0.38 & 0.18 &    0.15 & 0.13    \\
r2632\_11  & vmf131  & 13:10:30.42  & 32:27:16.6 & 0.29477 & 17.73 & 1.508    & -21.29  & -22.71 & 10.75 &  -10.37 & 0.15 &  -19.49 & 0.16    \\
r2632\_12  & vmf131  & 13:10:34.20  & 32:27:30.8 & 0.29464 & 19.21 & 1.024    & -20.25  & -21.22 & 10.02 &  -19.41 & 0.21 &  -27.83 & 0.33    \\
r2632\_17  & vmf131  & 13:10:47.14  & 32:27:56.8 & 0.29270 & 19.37 & 1.586    & -19.57  & -21.06 & 10.11 &   -4.03 & 0.30 &   -3.99 & 0.20    \\
r2641\_04  & vmf131  & 13:10:17.22  & 32:19:51.8 & 0.29577 & 18.90 & 1.903    & -19.78  & -21.56 & 10.36 &   -2.01 & 1.15 &  -12.32 & 0.90    \\
r2641\_05  & vmf131  & 13:10:14.02  & 32:23:33.6 & 0.29617 & 18.71 & 1.774    & -20.10  & -21.75 & 10.42 &   -0.16 & 1.23 &  -11.21 & 0.91    \\
r2641\_06  & vmf131  & 13:10:12.94  & 32:24:09.1 & 0.29574 & 18.59 & 0.864    & -21.02  & -21.84 & 10.21 &    2.78 & 0.52 &    1.22 & 0.30    \\
r2641\_07  & vmf131  & 13:10:10.66  & 32:21:44.2 & 0.29329 & 18.77 & 2.331    & -19.59  & -21.67 & 10.45 &   -5.09 & 0.38 &  -15.73 & 0.35    \\
r2641\_12  & vmf131  & 13:10:01.42  & 32:23:48.2 & 0.29614 & 17.98 & 1.500    & -21.07  & -22.47 & 10.65 &    .... & .... &   -0.36 & 0.39    \\
r2651\_08  & vmf131  & 13:10:47.60  & 32:20:11.4 & 0.29430 & 18.48 & 1.632    & -20.43  & -21.96 & 10.48 &   -2.87 & 0.25 &    2.55 & 0.18    \\
r2651\_17  & vmf131  & 13:11:13.98  & 32:19:10.5 & 0.29438 & 17.54 & 1.744    & -21.27  & -22.90 & 10.87 &   -4.27 & 0.17 &   -0.60 & 0.11    \\
r2651\_19  & vmf131  & 13:11:17.96  & 32:19:47.9 & 0.29408 & 18.15 & 1.511    & -20.86  & -22.28 & 10.58 &   -5.96 & 0.22 &   -8.38 & 0.15    \\
ba\_07     & vmf131  & 13:10:05.72  & 32:21:12.2 & 0.29651 & 18.32 & 1.105    & -21.08  & -22.13 & 10.42 &    5.07 & 0.80 &    0.24 & 0.60    \\
ba\_09     & vmf131  & 13:10:04.22  & 32:21:36.3 & 0.29003 & 19.00 & 1.615    & -19.89  & -21.39 & 10.25 &    3.28 & 0.81 &    2.95 & 0.70    \\
ba\_12     & vmf131  & 13:09:55.05  & 32:21:49.0 & 0.29382 & 18.47 & 1.684    & -20.39  & -21.97 & 10.49 &    3.51 & 0.55 &    3.18 & 0.36    \\
ba\_14     & vmf131  & 13:09:53.20  & 32:21:59.8 & 0.29125 & 18.87 & 1.681    & -19.97  & -21.53 & 10.31 &    4.67 & 0.68 &   -2.11 & 0.43    \\
ba\_18     & vmf131  & 13:10:11.38  & 32:22:02.3 & 0.29388 & 18.15 & 1.666    & -20.74  & -22.29 & 10.61 &    6.15 & 0.49 &    0.20 & 0.35    \\
ba\_25     & vmf131  & 13:09:51.54  & 32:22:17.8 & 0.29242 & 18.40 & 1.689    & -20.45  & -22.02 & 10.51 &    5.13 & 0.50 &    0.13 & 0.35    \\
ba\_28     & vmf131  & 13:09:56.11  & 32:22:16.8 & 0.29207 & 16.72 & 1.718    & -22.10  & -23.70 & 11.18 &    5.12 & 0.35 &    0.84 & 0.22    \\
ba\_30     & vmf131  & 13:09:58.50  & 32:22:31.3 & 0.29466 & 18.02 & 1.670    & -20.86  & -22.42 & 10.67 &    2.29 & 0.48 &    1.07 & 0.34    \\
ba\_36     & vmf131  & 13:10:00.18  & 32:22:59.4 & 0.29431 & 18.23 & 1.361    & -20.93  & -22.21 & 10.52 &   -6.82 & 0.38 &  -11.69 & 0.35    \\
ba\_37     & vmf131  & 13:09:56.33  & 32:23:10.8 & 0.28946 & 18.21 & 1.701    & -20.60  & -22.19 & 10.58 &   -6.66 & 0.77 &  -11.75 & 0.59    \\
ba\_39     & vmf131  & 13:09:57.68  & 32:23:13.0 & 0.29233 & 17.95 & 1.708    & -20.88  & -22.47 & 10.69 &   -2.66 & 0.46 &   -0.46 & 0.31    \\
r2611\_04  & vmf132  & 13:12:07.22  & 32:34:35.8 & 0.24552 & 18.99 & 1.422    & -19.61  & -20.95 & 10.04 &    6.99 & 0.54 &    2.91 & 0.32    \\
r2611\_13  & vmf132  & 13:11:51.74  & 32:33:29.2 & 0.24964 & 17.79 & 1.463    & -20.81  & -22.20 & 10.54 &    6.78 & 0.35 &    1.40 & 0.21    \\
r2611\_14  & vmf132  & 13:11:49.17  & 32:33:23.4 & 0.24656 & 18.11 & 1.488    & -20.42  & -21.85 & 10.41 &    0.91 & 0.40 &    1.31 & 0.23    \\
r2612\_02  & vmf132  & 13:12:27.01  & 32:32:06.6 & 0.24855 & 18.53 & 1.497    & -20.01  & -21.44 & 10.25 &  -15.32 & 0.23 &  -10.12 & 0.13    \\
r2612\_04  & vmf132  & 13:12:16.07  & 32:32:11.0 & 0.24771 & 19.32 & 1.137    & -19.60  & -20.67 &  9.85 &  -39.36 & 0.31 &  -58.97 & 0.39    \\
r2612\_06  & vmf132  & 13:12:10.39  & 32:30:03.0 & 0.24954 & 17.28 & 1.477    & -21.30  & -22.70 & 10.74 &    1.87 & 0.14 &    0.97 & 0.09    \\
r2612\_17  & vmf132  & 13:11:45.80  & 32:31:21.7 & 0.24628 & 17.93 & 1.579    & -20.50  & -22.02 & 10.50 &    0.97 & 0.22 &    1.33 & 0.12    \\
r2621\_03  & vmf132  & 13:11:01.05  & 32:30:41.6 & 0.24128 & 18.35 & 1.479    & -20.12  & -21.54 & 10.29 &    1.20 & 0.21 &    1.89 & 0.12    \\
r2621\_04  & vmf132  & 13:11:02.92  & 32:29:36.0 & 0.24061 & 19.36 & 1.538    & -19.04  & -20.52 &  9.89 &    0.66 & 0.40 &   -0.07 & 0.20    \\
r2621\_11  & vmf132  & 13:11:13.29  & 32:28:50.9 & 0.23976 & 18.63 & 1.575    & -19.73  & -21.24 & 10.19 &    4.32 & 0.27 &    1.15 & 0.13    \\
r2621\_13  & vmf132  & 13:11:17.64  & 32:28:11.0 & 0.24181 & 17.92 & 1.516    & -20.52  & -21.98 & 10.47 &    3.18 & 0.18 &    0.22 & 0.10    \\
r2621\_22  & vmf132  & 13:11:33.84  & 32:29:11.9 & 0.25032 & 18.17 & 1.536    & -20.36  & -21.83 & 10.41 &   -2.56 & 0.23 &   -4.01 & 0.13    \\
r2631\_21  & vmf132  & 13:10:43.33  & 32:27:04.1 & 0.24590 & 17.35 & 1.448    & -21.22  & -22.60 & 10.70 &   -3.50 & 0.33 &   -7.80 & 0.25    \\
r2631\_08  & vmf132  & 13:10:25.08  & 32:28:44.7 & 0.25008 & 19.30 & 1.361    & -19.41  & -20.70 &  9.92 &    .... & .... &   -1.35 & 0.28    \\
r2632\_07  & vmf132  & 13:10:25.93  & 32:30:15.4 & 0.24461 & 18.82 & 1.505    & -19.68  & -21.10 & 10.12 &   -0.94 & 0.83 &   -3.46 & 0.24    \\
r2632\_13  & vmf132  & 13:10:37.82  & 32:27:15.2 & 0.24566 & 18.34 & 1.486    & -20.19  & -21.60 & 10.31 &   -0.17 & 0.19 &    1.18 & 0.12    \\
r2641\_10  & vmf132  & 13:10:04.73  & 32:20:51.1 & 0.24818 & 17.47 & 1.469    & -21.10  & -22.50 & 10.66 &    9.17 & 1.56 &   -5.42 & 0.54    \\
ba\_29     & vmf132  & 13:09:49.99  & 32:22:41.0 & 0.24954 & 19.77 & 0.784    & -19.55  & -20.25 &  9.52 &  -53.59 & 1.20 &  -61.71 & 2.16    \\
r2811\_06  & vmf73   & 09:43:52.61  & 16:44:40.1 & 0.25382 & 17.82 & 1.544    & -20.73  & -22.20 & 10.56 &   -6.83 & 1.17 &   -3.35 & 0.36    \\
r2811\_16  & vmf73   & 09:43:58.38  & 16:41:09.6 & 0.25266 & 16.96 & 1.438    & -21.70  & -23.04 & 10.87 &   -0.16 & 0.65 &   -2.47 & 0.24    \\
r2811\_18  & vmf73   & 09:43:53.52  & 16:40:23.1 & 0.25161 & 18.32 & 1.553    & -20.21  & -21.68 & 10.36 &    0.80 & 1.17 &    0.43 & 0.33    \\
r2811\_19  & vmf73   & 09:43:58.81  & 16:40:02.3 & 0.25384 & 18.60 & 1.386    & -20.12  & -21.42 & 10.22 &   -1.26 & 2.62 &    0.70 & 0.67    \\
r2811\_20  & vmf73   & 09:44:03.21  & 16:39:48.5 & 0.25736 & 18.81 & 1.339    & -19.98  & -21.24 & 10.14 &    5.85 & 3.06 &   -1.53 & 0.88    \\
r2811\_24  & vmf73   & 09:44:01.37  & 16:38:01.1 & 0.25355 & 17.73 & 1.456    & -20.91  & -22.28 & 10.57 &    4.65 & 1.30 &    0.76 & 0.35    \\
r2811\_25  & vmf73   & 09:43:59.68  & 16:37:30.1 & 0.25423 & 18.10 & 1.536    & -20.47  & -21.93 & 10.46 &   -5.85 & 1.51 &   -2.06 & 0.46    \\
r2812\_05  & vmf73   & 09:44:05.00  & 16:38:34.3 & 0.25614 & 18.31 & 1.465    & -20.35  & -21.74 & 10.37 &    8.83 & 3.26 &   -0.05 & 0.45    \\
r2812\_09  & vmf73   & 09:44:00.32  & 16:40:11.5 & 0.24866 & 18.70 & 1.319    & -20.03  & -21.27 & 10.14 &    5.74 & 3.73 &   -1.70 & 0.82    \\
r2812\_12  & vmf73   & 09:43:59.37  & 16:41:09.9 & 0.25700 & 17.26 & 1.516    & -21.35  & -22.78 & 10.78 &    1.88 & 1.44 &   -2.53 & 0.26    \\
r2812\_14  & vmf73   & 09:43:53.57  & 16:41:43.2 & 0.25292 & 17.19 & 1.430    & -21.47  & -22.80 & 10.77 &    1.48 & 1.28 &    1.53 & 0.28    \\
r2821\_02  & vmf73   & 09:43:58.08  & 16:41:17.0 & 0.25200 & 16.67 & 1.469    & -21.95  & -23.33 & 11.00 &   -0.11 & 1.40 &    1.92 & 0.37    \\
r2821\_08  & vmf73   & 09:43:48.71  & 16:40:39.1 & 0.25456 & 17.87 & 0.823    & -21.45  & -22.22 & 10.33 &  -33.96 & 0.81 &  -69.22 & 1.50    \\
r2821\_12  & vmf73   & 09:43:43.00  & 16:40:34.5 & 0.25734 & 17.64 & 1.488    & -21.00  & -22.42 & 10.63 &   -3.06 & 1.44 &   -0.93 & 0.43    \\
r2821\_17  & vmf73   & 09:43:36.34  & 16:36:57.3 & 0.25693 & 17.58 & 1.503    & -21.04  & -22.46 & 10.65 &    3.38 & 2.29 &   -2.92 & 0.58    \\
r2821\_14  & vmf73   & 09:43:40.07  & 16:39:23.6 & 0.25781 & 18.45 & 1.460    & -20.22  & -21.62 & 10.31 &    4.93 & 3.98 &   -3.89 & 0.83    \\
r2821\_20  & vmf73   & 09:43:33.63  & 16:39:06.8 & 0.25295 & 17.80 & 1.467    & -20.83  & -22.19 & 10.54 &  -13.73 & 2.82 &    3.80 & 0.80    \\
r2821\_21  & vmf73   & 09:43:32.42  & 16:40:01.0 & 0.25389 & 18.55 & 1.477    & -20.08  & -21.47 & 10.26 &    0.92 & 2.23 &    1.63 & 0.74    \\
r2821\_27  & vmf73   & 09:43:23.53  & 16:39:46.4 & 0.25767 & 17.98 & 1.298    & -20.85  & -22.09 & 10.45 &   -0.29 & 1.31 &   -3.83 & 0.54    \\
r2821\_29  & vmf73   & 09:43:19.34  & 16:38:08.6 & 0.25733 & 17.87 & 1.430    & -20.83  & -22.19 & 10.53 &   -5.97 & 1.61 &    0.11 & 0.69    \\
r2822\_01  & vmf73   & 09:43:58.93  & 16:39:22.0 & 0.25607 & 18.91 & 1.398    & -19.81  & -21.14 & 10.11 &   -0.27 & 0.96 &    1.26 & 0.34    \\
r2822\_03  & vmf73   & 09:43:56.35  & 16:36:51.1 & 0.25523 & 17.57 & 1.363    & -21.19  & -22.49 & 10.64 &    2.69 & 0.64 &    1.52 & 0.36    \\
r2822\_04  & vmf73   & 09:43:55.89  & 16:40:36.0 & 0.25507 & 18.68 & 1.394    & -20.05  & -21.38 & 10.21 &    0.97 & 1.08 &    2.27 & 0.32    \\
r2822\_05  & vmf73   & 09:43:53.38  & 16:39:59.1 & 0.25108 & 17.73 & 1.382    & -20.98  & -22.29 & 10.56 &   -0.15 & 0.45 &   -2.38 & 0.17    \\
r2822\_06  & vmf73   & 09:43:51.72  & 16:41:45.0 & 0.25285 & 18.03 & 1.186    & -20.89  & -21.96 & 10.37 &   -2.82 & 0.50 &  -16.58 & 0.30    \\
r2822\_09  & vmf73   & 09:43:45.55  & 16:41:30.9 & 0.25137 & 18.83 & 0.973    & -20.32  & -21.21 & 10.01 &  -19.19 & 0.67 &  -22.61 & 0.55    \\
r2822\_14  & vmf73   & 09:43:38.75  & 16:38:55.5 & 0.25331 & 18.59 & 1.415    & -20.09  & -21.41 & 10.22 &    3.89 & 0.90 &    2.31 & 0.29    \\
r2822\_15  & vmf73   & 09:43:37.97  & 16:39:32.6 & 0.25742 & 17.00 & 1.493    & -21.64  & -23.06 & 10.89 &    4.35 & 0.56 &    1.01 & 0.16    \\
r2822\_16  & vmf73   & 09:43:36.80  & 16:41:02.7 & 0.25521 & 18.19 & 1.427    & -20.50  & -21.87 & 10.41 &    0.76 & 0.82 &    0.47 & 0.25    \\
r2822\_17  & vmf73   & 09:43:34.09  & 16:40:36.1 & 0.25003 & 19.08 & 1.328    & -19.67  & -20.92 & 10.01 &   -1.62 & 0.88 &   -0.11 & 0.35    \\
r2822\_19  & vmf73   & 09:43:30.57  & 16:38:56.0 & 0.25294 & 19.30 & 1.418    & -19.38  & -20.69 &  9.93 &    2.32 & 1.27 &    3.04 & 0.48    \\
r2822\_20  & vmf73   & 09:43:29.64  & 16:40:56.8 & 0.25775 & 19.36 & 1.101    & -19.66  & -20.71 &  9.86 &    1.58 & 0.65 &    0.46 & 0.40    \\
r2822\_22  & vmf73   & 09:43:25.34  & 16:39:07.2 & 0.25491 & 17.81 & 1.494    & -20.81  & -22.24 & 10.57 &    2.17 & 0.52 &    0.26 & 0.17    \\
r2822\_23  & vmf73   & 09:43:24.51  & 16:39:52.1 & 0.26135 & 18.46 & 1.281    & -20.44  & -21.65 & 10.28 &   13.14 & 1.26 &    1.88 & 0.50    \\
r2822\_25  & vmf73   & 09:43:22.06  & 16:39:07.9 & 0.25037 & 18.54 & 1.402    & -20.14  & -21.46 & 10.24 &   -1.78 & 0.67 &   -4.66 & 0.26    \\
r2831\_03  & vmf73   & 09:43:22.88  & 16:41:14.8 & 0.24829 & 18.39 & 1.205    & -20.47  & -21.60 & 10.25 &  -24.22 & 5.22 &  -16.09 & 1.77    \\
r2831\_10  & vmf73   & 09:43:08.02  & 16:42:45.4 & 0.25696 & 18.24 & 1.102    & -20.78  & -21.80 & 10.28 &  -10.66 & 1.02 &  -18.04 & 0.78    \\
r2831\_13  & vmf73   & 09:43:01.49  & 16:42:27.7 & 0.25601 & 17.15 & 1.416    & -21.56  & -22.90 & 10.81 &    4.92 & 1.29 &   -1.09 & 0.36    \\
r2841\_07  & vmf73   & 09:44:41.05  & 16:29:19.5 & 0.25003 & 18.64 & 1.176    & -20.27  & -21.37 & 10.14 &  -45.76 & 2.52 &  -21.50 & 0.73    \\
r2841\_10  & vmf73   & 09:44:36.52  & 16:27:31.5 & 0.25381 & 18.10 & 1.303    & -20.70  & -21.92 & 10.39 &  -14.14 & 0.75 &  -11.00 & 0.28    \\
r2841\_17  & vmf73   & 09:44:23.76  & 16:31:47.1 & 0.25035 & 18.12 & 1.383    & -20.57  & -21.88 & 10.39 &    1.61 & 0.98 &    1.43 & 0.22    \\
r2851\_04  & vmf73   & 09:43:50.60  & 16:28:20.5 & 0.25065 & 18.57 & 0.995    & -20.54  & -21.47 & 10.11 &  -12.13 & 0.56 &  -28.68 & 0.72    \\
r2851\_14  & vmf73   & 09:44:04.69  & 16:32:49.3 & 0.25340 & 19.02 & 1.446    & -19.63  & -20.98 & 10.05 &    3.75 & 2.19 &   -4.63 & 0.66    \\
r2851\_17  & vmf73   & 09:43:52.24  & 16:34:00.8 & 0.25195 & 17.70 & 1.405    & -20.99  & -22.30 & 10.57 &    1.31 & 0.55 &    0.78 & 0.18    \\
r2811\_01  & vmf74   & 09:43:44.47  & 16:46:05.3 & 0.17837 & 17.66 & 1.323    & -20.18  & -21.42 & 10.20 &    4.45 & 0.72 &    1.37 & 0.24    \\
r2811\_03  & vmf74   & 09:43:43.51  & 16:45:20.1 & 0.18029 & 18.27 & 1.278    & -19.68  & -20.87 &  9.97 &    2.27 & 0.97 &    0.85 & 0.33    \\
r2811\_05  & vmf74   & 09:43:44.49  & 16:44:54.2 & 0.18009 & 18.50 & 1.183    & -19.57  & -20.65 &  9.85 &  -21.52 & 1.18 &  -77.27 & 1.12    \\
r2811\_07  & vmf74   & 09:43:46.72  & 16:44:25.2 & 0.17917 & 18.19 & 1.198    & -19.84  & -20.92 &  9.96 &   -2.13 & 1.22 &   -2.40 & 0.40    \\
r2811\_08  & vmf74   & 09:43:45.15  & 16:44:05.6 & 0.17884 & 18.53 & 1.268    & -19.40  & -20.56 &  9.84 &   -8.10 & 1.85 &   -8.58 & 0.53    \\
r2811\_10  & vmf74   & 09:43:55.55  & 16:43:34.6 & 0.17876 & 18.04 & 1.286    & -19.86  & -21.05 & 10.04 &    0.93 & 1.33 &    0.97 & 0.34    \\
r2811\_11  & vmf74   & 09:43:49.12  & 16:43:21.2 & 0.18086 & 16.81 & 1.354    & -21.03  & -22.33 & 10.57 &    0.70 & 0.64 &   -1.23 & 0.18    \\
r2811\_13  & vmf74   & 09:43:53.02  & 16:42:48.2 & 0.17838 & 18.68 & 1.218    & -19.30  & -20.42 &  9.77 &    2.84 & 2.04 &    1.52 & 0.59    \\
r2811\_14  & vmf74   & 09:43:58.75  & 16:42:02.5 & 0.18250 & 17.82 & 1.305    & -20.11  & -21.34 & 10.16 &    5.52 & 1.71 &    1.56 & 0.36    \\
r2811\_22  & vmf74   & 09:43:43.31  & 16:39:18.5 & 0.17701 & 18.46 & 1.025    & -19.73  & -20.69 &  9.82 &  -21.35 & 2.32 &  -25.48 & 0.99    \\
r2811\_23  & vmf74   & 09:43:59.52  & 16:38:29.8 & 0.17833 & 17.98 & 1.123    & -20.12  & -21.14 & 10.02 &   -2.49 & 0.96 &   -6.81 & 0.28    \\
r2812\_02  & vmf74   & 09:43:43.06  & 16:37:36.0 & 0.18249 & 17.68 & 1.344    & -20.20  & -21.47 & 10.22 &   -0.22 & 1.36 &   -0.52 & 0.21    \\
r2812\_16  & vmf74   & 09:44:01.02  & 16:42:04.1 & 0.17750 & 17.95 & 1.203    & -20.03  & -21.16 & 10.07 &  -16.80 & 4.23 &  -10.69 & 0.60    \\
r2812\_17  & vmf74   & 09:43:45.13  & 16:42:46.3 & 0.18134 & 17.26 & 1.307    & -20.66  & -21.90 & 10.39 &    6.09 & 1.68 &   -4.01 & 0.24    \\
r2812\_22  & vmf74   & 09:43:43.45  & 16:44:31.8 & 0.18096 & 17.77 & 1.289    & -20.18  & -21.39 & 10.18 &    8.73 & 3.08 &    0.07 & 0.36    \\
r2812\_21  & vmf74   & 09:43:44.86  & 16:44:02.2 & 0.17932 & 17.70 & 1.340    & -20.14  & -21.39 & 10.19 &   -6.10 & 1.10 &   -9.78 & 0.25    \\
r2821\_03  & vmf74   & 09:43:56.38  & 16:39:57.5 & 0.17992 & 18.21 & 1.239    & -19.77  & -20.92 &  9.97 &   10.57 & 3.86 &   -0.03 & 0.71    \\
r2821\_06  & vmf74   & 09:43:51.14  & 16:37:26.3 & 0.17486 & 18.72 & 0.891    & -19.70  & -20.37 &  9.55 &  -23.75 & 2.87 &  -31.60 & 2.05    \\
r2821\_07  & vmf74   & 09:43:50.04  & 16:39:54.7 & 0.17926 & 18.60 & 1.319    & -19.27  & -20.49 &  9.83 &   -3.49 & 4.05 &    2.84 & 0.91    \\
r2821\_09  & vmf74   & 09:43:47.69  & 16:39:10.2 & 0.17974 & 18.67 & 1.258    & -19.29  & -20.45 &  9.79 &   -5.83 & 6.46 &    0.17 & 1.12    \\
r2821\_10  & vmf74   & 09:43:46.04  & 16:39:54.5 & 0.17763 & 18.76 & 1.285    & -19.11  & -20.33 &  9.76 &  -16.78 & 7.64 &    1.04 & 1.10    \\
r2821\_11  & vmf74   & 09:43:44.53  & 16:39:19.8 & 0.17656 & 17.73 & 1.256    & -20.17  & -21.36 & 10.15 &   -6.26 & 2.14 &   -0.59 & 0.49    \\
r2822\_07  & vmf74   & 09:43:49.72  & 16:40:51.4 & 0.18048 & 17.63 & 1.306    & -20.28  & -21.51 & 10.23 &    4.17 & 0.58 &    1.07 & 0.15    \\
r2822\_10  & vmf74   & 09:43:44.15  & 16:40:47.1 & 0.17948 & 18.43 & 1.307    & -19.46  & -20.67 &  9.90 &   -0.03 & 0.80 &   -0.40 & 0.20    \\
r2822\_13  & vmf74   & 09:43:39.73  & 16:37:22.5 & 0.17946 & 19.23 & 1.161    & -18.85  & -19.89 &  9.54 &    1.35 & 1.87 &  -14.50 & 0.45    \\
r2831\_09  & vmf74   & 09:43:08.92  & 16:41:44.8 & 0.18137 & 18.21 & 1.415    & -19.57  & -20.93 & 10.03 &  -14.13 & 2.55 &  -19.38 & 0.88    \\
r2831\_20  & vmf74   & 09:42:44.14  & 16:45:34.9 & 0.17989 & 18.08 & 1.173    & -19.99  & -21.06 & 10.00 &   -5.48 & 1.19 &   -2.32 & 0.60    \\
r2841\_03  & vmf74   & 09:44:54.15  & 16:28:00.6 & 0.18333 & 19.13 & 1.096    & -19.07  & -20.07 &  9.58 &  -27.04 & 8.74 &  -17.81 & 0.82    \\
r2841\_08  & vmf74   & 09:44:40.09  & 16:30:59.7 & 0.18478 & 19.32 & 1.081    & -18.93  & -19.93 &  9.52 &  -21.39 & 1.97 &  -28.51 & 0.92    \\
r2841\_20  & vmf74   & 09:44:16.29  & 16:28:47.4 & 0.18157 & 19.46 & 0.910    & -18.97  & -19.78 &  9.39 &  -28.28 & 2.91 &  -13.70 & 1.07    \\
r2851\_09  & vmf74   & 09:43:50.52  & 16:30:28.0 & 0.18124 & 19.09 & 1.184    & -19.00  & -20.10 &  9.63 &    0.54 & 2.14 &   -0.59 & 0.77    \\
r2851\_10  & vmf74   & 09:43:48.57  & 16:30:39.7 & 0.17962 & 18.58 & 0.794    & -19.95  & -20.62 &  9.64 &  -29.97 & 0.65 &  -41.39 & 0.65    \\
r2851\_11  & vmf74   & 09:43:52.55  & 16:31:20.4 & 0.18176 & 19.32 & 1.128    & -18.84  & -19.87 &  9.52 &  -39.42 & 1.39 &  -65.98 & 1.24    \\
r2851\_19  & vmf74   & 09:43:59.00  & 16:35:02.8 & 0.18102 & 19.39 & 0.904    & -19.05  & -19.86 &  9.42 &   -5.48 & 1.62 &  -17.97 & 0.64    \\
r2221\_09	& group1 & 17:26:29.54	& 74:29:34.7	& 0.05308	& 16.30	& 1.114	   & -18.89 & -19.96 &  9.52 &     .... & .... &  -13.81 & 0.16  \\ 
r2222\_11	& group1 & 17:26:18.59	& 74:30:54.1	& 0.05278	& 18.65	& 0.844	   & -16.97 & -17.74 &  8.53 &     .... & .... &    .... & ....  \\ 
r2231\_18	& group1 & 17:25:26.20	& 74:30:15.5	& 0.05219	& 17.48	& 0.835	   & -18.12 & -18.89 &  8.98 &     .... & .... &   -9.51 & 0.12  \\ 
r2222\_20	& group1 & 17:26:15.09	& 74:25:49.2	& 0.05292	& 17.89	& 0.781	   & -17.81 & -18.54 &  8.79 &     .... & .... &  -38.39 & 0.13  \\ 
r2221\_08	& group2 & 17:27:07.02	& 74:30:29.4	& 0.04385	& 18.36	& 0.940	   & -16.73 & -17.57 &  8.48 &     .... & .... &  -41.37 & 0.46  \\ 
r2221\_11	& group2 & 17:27:07.97	& 74:28:40.5	& 0.04208	& 16.99	& 0.866	   & -18.08 & -18.88 &  8.97 &     .... & .... &  -21.11 & 0.26  \\ 
r2222\_18	& group2 & 17:26:45.45	& 74:27:01.2	& 0.04405	& 19.21	& 0.657	   & -16.34 & -16.92 &  8.06 &     .... & .... &  -51.86 & 0.39  \\ 
r2231\_08	& group2 & 17:24:25.65	& 74:28:34.5	& 0.04191	& 17.38	& 0.734	   & -17.87 & -18.56 &  8.77 &     .... & .... &  -65.83 & 0.15  \\ 
r2231\_19	& group2 & 17:25:34.22	& 74:28:54.1	& 0.04133	& 17.43	& 0.962	   & -17.42 & -18.34 &  8.81 &     .... & .... &   -8.52 & 0.11  \\ 
r2211\_11	& group3 & 17:29:41.54	& 74:42:31.7	& 0.24259	& 18.30	& 1.285	   & -20.41 & -21.62 & 10.27 &   -14.98 & 0.38 &  -20.88 & 0.24  \\ 
r2211\_13	& group3 & 17:29:48.60	& 74:42:15.0	& 0.24404	& 17.93	& 1.503	   & -20.56 & -21.99 & 10.47 &     3.07 & 1.35 &    2.55 & 0.38  \\ 
r2211\_17	& group3 & 17:30:02.07	& 74:41:43.3	& 0.24405	& 17.85	& 1.333	   & -20.83 & -22.08 & 10.46 &    -5.01 & 0.70 &   -9.60 & 0.30  \\ 
r2211\_19	& group3 & 17:30:09.69	& 74:42:44.1	& 0.24513	& 18.45	& 1.431	   & -20.14 & -21.49 & 10.25 &    -6.01 & 0.45 &   -4.50 & 0.21  \\ 
r2211\_20	& group3 & 17:30:13.48	& 74:41:02.5	& 0.24471	& 17.89	& 1.476	   & -20.65 & -22.04 & 10.48 &     7.76 & 1.19 &   -0.52 & 0.28  \\ 
r2211\_21	& group3 & 17:30:16.74	& 74:42:26.8	& 0.24230	& 18.69	& 1.510	   & -19.76 & -21.21 & 10.17 &     5.14 & 0.64 &    3.05 & 0.24  \\ 
r2212\_11	& group3 & 17:29:40.88	& 74:41:23.3	& 0.24196	& 18.12	& 1.467	   & -20.38 & -21.78 & 10.38 &     4.51 & 1.89 &    1.80 & 0.64  \\ 
r2242\_18	& group3 & 17:24:31.58	& 74:37:39.2	& 0.24194	& 18.17	& 1.709	   & -20.06 & -21.72 & 10.41 &   -13.17 & 1.00 &   -2.97 & 0.21  \\ 
r2621\_01	& group4 & 13:10:52.83	& 32:30:58.0	& 0.18687	& 19.30	& 1.355	   & -18.63 & -20.03 &  9.68 &    -8.00 & 0.40 &  -19.11 & 0.22  \\ 
r2631\_16	& group4 & 13:10:37.26	& 32:26:37.6	& 0.18599	& 17.17	& 1.017	   & -21.14 & -22.14 & 10.26 &   -20.02 & 0.25 &  -27.48 & 0.30  \\ 
r2631\_11	& group4 & 13:10:30.08	& 32:29:10.8	& 0.18727	& 16.42	& 1.478	   & -21.43 & -22.93 & 10.89 &     0.63 & 0.28 &    1.87 & 0.14  \\ 
r2631\_01	& group4 & 13:10:09.80	& 32:29:44.2	& 0.18593	& 18.05	& 1.416	   & -19.91 & -21.28 & 10.20 &     2.90 & 0.38 &    0.17 & 0.18  \\ 
r2632\_15	& group4 & 13:10:39.83	& 32:28:42.5	& 0.18635	& 17.39	& 1.538	   & -20.47 & -21.96 & 10.43 &    -0.51 & 0.19 &   -2.62 & 0.08  \\ 
r2211\_01	& field	& 17:28:35.55	& 74:43:18.4	& 0.32074	& 18.43	& 1.319	   & -21.01 & -22.24 & 10.52 &   -10.68 & 0.38 &  -15.00 & 0.40  \\ 
r2211\_02	& field	& 17:28:46.44	& 74:42:38.1	& 0.19443	& 18.45	& 0.983	   & -20.01 & -20.93 &  9.90 &   -10.03 & 0.38 &  -13.96 & 0.25  \\ 
r2211\_04	& field	& 17:29:01.00	& 74:40:07.9	& 0.27259	& 18.76	& 1.594	   & -19.96 & -21.47 & 10.28 &     6.43 & 0.75 &   -6.25 & 0.30  \\ 
r2211\_12	& field	& 17:29:45.21	& 74:40:22.0	& 0.15859	& 16.18	& 1.330	   & -21.36 & -22.63 & 10.68 &     .... & .... &    .... & ....  \\ 
r2211\_15	& field	& 17:29:53.77	& 74:39:44.1	& 0.15745	& 17.39	& 1.226	   & -20.31 & -21.42 & 10.15 &     .... & .... &   -6.20 & 0.25  \\ 
r2211\_16	& field & 17:29:59.01	& 74:41:08.1	& 0.33182	& 18.09	& 1.178	   & -21.57 & -22.66 & 10.65 &   -29.21 & 0.52 &  -30.22 & 1.14  \\ 
r2211\_18	& field	& 17:30:05.41	& 74:40:00.7	& 0.15708	& 18.60	& 1.324	   & -18.97 & -20.17 &  9.69 &     .... & .... &   -3.98 & 0.41  \\ 
r2211\_22	& field	& 17:30:22.64	& 74:39:20.9	& 0.34028	& 18.94	& 1.821	   & -20.31 & -21.95 & 10.49 &     2.62 & 0.59 &   -0.04 & 0.45  \\ 
r2211\_24	& field	& 17:30:45.34	& 74:41:21.3	& 0.31546	& 18.73	& 1.589	   & -20.44 & -21.91 & 10.45 &     0.20 & 0.46 &    2.05 & 0.41  \\ 
r2212\_12	& field	& 17:29:43.86	& 74:41:45.9	& 0.15985	& 18.38	& 0.975	   & -19.63 & -20.51 &  9.71 &   -30.89 & 0.95 &    .... & ....  \\ 
r2212\_14	& field	& 17:29:50.08	& 74:42:24.7	& 0.24585	& 17.73	& 1.429	   & -20.86 & -22.22 & 10.55 &    -6.09 & 0.50 &   -5.65 & 0.25  \\ 
r2212\_16	& field	& 17:30:00.75	& 74:42:09.0	& 0.27302	& 19.24	& 1.511	   & -19.56 & -21.00 & 10.08 &    -2.40 & 1.02 &    3.05 & 0.81  \\ 
r2212\_19	& field	& 17:30:18.14	& 74:41:44.4	& 0.33812	& 18.31	& 1.818	   & -20.91 & -22.54 & 10.73 &     2.78 & 0.72 &   -3.66 & 0.65  \\ 
r2221\_04	& field	& 17:27:06.87	& 74:32:15.8	& 0.28072	& 18.32	& 0.985	   & -21.03 & -22.01 & 10.33 &     .... & .... &  -46.66 & 1.04  \\ 
r2221\_13	& field	& 17:26:25.30	& 74:27:56.1	& 0.22819	& 18.78	& 1.348	   & -19.70 & -20.96 & 10.03 &    -8.97 & 1.04 &  -15.25 & 0.47  \\ 
r2222\_01	& field	& 17:26:43.68	& 74:35:29.1	& 0.50080	& 19.36	& 1.680	   & -21.34 & -22.69 & 10.71 &    -2.40 & 0.33 &    .... & ....  \\ 
r2222\_02	& field	& 17:27:26.70	& 74:34:42.8	& 0.18050	& 19.03	& 1.057	   & -19.21 & -20.16 &  9.60 &   -70.67 & 1.25 &  -13.85 & 0.18  \\ 
r2222\_04	& field	& 17:26:24.13	& 74:33:53.7	& 0.28905	& 19.12	& 1.227	   & -20.12 & -21.29 & 10.12 &    -7.08 & 0.30 &  -22.98 & 0.37  \\ 
r2222\_05	& field	& 17:27:22.22	& 74:32:27.3	& 0.24178	& 19.18	& 0.881	   & -19.94 & -20.76 &  9.78 &   -18.68 & 0.31 &  -19.77 & 0.22  \\ 
r2222\_06	& field	& 17:26:46.66	& 74:32:13.2	& 0.55816	& 19.24	& 1.698	   & -21.80 & -23.14 & 10.88 &   -10.19 & 0.27 &    .... & ....  \\ 
r2222\_08	& field	& 17:26:54.96	& 74:31:35.9	& 0.27059	& 18.07	& 1.020	   & -21.16 & -22.14 & 10.41 &    -0.55 & 0.24 &   10.24 & 0.62  \\ 
r2222\_09	& field	& 17:27:09.38	& 74:31:12.9	& 0.65900	& 19.26	& 2.406	   & -22.20 & -23.83 & 11.23 &    -5.89 & 0.68 &    0.00 & 0.00  \\ 
r2222\_12	& field	& 17:26:53.67	& 74:29:57.5	& 0.54749	& 19.25	& 2.087	   & -21.61 & -23.16 & 10.95 &    -3.17 & 0.47 &    .... & ....  \\ 
r2222\_13	& field	& 17:26:33.10	& 74:29:40.5	& 0.29010	& 18.91	& 1.662	   & -19.93 & -21.49 & 10.30 &    -5.43 & 0.36 &    2.56 & 0.24  \\ 
r2222\_15	& field	& 17:26:10.37	& 74:29:09.5	& 0.18043	& 18.17	& 1.046	   & -20.08 & -21.02 &  9.94 &   -25.86 & 0.34 &  -29.13 & 0.18  \\ 
r2222\_17	& field	& 17:26:59.73	& 74:27:27.3	& 0.41500	& 19.26	& 1.977	   & -20.60 & -22.26 & 10.62 &     1.51 & 0.37 &  -10.05 & 0.41  \\ 
r2231\_06	& field	& 17:24:14.39	& 74:29:43.0	& 0.21837	& 19.06	& 0.920	   & -19.78 & -20.61 &  9.73 &   -25.41 & 0.51 &  -26.11 & 0.43  \\ 
r2231\_07	& field	& 17:24:20.53	& 74:28:13.3	& 0.54921	& 19.14	& 1.646	   & -21.89 & -23.17 & 10.87 &    -8.90 & 0.25 &    .... & ....  \\ 
r2231\_11	& field	& 17:24:37.73	& 74:30:13.6	& 0.18489	& 19.49	& 1.168	   & -18.65 & -19.75 &  9.49 &   -48.04 & 0.78 &  -32.86 & 0.40  \\ 
r2231\_13	& field	& 17:24:57.93	& 74:29:05.6	& 0.43546	& 18.32	& 1.913	   & -21.77 & -23.33 & 11.03 &     0.84 & 0.23 &    .... & ....  \\ 
r2231\_16	& field	& 17:25:12.29	& 74:29:40.6	& 0.29537	& 19.17	& 1.210	   & -20.12 & -21.26 & 10.11 &   -19.73 & 0.31 &  -48.69 & 0.45  \\ 
r2231\_23	& field	& 17:25:54.24	& 74:26:49.6	& 0.34361	& 18.36	& 1.546	   & -21.11 & -22.52 & 10.67 &    -1.55 & 0.29 &   -8.67 & 0.65  \\ 
r2241\_02	& field	& 17:23:27.27	& 74:46:43.8	& 0.44623	& 18.36	& 2.172	   & -21.68 & -23.40 & 11.09 &     4.18 & 0.47 &    .... & ....  \\ 
r2241\_04	& field	& 17:24:15.13	& 74:45:07.0	& 0.43284	& 18.39	& 1.686	   & -21.81 & -23.21 & 10.94 &    -7.59 & 0.32 &    .... & ....  \\ 
r2241\_11	& field	& 17:23:55.07	& 74:42:40.0	& 0.33892	& 18.45	& 1.800	   & -20.79 & -22.40 & 10.66 &    -9.00 & 0.91 &    0.97 & 0.44  \\ 
r2241\_14	& field	& 17:24:12.80	& 74:40:47.1	& 0.18050	& 17.48	& 1.381	   & -20.33 & -21.65 & 10.31 &    -5.11 & 2.25 &    0.67 & 0.27  \\ 
r2241\_20	& field	& 17:23:50.33	& 74:38:39.8	& 0.04373	& 18.69	& 1.516	   & -15.32 & -16.93 &  8.48 &     .... & .... &  -11.48 & 0.19  \\ 
r2241\_21	& field	& 17:23:54.90	& 74:38:24.1	& 0.26848	& 17.01	& 0.754	   & -22.46 & -23.19 & 10.70 &   -17.10 & 0.32 &  -36.41 & 0.29  \\ 
r2241\_22	& field	& 17:23:47.22	& 74:38:01.9	& 0.24012	& 17.91	& 1.212	   & -20.84 & -21.98 & 10.39 &     2.47 & 0.58 &    1.78 & 0.18  \\ 
r2242\_03	& field	& 17:23:39.55	& 74:45:22.9	& 0.42995	& 19.34	& 1.555	   & -20.90 & -22.23 & 10.52 &   -13.05 & 0.48 &  -50.55 & 1.95  \\ 
r2242\_07	& field	& 17:23:37.38	& 74:43:40.0	& 0.29567	& 18.02	& 1.610	   & -20.93 & -22.43 & 10.66 &     1.50 & 0.67 &    0.67 & 0.25  \\ 
r2242\_10	& field	& 17:23:51.86	& 74:42:56.3	& 0.20966	& 19.04	& 1.696	   & -18.79 & -20.45 &  9.90 &     3.91 & 2.31 &   -0.35 & 0.28  \\ 
r2242\_11	& field	& 17:24:02.24	& 74:42:37.0	& 0.33906	& 18.85	& 1.709	   & -20.46 & -22.00 & 10.49 &    -0.95 & 0.99 &   -3.18 & 0.40  \\ 
r2242\_12	& field	& 17:23:13.33	& 74:42:15.4	& 0.29681	& 18.90	& 1.691	   & -19.99 & -21.57 & 10.33 &    -0.22 & 0.90 &    1.00 & 0.36  \\ 
r2242\_13	& field	& 17:23:17.51	& 74:40:37.5	& 0.18069	& 18.01	& 1.095	   & -20.18 & -21.18 & 10.02 &    -5.53 & 0.39 &  -21.72 & 0.21  \\ 
r2251\_02	& field	& 17:23:53.84	& 74:23:11.3	& 0.19508	& 18.20	& 1.284	   & -19.89 & -21.15 & 10.09 &     4.56 & 2.45 &    0.26 & 0.24  \\ 
r2251\_06	& field	& 17:23:50.15	& 74:20:50.9	& 0.05868	& 18.68	& 0.877	   & -17.14 & -17.94 &  8.61 &     .... & 0.00 &  -26.72 & 0.30  \\ 
r2251\_08	& field	& 17:24:12.25	& 74:19:17.4	& 0.22829	& 18.58	& 0.903	   & -20.39 & -21.21 &  9.96 &   -20.21 & 0.80 &  -27.68 & 0.36  \\ 
r2251\_10	& field	& 17:24:04.79	& 74:18:39.2	& 0.44273	& 18.39	& 0.822	   & -22.37 & -23.16 & 10.67 &    -5.02 & 0.14 &  -21.15 & 0.35  \\ 
r2251\_11	& field	& 17:25:05.54	& 74:17:51.8	& 0.04355	& 16.66	& 1.012	   & -18.29 & -19.21 &  9.17 &     .... & 0.00 &  -10.18 & 0.16  \\ 
r2251\_15	& field	& 17:24:45.55	& 74:15:44.8	& 0.06325	& 18.35	& 0.743	   & -17.81 & -18.51 &  8.77 &     .... & 0.00 &  -11.96 & 0.30  \\ 
r2251\_16	& field	& 17:24:54.82	& 74:15:13.7	& 0.29141	& 16.71	& 1.161	   & -22.60 & -23.67 & 11.03 &   -12.12 & 0.75 &   -1.42 & 0.25  \\ 
r2611\_02	& field	& 13:12:15.56	& 32:33:06.4	& 0.26403	& 18.93	& 1.349	   & -20.04 & -21.39 & 10.17 &    -8.30 & 0.40 &  -23.45 & 0.37  \\ 
r2611\_03	& field	& 13:12:10.01	& 32:32:04.4	& 0.13084	& 18.81	& 1.000	   & -18.62 & -19.70 &  9.35 &   -24.90 & 0.61 &  -28.72 & 0.26  \\ 
r2611\_07	& field	& 13:12:02.58	& 32:31:35.8	& 0.43572	& 18.12	& 2.142	   & -21.96 & -23.59 & 11.15 &     1.37 & 0.36 &   -1.01 & 0.55  \\ 
r2611\_11	& field	& 13:11:54.78	& 32:30:38.6	& 0.15710	& 18.72	& 0.935	   & -19.24 & -20.21 &  9.49 &   -13.09 & 0.37 &  -10.49 & 0.23  \\ 
r2611\_16	& field	& 13:11:46.33	& 32:32:54.9	& 0.35143	& 18.08	& 1.796	   & -21.23 & -22.78 & 10.83 &     3.52 & 0.33 &    3.01 & 0.31  \\ 
r2612\_01	& field	& 13:12:28.82	& 32:30:27.7	& 0.26270	& 19.01	& 1.600	   & -19.73 & -21.25 & 10.23 &     0.00 & 0.27 &   -0.05 & 0.18  \\ 
r2612\_03	& field	& 13:12:18.72	& 32:30:40.0	& 0.49186	& 19.47	& 1.650	   & -21.44 & -22.44 & 10.39 &   -12.58 & 0.18 &    .... & ....  \\ 
r2612\_09	& field	& 13:12:04.96	& 32:31:56.8	& 0.36498	& 19.11	& 1.346	   & -20.85 & -22.13 & 10.47 &   -11.92 & 0.19 &  -14.38 & 0.26  \\ 
r2612\_11	& field	& 13:12:01.09	& 32:31:31.0	& 0.43522	& 19.02	& 2.049	   & -21.37 & -22.91 & 10.92 &     1.11 & 0.22 &    1.50 & 0.37  \\ 
r2612\_12	& field	& 13:11:59.85	& 32:31:24.2	& 0.43773	& 18.74	& 2.075	   & -21.56 & -23.13 & 10.93 &     1.68 & 0.20 &    0.00 & 0.33  \\ 
r2612\_14	& field	& 13:11:52.16	& 32:32:12.1	& 0.44078	& 18.83	& 2.091	   & -21.50 & -22.91 & 10.86 &     2.98 & 0.21 &   -5.77 & 0.32  \\ 
r2612\_19	& field	& 13:11:42.75	& 32:32:26.2	& 0.14883	& 16.72	& 1.208	   & -21.00 & -22.09 & 10.33 &   -13.62 & 0.15 &  -20.03 & 0.08  \\ 
r2621\_02	& field	& 13:10:57.07	& 32:28:05.7	& 0.12024	& 18.68	& 1.209	   & -18.33 & -19.57 &  9.51 &    -1.12 & 0.60 &  -10.26 & 0.13  \\ 
r2621\_05	& field	& 13:11:04.39	& 32:32:09.8	& 0.15807	& 18.69	& 0.962	   & -19.38 & -20.34 &  9.36 &     .... & .... &   -1.37 & 0.13  \\ 
r2621\_06	& field	& 13:11:05.84	& 32:29:57.6	& 0.30678	& 18.87	& 1.875	   & -19.98 & -21.76 & 10.45 &    -3.45 & 0.27 &   -1.38 & 0.17  \\ 
r2621\_09	& field	& 13:11:10.48	& 32:29:39.1	& 0.23464	& 19.41	& 1.383	   & -19.30 & -20.53 &  9.82 &    -2.30 & 0.30 &   -7.58 & 0.19  \\ 
r2621\_10	& field	& 13:11:12.39	& 32:32:06.0	& 0.30175	& 18.37	& 1.359	   & -21.15 & -22.32 & 10.49 &    -9.52 & 0.17 &  -12.55 & 0.16  \\ 
r2621\_12	& field	& 13:11:14.83	& 32:30:59.0	& 0.30721	& 18.38	& 1.630	   & -21.00 & -22.43 & 10.66 &    -3.89 & 0.15 &   -3.95 & 0.13  \\ 
r2621\_18	& field	& 13:11:28.25	& 32:28:06.1	& 0.43404	& 18.58	& 2.144	   & -21.45 & -23.28 & 11.03 &    -4.27 & 0.23 &    .... & ....  \\ 
r2631\_02	& field	& 13:10:15.66	& 32:27:22.0	& 0.18490	& 16.26	& 1.483	   & -21.62 & -23.04 & 10.87 &    -4.72 & 0.21 &   -0.55 & 0.09  \\ 
r2631\_03	& field	& 13:10:16.97	& 32:29:06.1	& 0.12306	& 16.99	& 1.228	   & -20.25 & -21.28 & 10.02 &     .... & .... &  -11.70 & 0.27  \\ 
r2631\_04	& field	& 13:10:18.44	& 32:27:39.6	& 0.12364	& 17.01	& 1.292	   & -19.89 & -21.28 & 10.08 &     .... & .... &  -25.65 & 0.19  \\ 
r2631\_05	& field	& 13:10:19.66	& 32:29:34.4	& 0.25985	& 18.90	& 1.514	   & -20.00 & -21.00 &  9.75 &   -57.95 & 1.65 &   -8.50 & 0.40  \\ 
r2631\_10	& field	& 13:10:28.21	& 32:26:18.3	& 0.15738	& 18.66	& 0.987	   & -19.22 & -20.26 &  9.63 &   -31.05 & 0.58 &    0.25 & 0.35  \\ 
r2632\_01	& field	& 13:10:09.53	& 32:26:28.6	& 0.12557	& 18.07	& 1.184	   & -19.04 & -20.25 &  9.73 &     .... & .... &    .... & ....  \\ 
r2632\_04	& field	& 13:10:20.31	& 32:27:23.8	& 0.12333	& 19.12	& 0.829	   & -18.20 & -19.11 &  9.14 &     .... & .... &  -23.07 & 0.18  \\ 
r2632\_05	& field	& 13:10:23.46	& 32:29:52.4	& 0.40813	& 19.41	& 1.099	   & -20.92 & -21.75 & 10.00 &   -36.95 & 0.27 &    .... & ....  \\ 
r2632\_06	& field	& 13:10:24.87	& 32:27:36.1	& 0.18508	& 17.35	& 1.446	   & -20.46 & -21.94 & 10.48 &    -1.76 & 0.27 &   -2.10 & 0.11  \\ 
r2632\_14	& field	& 13:10:38.86	& 32:28:04.1	& 0.30783	& 19.37	& 1.571	   & -19.69 & -21.14 & 10.06 &   -23.37 & 1.36 &  -16.20 & 1.24  \\ 
r2632\_18	& field	& 13:10:50.60	& 32:30:19.7	& 0.40243	& 19.13	& 1.944	   & -20.62 & -22.54 & 10.79 &    -1.45 & 0.23 &    .... & ....  \\ 
r2641\_11	& field	& 13:10:03.30	& 32:21:30.2	& 0.28419	& 18.66	& 1.508	   & -20.38 & -21.83 & 10.44 &   -13.38 & 1.20 &  -39.16 & 2.26  \\ 
r2651\_01	& field	& 13:10:32.47	& 32:19:15.3	& 0.28500	& 19.42	& 1.513	   & -19.47 & -20.90 & 10.10 &     1.53 & 0.34 &    2.22 & 0.30  \\ 
r2651\_03	& field	& 13:10:40.07	& 32:20:47.8	& 0.55177	& 18.00	& 1.669	   & -23.20 & -24.32 & 11.22 &   -14.19 & 0.11 &    .... & ....  \\ 
r2651\_04	& field	& 13:10:42.49	& 32:16:34.9	& 0.12691	& 16.04	& 1.423	   & -20.79 & -22.27 & 10.59 &    -0.13 & 0.19 &   -3.65 & 0.04  \\ 
r2651\_07	& field	& 13:10:46.66	& 32:21:16.7	& 0.30771	& 18.66	& 1.317	   & -20.69 & -22.01 & 10.44 &     0.53 & 0.18 &    0.38 & 0.17  \\ 
r2651\_10	& field	& 13:10:56.84	& 32:17:15.9	& 0.14591	& 18.12	& 1.154	   & -19.38 & -20.58 &  9.86 &     .... & .... &   -0.87 & 0.11  \\ 
r2651\_11	& field	& 13:10:59.78	& 32:18:39.0	& 0.40701	& 18.66	& 1.716	   & -21.32 & -22.73 & 10.76 &    -3.53 & 0.18 &    .... & ....  \\ 
r2651\_12	& field	& 13:11:01.13	& 32:18:08.6	& 0.40683	& 17.94	& 2.069	   & -21.73 & -23.53 & 11.15 &     2.65 & 0.18 &    .... & ....  \\ 
r2651\_13	& field	& 13:11:07.05	& 32:17:24.8	& 0.40657	& 18.66	& 2.031	   & -21.48 & -22.96 & 10.85 &     2.88 & 0.24 &    .... & ....  \\ 
r2651\_16	& field	& 13:11:12.51	& 32:18:42.8	& 0.60652	& 19.00	& 1.880	   & -22.22 & -23.46 & 10.78 &   -10.59 & 0.22 &    .... & ....  \\ 
r2811\_12a	& field	& 09:43:58.66	& 16:43:04.5	& 0.16614	& 18.56	& 0.943	   & -19.73 & -20.44 &  9.39 &     .... & .... &  -45.42 & 1.74  \\ 
r2812\_19	& field	& 09:43:51.27	& 16:43:11.2	& 0.15433	& 18.95	& 0.659	   & -18.88 & -19.72 &  9.52 &   -26.92 & 1.21 &  -34.63 & 0.75  \\ 
r2812\_23	& field	& 09:43:55.30	& 16:44:48.9	& 0.16487	& 18.46	& 1.398	   & -19.09 & -20.51 &  9.91 &    12.44 & 7.53 &    4.62 & 0.50  \\ 
r2812\_25	& field	& 09:43:55.77	& 16:45:33.8	& 0.15788	& 19.10	& 1.155	   & -18.61 & -19.84 &  9.61 &    -1.11 & 7.84 &    .... & ....  \\ 
r2812\_26	& field	& 09:43:51.93	& 16:45:45.2	& 0.21565	& 19.33	& 1.746	   & -18.67 & -20.54 &  9.73 &     .... & .... &   -0.58 & 1.05  \\ 
r2821\_05	& field	& 09:43:53.29	& 16:41:20.1	& 0.07725	& 17.93	& 0.764	   & -16.26 & -17.16 &  8.22 &     .... & .... &  -69.03 & 0.65  \\ 
r2821\_23	& field	& 09:43:28.76	& 16:37:53.2	& 0.18995	& 18.08	& 1.182	   & -20.06 & -21.27 & 10.19 &     0.77 & 2.50 &    3.92 & 1.62  \\ 
r2821\_28	& field	& 09:43:22.54	& 16:38:53.6	& 0.27333	& 17.90	& 1.451	   & -21.09 & -22.48 & 10.67 &    -5.03 & 1.22 &  -19.08 & 0.83  \\ 
r2822\_24	& field	& 09:43:23.20	& 16:40:38.3	& 0.16715	& 18.83	& 1.230	   & -18.89 & -20.24 &  9.70 &   -21.56 & 1.01 &  -28.55 & 0.41  \\ 
r2831\_05	& field	& 09:43:19.89	& 16:42:29.8	& 0.16724	& 17.58	& 1.149	   & -20.30 & -21.52 & 10.06 &   -11.15 & 2.29 &  -20.93 & 0.78  \\ 
r2831\_08	& field	& 09:43:12.49	& 16:44:30.0	& 0.17047	& 17.87	& 1.241	   & -19.84 & -21.31 & 10.25 &     .... & .... &   -8.77 & 0.62  \\ 
r2831\_14	& field	& 09:43:00.57	& 16:42:17.0	& 0.14972	& 16.86	& 1.207	   & -20.65 & -21.96 & 10.44 &     3.67 & 1.17 &   -7.81 & 0.24  \\ 
r2831\_16	& field	& 09:42:56.10	& 16:41:13.4	& 0.23116	& 18.63	& 1.252	   & -20.17 & -21.21 &  9.79 &    -9.85 & 1.60 &  -20.22 & 0.95  \\ 
r2831\_17	& field	& 09:42:52.23	& 16:42:47.2	& 0.16275	& 17.88	& 1.043	   & -20.13 & -21.18 &  9.89 &   -21.33 & 1.13 &   -7.25 & 0.45  \\ 
r2831\_18	& field	& 09:42:51.23	& 16:41:08.8	& 0.23352	& 17.07	& 1.118	   & -21.63 & -22.86 & 10.70 &    -5.96 & 0.39 &  -22.38 & 0.31  \\ 
r2841\_01	& field	& 09:44:58.57	& 16:30:07.6	& 0.22351	& 19.41	& 1.424	   & -18.65 & -20.22 &  9.86 &     0.31 & 3.79 &    1.67 & 0.85  \\ 
r2841\_09	& field	& 09:44:38.51	& 16:27:52.5	& 0.23310	& 17.25	& 1.330	   & -21.14 & -22.58 & 10.74 &     1.08 & 0.92 &   -3.37 & 0.23  \\ 
r2841\_12	& field	& 09:44:33.83	& 16:30:53.1	& 0.27388	& 19.33	& 0.753	   & -19.68 & -20.73 &  9.93 &   -39.60 & 1.62 &  -51.65 & 3.45  \\ 
r2841\_13	& field	& 09:44:32.35	& 16:28:34.4	& 0.15980	& 18.91	& 1.039	   & -18.87 & -20.08 &  9.47 &    -8.75 & 1.07 &    .... & ....  \\ 
r2841\_18	& field	& 09:44:21.95	& 16:30:48.2	& 0.23268	& 19.19	& 1.459	   & -19.13 & -20.61 &  9.85 &     2.60 & 1.96 &   -2.44 & 0.43  \\ 
r2841\_19a	& field	& 09:44:20.04	& 16:30:28.5	& 0.23338	& 17.51	& 1.290	   & -21.03 & -22.49 & 10.63 &    -2.40 & 0.47 &   -3.95 & 0.15  \\ 
r2851\_01b	& field	& 09:43:57.19	& 16:25:17.4	& 0.22289	& 19.09	& 1.099	   & -19.41 & -20.69 &  9.99 &   -26.49 & 0.94 &  -33.61 & 0.94  \\ 
r2851\_06	& field	& 09:44:07.29	& 16:29:22.8	& 0.23247	& 17.24	& 1.355	   & -21.25 & -22.70 & 10.73 &     1.15 & 0.69 &   -1.32 & 0.17  \\ 
r2851\_13	& field	& 09:43:58.24	& 16:32:18.3	& 0.22294	& 18.11	& 1.310	   & -20.32 & -21.94 & 10.40 &     0.25 & 1.08 &    0.83 & 0.29  \\

\end{longtable}
\end{scriptsize}
}

\end{document}